\documentclass[twocolumn]{aastex631}
\usepackage[dvipsnames]{xcolor}
\usepackage{amsmath}
\usepackage{booktabs}
\usepackage{float}
\usepackage{CJK}

\newcommand{\astrid}{\textsc{ASTRID}}
\newcommand{\Msun}{$M_{\odot}$}
\newcommand{\revise}[1]{{#1}}

\begin{document}
\begin{CJK*}{UTF8}{bsmi}

\title{The ASTRID Simulation at $z=0$: From Massive Black Holes to Large-scale Structure}

\author[0000-0002-8828-8461]{Yihao Zhou (周亦豪)}
\affiliation{McWilliams Center for Cosmology, Department of Physics, Carnegie Mellon University, Pittsburgh, PA 15213, USA}

\author[0000-0002-6462-5734]{Tiziana Di Matteo}
\affiliation{McWilliams Center for Cosmology, Department of Physics, Carnegie Mellon University, Pittsburgh, PA 15213, USA}

\author[0000-0001-5803-5490]{Simeon Bird}
\affiliation{Department of Physics \& Astronomy, University of California, Riverside, 900 University Avenue, Riverside, CA 92521, USA}

\author[0000-0003-0697-2583]{Rupert Croft}
\affiliation{McWilliams Center for Cosmology, Department of Physics, Carnegie Mellon University, Pittsburgh, PA 15213, USA}

\author[0000-0001-7899-7195]{Yueying Ni}
\affiliation{Center for Astrophysics $\vert$ Harvard \& Smithsonian, Cambridge, MA 02138, US}

\author[0000-0001-6221-6024]{Yanhui Yang}
\affiliation{Department of Physics \& Astronomy, University of California, Riverside, 900 University Ave., Riverside, CA 92521, USA}

\author[0000-0001-6627-2533]{Nianyi Chen}
\affiliation{School of Natural Sciences, Institute for Advanced Study, Princeton, NJ 08540, USA}

\author[0009-0006-7511-0329]{Patrick Lachance}
\affiliation{McWilliams Center for Cosmology, Department of Physics, Carnegie Mellon University, Pittsburgh, PA 15213, USA}

\author[0009-0008-0370-6021]{Xiaowen Zhang}
\affiliation{McWilliams Center for Cosmology, Department of Physics, Carnegie Mellon University, Pittsburgh, PA 15213, USA}

\author{Fatemeh Hafezianzadeh}
\affiliation{McWilliams Center for Cosmology, Department of Physics, Carnegie Mellon University, Pittsburgh, PA 15213, USA}

\correspondingauthor{Yihao Zhou}
\email{yihaoz@andrew.cmu.edu}



\begin{abstract}

We present the $z=0$ results for the cosmological simulation \astrid. 
Hosting $2\times 5500^3\approx$ 0.33 trillion particles in a box of $370\, {\rm Mpc}$ per side, \astrid\ is one of the largest cosmological hydrodynamic simulations evolved to $z=0$. 
\astrid\ features a large population of massive black holes (MBHs), covering a wide mass range $4\times10^{4}\sim 2\times 10^{11}$~\Msun. 
The adopted dynamical friction model provides a relatively accurate description of MBH dynamics, making \astrid\ a powerful tool to study MBH growth and mergers in a cosmological context.
\astrid\ successfully captures the coevolution of MBHs and their host galaxies, producing $M_{\rm BH}-M_{\star}$ and $M_{\rm BH}-\sigma$ relations in good agreement with observations. 
Notably, \astrid\ generates scatter in these relations that is more consistent with observations than previous simulations, indicating a more realistic MBH diversity. 
The galaxy stellar mass function at $z=0$ is generally consistent with observational constraints. 
When dust attenuation is applied, the galaxy luminosity function also agrees well with observations, and the bimodality in galaxy colors is reproduced as well. 
\astrid\ hosts a large population of massive galaxy groups and clusters: 7 halos have $M_{\rm 200c}>10^{15}$~\Msun, and 9709 halos have $M_{\rm 200c}>10^{13}$~\Msun.
We quantify the stellar mass content in these halos, 
and find that the correlations between the stellar and halo mass match well with observational constraints.
Finally, we present the $z=0$ power spectra of MBH and galaxies, as well as their bias with respect to the matter power spectrum. We find that MBHs with $M_{\rm BH}\geq 10^{8}$~\Msun\ and galaxies with $M_{\star}\geq 10^{10.5}$~\Msun\ serve as good tracers of large-scale structure. 

\end{abstract}

\keywords{}

\section{Introduction} \label{sec:intro}

Large-scale cosmological simulations are continually pushing computational frontiers to understand galaxy formation in a cosmological context.
Many simulations have successfully produced realistic galaxy populations on cosmological scales, such as Illustris \citep{Vogelsberger2014, Vogelsberger2014a}, 
EAGLE \citep{Schaye2015}, 
MassiveBlack-II \citep{Khandai2015_massivebhII},
BlueTides \citep{Feng2016_bluetides}, 
Horizon-AGN \citep{Dubois2016}, 
Romulus \citep{Tremmel2017},
Magneticum \citep{Dolag2017},
IllustrisTNG \citep{Pillepich2018, Nelson2018, Springel2018, Marinacci2018, Naiman2018}, SIMBA \citep{Dave2019}, NewHorizon \citep{Dubois2021}, and
FLAMINGO \citep{Schaye2023} (see \citet{Vogelsberger2020} for a review).  
Compared to dark-matter-only simulations, these large-volume hydrodynamical simulations 
offer a more complete description of cosmic structures by including a variety of gas, stellar, and active galactic nucleus (AGN) physics.
With the inclusion of well-calibrated subgrid models, cosmological simulations provide detailed information about the origins and statistics of the galaxy population. Specific observational features that have been reproduced include clustering bias \citep{Artale2017}, two-point correlation functions \citep{Springel2018}, galaxy colors \citep{Nelson2018}, galaxy morphologies \citep{Pillepich2019}, galaxy chemical history \citep{Torrey2018}, merger history \citep{Eisert2023}.

In addition to galaxy properties, a cosmological simulation is also a powerful tool for studying the population of massive black holes (MBHs), and the subset of them visible as AGN. 
A variety of observational evidence confirms that massive black holes are ubiquitous in the Universe, residing at the center of almost all massive galaxies \citep{Tremaine2002, Kormendy2013}. 
MBHs coevolve with their local environment via gas accretion and energetic feedback, and play a pivotal role in shaping the cosmic landscape we observe today. MBHs are usually treated as collisionless sink particles in the simulations mentioned above, particles which are able to accrete surrounding gas, release energy into neighboring particles, and merge with other MBHs \citep{DiMatteo2005_BH_model, Weinberger2018}. 
In large-volume simulations, due to the limits of achievable resolution, the MBH accretion disk cannot be fully resolved. Hence the aforementioned accretion, feedback, and merging are all accomplished through subgrid models. There is still no consensus on the details of these MBH models, and models adopted in different simulations vary widely (see \citet{Habouzit2022} for a recent review).

The next decade promises an unprecedented influx of observational data related to MBHs. 
With the James Webb Space Telescope (JWST) opening up a new frontier for high-z AGN science, MBHs formed in the first billion years after the Big Bang are now being observed \citep{Ubler2023_highz_QSO, Matthee2024_highzqso}. 
This data provides valuable insights into theories for how  MBHs are seeded (e.g.,
\citealt{jeon202}).
Scaling relations derived from local galaxies show that these MBHs are overmassive compared to their hosts, posing challenges to the existing theory of the formation and growth of MBHs. 
The next generation of X-ray missions, e.g., Athena \citep{Nandra2013}, AXIS \citep{Mushotzky2018}, and LynX \citep{TheLynxTeam2018}, will increase the current X-ray flux sensitivity by at least 1 order of magnitude, and aim to explore larger areas and the deeper Universe.  
They will be able to identify compact MBH binaries at small separations. Moreover, upcoming surveys such as the Roman High Latitude Survey \citep{Wang2022_HLS} and the Vera Rubin Observatory Legacy Survey of Space and Time \citep[LSST:][]{Ivezic2019} will reveal a broader and fainter MBH population.

The recently inaugurated era of gravitational-wave (GW) astronomy has opened up new avenues for understanding the intricate nature of MBH formation and interactions.
Recent reports by the Pulsar Timing Array (PTA) collaborations (NANOGrav \citep{Agazie2023_nanograv_pta}, EPTA+InPTA \citep{EPTACollaboration2023_EPTA_GWB}, and CPTA \citep{Xu2023_CPTA}) provide evidence for a stochastic gravitational-wave background (GWB) at nanohertz frequencies.
The primary source of the GWB at these frequencies is believed to be the population of MBH mergers with mass $M_{\mathrm{BH}}\gtrsim10^{8}$~\Msun\ \citep{Burke-Spolaor2019}. 
Continuous waves (CWs) from individual MBH binaries are expected to be observed in the next few years \citep{Rosado2015, Kelley2018_SS, Mingarelli2017}.
Additionally, the Laser Interferometer Space Antenna (LISA) \citep{Amaro-Seoane2017_LISA}, expected to launch in the mid-2030s, is set to explore GW signals emitted at millihertz frequencies, targeting  MBH mergers at the lower-mass end ($M_{\mathrm{BH}}\lesssim 10^{8}$~\Msun) \citep{Katz2020, Wang2025}.

Fully leveraging the wealth of GW and EM observations in the near future requires a cosmological simulation that hosts a large population of MBH mergers, spans a wide mass range and various galactic environments, and incorporates improved models for MBH dynamics. 
In this work, we present the \astrid\ simulation at $z=0$. Below, we briefly list some of \astrid 's key features. 

\begin{enumerate}
    \item \texttt{Large particle number}: The simulation initial conditions include  $3.3\times 10^{11}$ particles: $5500^{3}$ DM particles and $5500^{3}$ smoothed particle hydrodynamics (SPH) mass elements. 
    Evolving the simulation from $z=99$ to $z=0$ consumed about 896M CPU hours on the Frontera facility at the Texas Advanced Computing Center.
    This makes \astrid\ one of the largest cosmological hydrodynamical simulations in terms of particle number that have been evolved to $z=0$. 

    \item \texttt{Large MBH population}: 
    \astrid\ covers a BH mass range spanning almost 7 orders of magnitude, and hosts over 3 million MBH merger events.
    On the one hand, \astrid\ adopts a relatively small BH seed mass, down to $3\times10^{4}\ M_{\odot}/h$, enabling us to study  MBH seed mergers, the main target of the upcoming LISA mission. On the other hand, the relatively large simulation volume samples the sparse massive end of the MBH mass function.
    The most-massive black hole in \astrid\ is $2\times 10^{11}\ M_{\odot}$.
    These MBHs undergo mergers which will contribute to the PTA GW background and are likely to be detected as PTA CW sources \citep{Chen2025, Zhou2025_PTA_CW}.

\begin{figure*}
    \centering
    \includegraphics[width=1\linewidth]{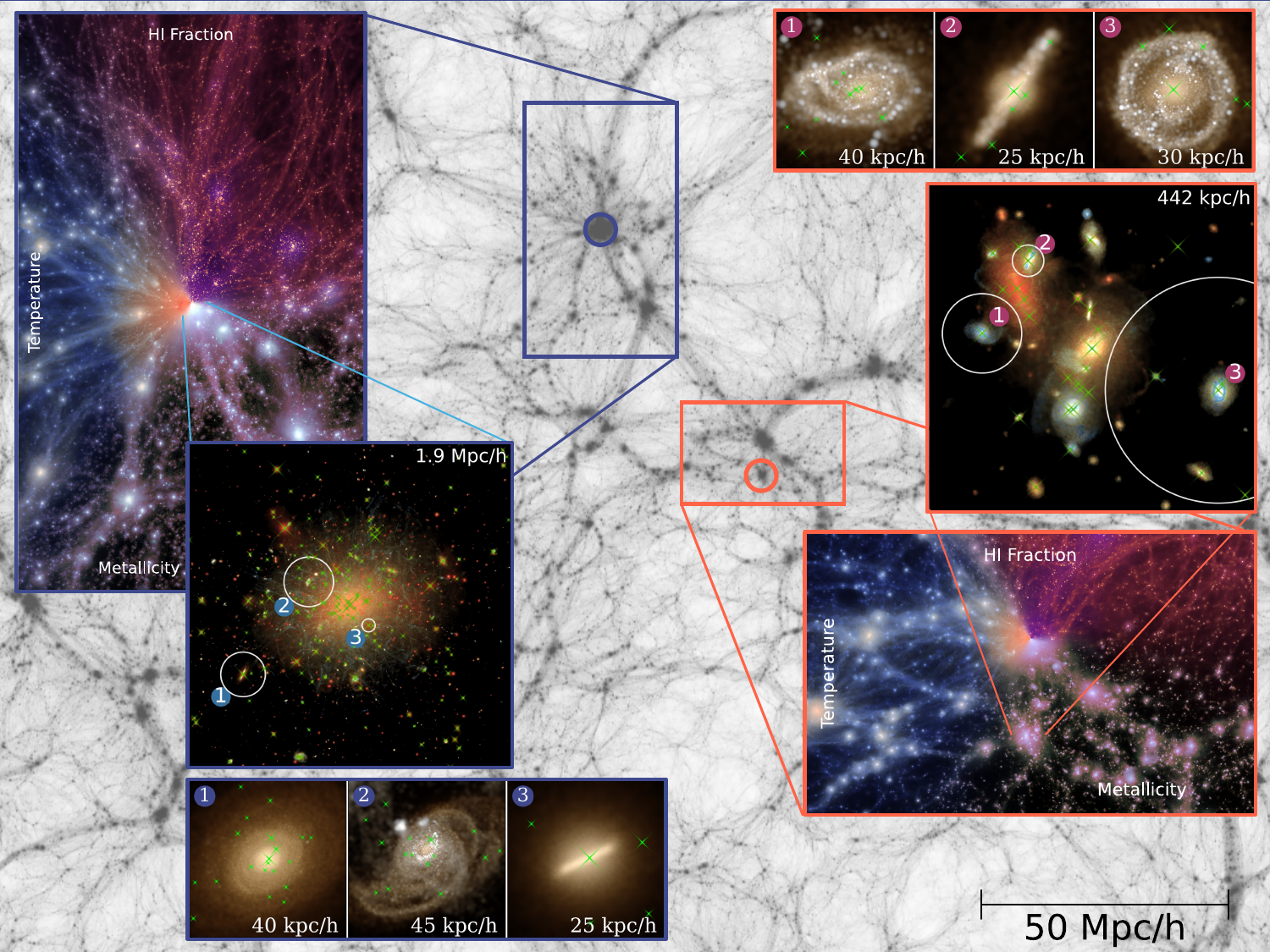}
    \caption{Visualization of the \astrid\ simulation at $z=0$. The underlying background is the dark matter density through a slab $25\, {\rm cMpc}\,h^{-1}$ thick (i.e., 10\% of the simulation box length). The panel spans $250\times 187.5\ {\rm cMpc}^{2}\,h^{-2}$, covering 75\% of the full $250\times250\,{\rm cMpc}^{2}\,h^{-2}$ face. 
    We further zoom in on two regions within this slice. 
    \textit{Blue zoom (left):} region centered on the most-massive halo in \astrid\ ($\log M_{\rm 200c}/M_{\odot}=15.3$), whose position is marked by the blue circle. 
    The leftmost inset shows a trichromatic gas map in which hue encodes temperature (left), neutral hydrogen fraction (top), and metallicity (bottom).
    The left-middle inset zooms to a $1.9\ {\rm cMpc}\,h^{-1}$ box and displays the stellar density colored by stellar age: red hues correspond to older stellar populations, and blue hues represent the younger ones.  
    Green crosses indicate the black holes with $M_{\rm BH}\geq 8\times 10^{6}$~\Msun. 
    Three example galaxies are highlighted: their virial radii are denoted by the white circles, and their mock observations for the Hubble Space Telescope (HST) Wide Field Camera (WFC3) are shown in the lower row. 
    The red, green, and blue channels show the F625W, F475W, and F390W filters, respectively.
    \textit{Red zoom (right):} region around a $\log M_{\rm 200c} = 14.8$~\Msun\ halo. The red circle marks a nearby star-forming halo with $\log M_{\rm 200c}=14.0$~\Msun. 
    As for the blue zoom, we show the gas tricolor map (right-bottom inset), the stellar age field (right-middle inset), and mock HST images of three member galaxies (right-top inset) within the star-forming halo. 
    }\label{fig:large_scale}
\end{figure*}

    \item \texttt{Better MBH Dynamics}: Instead of adopting the widely used BH repositioning algorithm \citep{Weinberger2018,Crain2015, Dave2019, Sijacki2015}, \astrid\ includes a subgrid model to compensate for the dynamical friction contributed by those stars and BHs which surround the MBH particles, but are below the resolution limit of the simulation. 
    This model has been validated against both semianalytical predictions and high-resolution simulations \citep{Genina2024, Zhou2025}. 
    It not only provides a better description of MBH dynamics, but also leads to realistic MBH velocities and trajectories. Because of this, the MBH orbital eccentricity, an important indicator of the binary hardening timescale, can be measured for MBH binaries \citep{Chen2023_dualAGN, Wang2025}.
\end{enumerate}

\revise{A variety of papers have used \astrid\ simulation data at high redshift. 
\citet{Ni2022_astrid} presented the \astrid\ MBH population at $z\geq 3$, and \citet{Ni2024} specifically showed the influence of AGN feedback on galaxy evolution. 
\citet{Chen2023_dual} examined the properties and cosmic evolution of dual AGNs.
\citet{lachance2024} investigated the galaxy morphology at $3\leq z\leq 6$.
Using the galaxy populations at $5\leq z \leq 8$, 
\citet{lachance2025} identified the counterparts to the ``little red dot'' (LRD) objects detected by JWST. 
The \astrid\ MBH binary population down to $z=0$ has been used to make predictions for future GW detectors. 
\citet{Chen2025} estimated the GWB signals and found that \astrid\ produces a GWB lower than the constraints given by NANOGrav 15 yr data \citep{Agazie2023_nanograv_pta}, and consistent with the prediction based on local EM observations \citep[e.g][]{Sato-Polito2025}.
\citet{Zhou2025_PTA_CW} searched for PTA CW sources in \astrid, and found that all high-detectability CW sources arise from the central galaxies of massive galaxy clusters. 
\citet{Wang2025} calculated the LISA signal for each MBH merger in \astrid\ and estimated the LISA detection rate. 
}

\revise{This paper focuses on the $z=0$ results in \astrid, and is structured as follows.}
Section~\ref{sec:models} summarizes the adopted simulation models. 
In Section~\ref{sec:MBH} we present the MBH population in \astrid. We analyze the properties of galaxies and galaxy clusters in Sections~\ref{sec:galaxy} and \ref{sec:groups}, respectively. In Section~\ref{sec:clustering} we study the matter clustering in \astrid. We conclude in Section~\ref{sec:conclusion}.

\section{\astrid\ simulation} 
\label{sec:models}

\astrid\ \citep{Bird2022,Ni2022_astrid,Ni2024} is a large-volume cosmological hydrodynamical simulation performed using the SPH code \textsc{MP-GADGET} \citep{Feng2018}. The simulation evolves a cube of 250~$h^{-1}$cMpc per side with $2\times5500^3$ initial tracer particles comprising dark matter and baryons. The cosmological parameters used are from  the \cite{Planck} cosmology, with $h=0.6774$, $\Omega_{0}=0.3089$, $\Omega_{\mathrm\Lambda}=0.6911$, $\Omega_{\mathrm{b}} = 0.0486$, $\sigma_{\mathrm{8}}=0.816$, and $n_{\mathrm{s}}=0.9667$. 
The initial conditions are set at $z=99$, and have a mass resolution of $M_{\rm DM} = 6.7 \times 10^6$~$h^{-1}M_\odot$  and $M_{\rm gas} = 1.3 \times 10^6$~$h^{-1}M_{\odot}$.
The gravitational softening length is $\epsilon_{\rm g} = 1.5\ h^{-1}{\rm ckpc}$ for both DM and gas particles. 

\astrid\ includes a full-physics subgrid treatment for modeling galaxy formation, MBHs, stellar and AGN feedback, and inhomogeneous hydrogen and helium reionization. 
In Figure~\ref{fig:large_scale}, we 
present a visual overview of the \astrid\ simulation at $z=0$. 
The underlying background shows the projected DM density through a $25\, {\rm cMpc}\,h^{-1}$ thick slab (10\% of the simulation box length) over a $250\times 187.5\ {\rm cMpc}^{2}\,h^{-2}$ area (75\% of the full $250\times250\,{\rm cMpc}^{2}\,h^{-2}$ face).
We highlight two regions: the left insets zoom into the most-massive halo in \astrid\ with $\log M_{\rm 200c}/M_{\odot}=15.3$), and the right insets show a star-forming halo with $M_{\rm 200c}=10^{14.7}$~\Msun.
For both halos, we represent their surrounding gas density field colored by temperature, neutral hydrogen fraction,  and metallicity, stellar density field colored by stellar age, and the mock Hubble Space Telescope (HST) images for three member galaxies.
We use the green spikes to mark the black holes with $M_{\rm BH}\geq 8\times 10^{6}$~\Msun. It can be seen that both halos host a population of MBHs, and some of them are wandering MBH located at the outskirts of the galaxies.

In the following part of this section, we briefly list the subgrid models relevant to MBH evolution and the prescription used to identify halo structures. We refer the reader to \cite{Bird2022} and \cite{Ni2022_astrid} for more details of the models implemented.

\begin{figure}
    \includegraphics[width=1\linewidth]{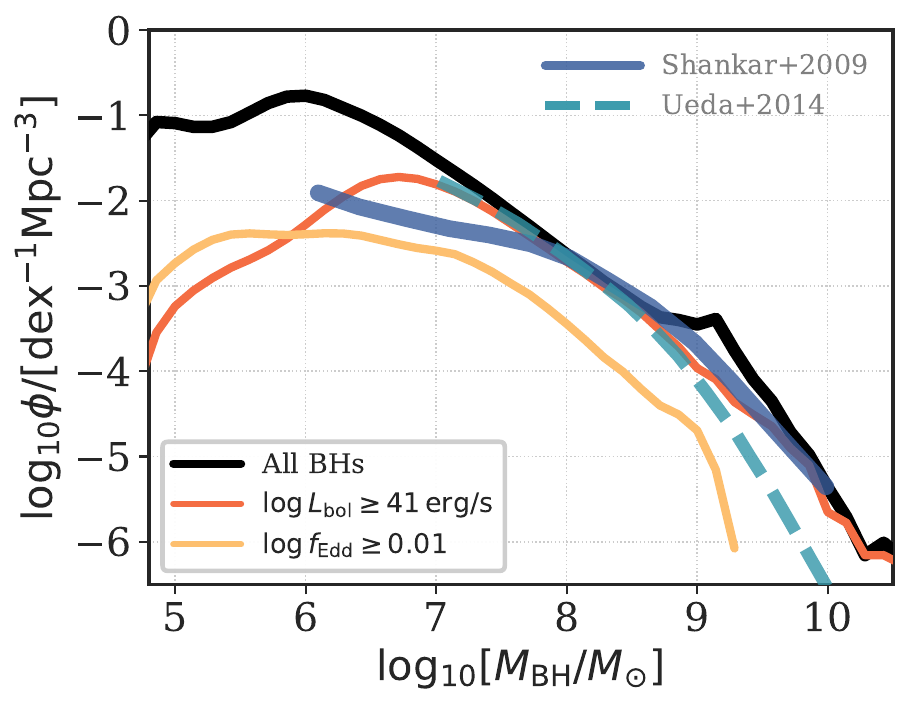}
    \caption{The black hole mass function in \astrid\ at $z=0$. The black solid curve includes all the BHs. The orange and yellow solid curves show the contribution from the BH population with $L_{\mathrm{bol}}\geq 10^{41}\,$erg/s
    and $f_{\rm Edd}\geq 0.01$, respectively. 
    The blue solid and dashed lines show the observational constraints given by \citet{Shankar2009} and \citet{Ueda2014}.
    }\label{fig:BHMF}
\end{figure}

\subsection{BH Models}
\subsubsection{BH Seeding}

MBHs in \astrid\ are seeded in halos with total mass $M_{\rm halo,FOF} > 5 \times 10^9~h^{-1}\rm{M_\odot}$ and  stellar mass $M_{\rm *,FOF} > 2 \times 10^6$~$h^{-1}\rm{M_\odot}$. 
The threshold for the halo stellar mass ensures that BHs are seeded only in halos with enough cold dense gas to form stars. 
Considering the complex astrophysical process involving BH seed formation, we allow halos with the same mass to host different BH seeds. Hence, rather than adopting a uniform seed mass, we probe a range of BH seed masses by  stochastically drawing from the range $3\times10^{4}~h^{-1}\rm{M_\odot}$ to $3\times10^{5} ~h^{-1}\rm{M_\odot}$.

\subsubsection{BH Dynamics}

\begin{figure*}[htbp]
    \centering
    \includegraphics[width=1\linewidth]{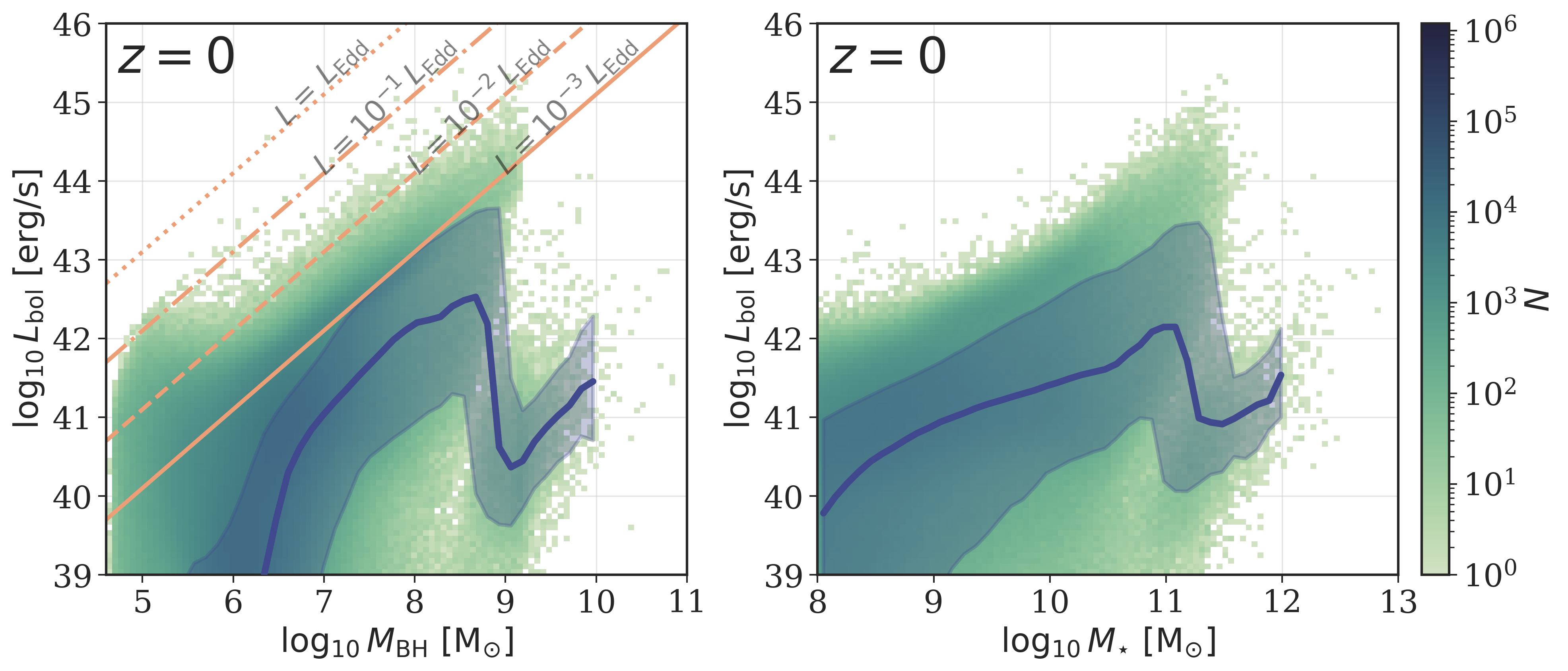}
    \caption{\textit{Left:} Relation between BH bolometric luminosity $L_{\rm bol}$ and BH mass $M_{\rm BH}$ at $z=0$ in \astrid. The underlying background shows a 2D distribution of $L_{\mathrm{bol}}-M_{\mathrm{BH}}$. The blue solid line gives the median AGN luminosity for each $M_{\rm BH}$ bin, and the shaded area depicts the 16th-84th percentiles. The orange dotted/dot-dashed/dashed/solid lines mark the $10^{0}$/$10^{-1}$/$10^{-2}$/$10^{-3} \times L_{\rm Edd}$, respectively. 
    \textit{Right:} The relation between BH bolometric luminosity $L_{\rm bol}$ and the stellar mass of their host galaxies $M_{\star}$. 
    }\label{fig:Lbol_Mbh}
\end{figure*}

A subgrid model is applied to estimate dynamical friction on the MBHs following \cite{Tremmel2015} and \cite{Chen2022_DF}.
Dynamical friction is the drag force experienced by the MBHs when they move through a medium of smaller collisionless particles. It dissipates the momentum of the MBHs, causing their orbits to decay toward the galactic center, and hence plays an important role in describing the coevolution of BHs and their host galaxies, and predicting MBH merger rates. 
Compared to the widely used implementation that directly repositions BHs to the local potential minimum \citep{Weinberger2017_tng}, this dynamical friction model not only gives a better estimation of the MBH hardening timescale, but also provides well-defined MBH trajectories and velocities.
We save the position, velocity, and accretion rate for each BH in \astrid\ across the entire simulation on the fly. This means we have the full history of all BHs and are able to trace BH activity over cosmic time. 
The available MBH trajectories allow us to impose a more physical criterion for BH mergers. Following \citet{Bellovary2011}  and \citet{Tremmel2017}, we merge two BHs based on their separation and whether they are gravitationally bound:
\begin{equation}
\left\{
\begin{aligned}
|\Delta r| &< 2\,\epsilon_g,\\
\tfrac12\,|\Delta v|^{2} &< \Delta a \cdot \Delta r ,
\end{aligned}
\right.
\end{equation}
where $\Delta a$, $\Delta v$, and $\Delta r$ are the relative acceleration, velocity, and position of the two BHs. $\epsilon_{\rm g}=1.5\,h^{-1} {\rm ckpc}$ is the gravitational softening length. 

We ensure stability of the dynamical friction model immediately after black hole seeding by adopting a BH mass tracer, the dynamical mass $M_{\rm dyn}$. This is necessary, as the minimum BH seed mass is smaller than that of the stellar and DM particles, and would otherwise experience significant artificial dynamical heating. $M_{\rm dyn}$ is used only for calculating the gravitational force and the dynamical friction. We use the intrinsic BH mass $M_{\rm BH}$ to account for BH accretion and AGN feedback. 
When a new BH is seeded, we initialize the corresponding $M_{\rm dyn}=10^{7}\,h^{-1}$~\Msun, which is $1.5$ times the mass of DM particles. $M_{\rm dyn}$ is kept until $M_{\rm BH}$ grows above $M_{\rm dyn}$. After that, $M_{\rm BH}$ is used to calculate the dynamical friction force applied to the BHs. As shown by \citet{Chen2022_DF}, this can effectively alleviate dynamical heating and stabilize the BH motion in the early growth phase.

\subsubsection{BH Accretion}
\label{sec:BHaccretion}
The BH accretion rate $\dot{M}_{\mathrm{BH}}$ is estimated using the Bondi-Hoyle formalism \citep{BondiHoyle1944, DiMatteo2005_BH_model}, which is based on the local properties of nearby gas particles:
\begin{equation}    
\label{equ:Mdot}
\dot{M}_{\mathrm{B}}=4\pi\alpha \,G^{2}\,M_{\mathrm{BH}}^{2}\,\rho_{\mathrm{BH}}\left(c_{\mathrm{s}}^{2}+v^{2}_{\mathrm{vel}}\right)^{-3/2},
\end{equation}
where $c_{\mathrm{s}}$ is the local sound speed, $\rho_{\mathrm{BH}}$ is the gas density around the BH, and $v_{\mathrm{vel}}$ is the velocity of the black hole relative to the surrounding gas. 
The dimensionless boost $\alpha=100$ is adopted to account for the underestimation of the accretion rate due to the unresolved interstellar medium. 
Super-Eddington accretion is allowed with an upper limit of twice the Eddington accretion rate $\dot{M}_{\mathrm{Edd}}$. Therefore, the black hole accretion rate $\dot{M}_{\mathrm{BH}}$ is determined by $\dot{M}_{\mathrm{BH}} = \min\left(\dot{M}_{\mathrm{B}}, 2\dot{M}_{\mathrm{Edd}}\right)$.

In this work, we compute the AGN luminosity following the model of \citet{Churazov2005}, which is also used by \citet{Habouzit2021, Habouzit2022} and \citet{Weinberger2017_tng}. 
BHs are distinguished as radiatively efficient or radiatively inefficient based on the Eddington ratio $f_{\rm Edd}=\dot{M}_{\rm BH}/\dot{M}_{\rm Edd}$: 

\begin{equation}
    L_{\rm bol} = 
    \begin{cases}
    \eta\,\dot{M}_{\rm BH}\,c^{2}, & f_{\rm Edd} >0.1;\\
    10 f_{\rm Edd}\, \eta\, \dot{M}_{\rm BH}\,c^{2} , & f_{\rm Edd} \leq 0.1,
    \end{cases}
\end{equation}
where $\eta=0.1$ is the radiative efficiency \citep{Shakura1973_BH}.
To better compare with observational data, in this work we present the hard X-ray luminosity of BHs ($L_{\rm X}$), which is converted from $L_{\rm bol}$ according to the bolometric correction proposed by \citet{Hopkins2007}.

\subsubsection{AGN Feedback}
\astrid\ adopts two-mode AGN feedback: 
high-accretion mode (or thermal feedback) and low-accretion mode (or kinetic feedback). 
The low-accretion mode is activated at $z=2.3$ and is implemented for BHs with $f_{\rm Edd}<\chi_{\rm thr}$.
The high-accretion mode is applied to BHs with $f_{\rm Edd}\geq \chi_{\rm thr}$ for $z\leq 2.3$, and the entire BH population for $z>2.3$ \footnote{When the high-accretion mode is activated at $z=2.3$, only 543 MBHs are found to be in the kinetic feedback regime. Among them, 127 have masses $M_{\rm BH}\geq10^{8}$~\Msun, which is  $<4\%$ of the total MBHs population above this mass threshold in \astrid\ at the same redshift. This suggests that the low-accretion mode plays a negligible role at $z\gtrsim 2.3$.}.
The Eddington threshold $\chi_{\rm thr}$ is dependent on the BH mass: 
\begin{equation}
    \chi_{\rm thr} = \min \left[0.002\left(\frac{M_{\rm BH}}{M_{\rm crit}}\right)^{2},\  0.05\right],
    \label{equ:chi_thr}
\end{equation}
where $M_{\rm crit}=5\times 10^{8}\,h^{-1}$~\Msun\ is the critical mass for AGN feedback.

In the high-accretion mode, 5\% of the radiated energy $\Delta \dot{E}_{\rm high}=0.05\,\eta\, \dot{M}_{\rm BH}c^{2}$ is thermally injected into the gas within twice the radius of the SPH smoothing kernel of the BH. The injection is done istropically, distributing energy according to the SPH kernel weight.
The low-accretion mode is implemented following the prescription described by \citet{Weinberger2017_tng}, albeit with a different set of parameters. The AGN feedback is deposited as $\Delta \dot{E}_{\rm low}=\epsilon_{\rm f,kin}\,\dot{M}_{\rm BH}\, c^{2}$. The feedback efficiency $\epsilon_{\rm f,kin}$ is scaled by the BH local gas density and has a maximum value capped at $0.05$:
\begin{equation}
    \epsilon_{\rm f,kin} = {\rm min}\left(\frac{\rho_{\rm BH}}{0.05\, \rho_{\rm SF,thres}},\ 0.05 \right),
\end{equation}
where $\rho_{\rm SF,thres}$ is the density threshold for star formation. 

\subsection{Identifying Bound Structures}

We identify halos using the friends-of-friends (FOF) algorithm \citep{Davis1985} with a linking length of 0.2 times the mean particle separation.
Each halo hosts at least  32 DM particles. 
Subsequently, we post-process these FOF groups with the \textsc{SUBFIND} algorithm \citep{Springel2001} to hierarchically identify the substructures. 
The halo samples we study in this work are FOF halos, and the galaxies are the \textsc{SUBFIND} substructures with nonzero stellar mass. 
In this work, we limit our galaxy population to those whose stellar mass within twice the stellar half-mass-radius is larger than $10^{8}\ M_{\odot}$.
The halo size is characterized by $R_{200c}$ or $R_{500c}$, which is defined as the radius within which the mean enclosed mass density is 200 or 500 times the critical value $\rho_{\rm c}$.
We adopt the spherical-overdensity mass $M_{\rm 200c}$ or $M_{\rm 500c}$ for the halo mass, i.e., the total mass enclosed within $R_{\rm 200c}$ or $R_{\rm 500c}$. 
If a galaxy resides at the potential minimum of a halo, we call it the central galaxy of this halo. All other galaxies within $R_{200c}$ are defined as satellite galaxies.

\begin{figure}
    \includegraphics[width=1\linewidth]{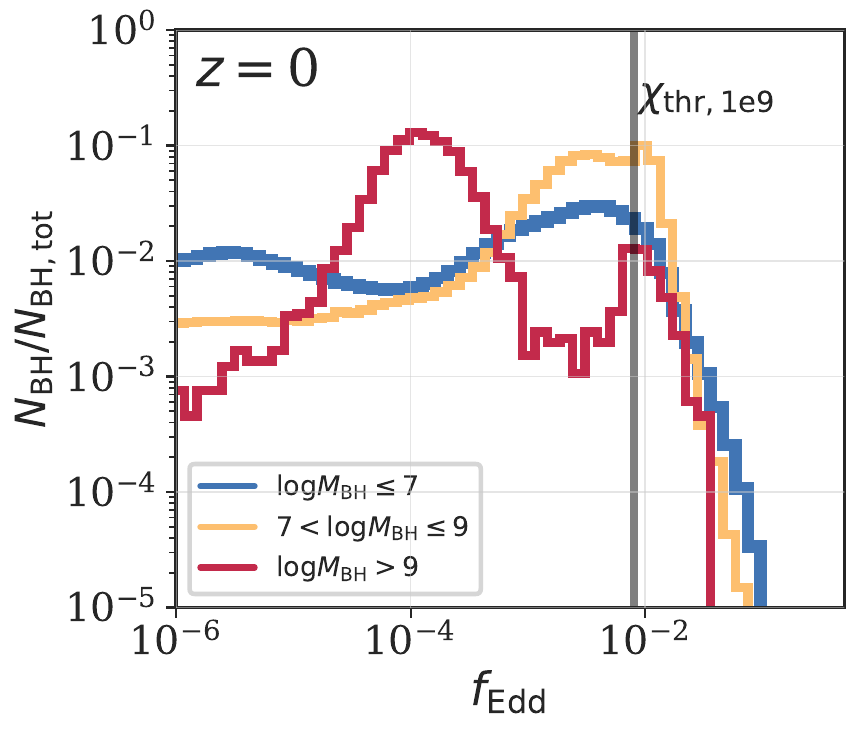}
    \caption{The Eddington ratio distributions for BH population in \astrid\ at $z=0$. 
    We show the BH in three mass bins: BHs with mass $M_{\rm BH}\leq 10^{7}$~\Msun\ (blue), $M_{\rm BH}=10^{7}-10^{9}$~\Msun\ (yellow), and those with $M_{\rm BH} > 10^{9}$~\Msun\ (red). The three distributions are normalized to the BH number in the corresponding mass bin. 
    The vertical line marks $\chi_{\rm thr}$ (the Eddington threshold for the AGN kinetic feedback mode; see Equation~\ref{equ:chi_thr}) for $M_{\rm BH}=10^{9}\,h^{-1}$~\Msun. 
    In both panels, we only include the central BH, i.e., the most-massive BH, of each galaxy. 
    }\label{fig:fedd_dist}
\end{figure}

\begin{figure*}[htbp]
    \centering
    \includegraphics[width=1\linewidth]{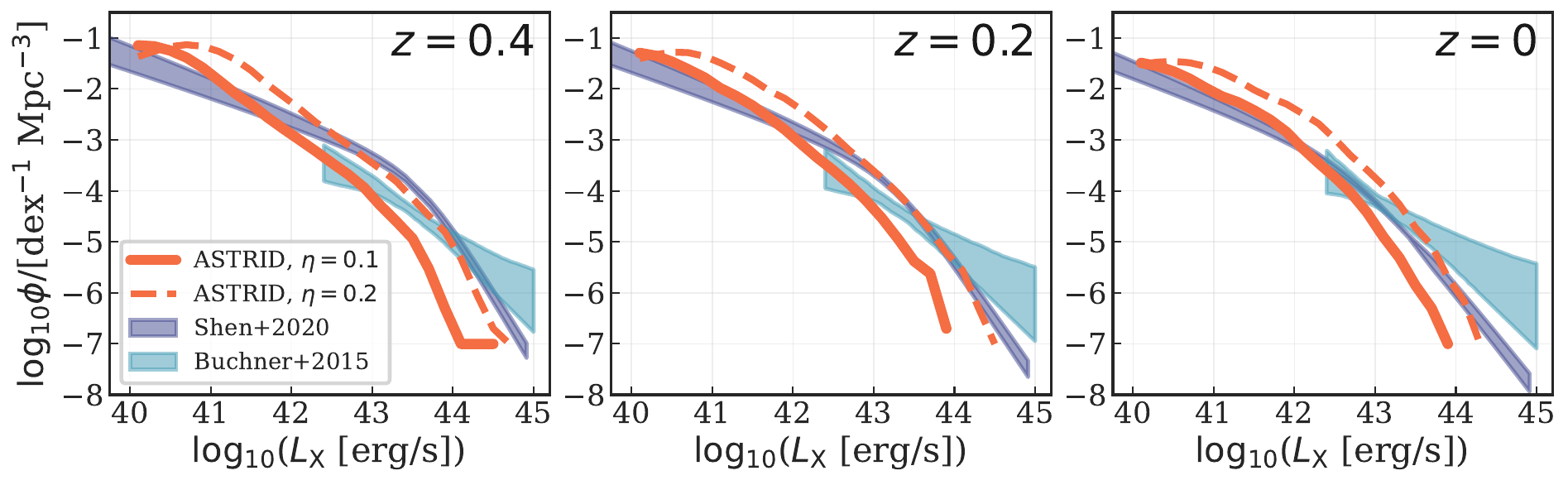}
    \caption{AGN hard X-ray (2-10 keV) luminosity function in \astrid\ at z=0.4 (left), z=0.2 (middle), and z=0 (right).
    The orange solid line presents the results based on the radiative efficiency $\eta=0.1$, and the orange dashed line corresponds to $\eta=0.2$. 
    In each panel, the purple region shows the observational constraints compiled by \citet{Shen2020}, and the blue region is the observation from \citet{Buchner2015}. 
    }\label{fig:Lx_func}
\end{figure*}

\section{MBH populations} 
\label{sec:MBH}

\subsection{Black Hole Mass Function}
We start our analysis by presenting an overview of the MBH population in \astrid. 
In Figure~\ref{fig:BHMF}, we show the black hole mass function at $z=0$. The black solid curve includes all  BHs. The orange and yellow solid curves show the contribution from the BH population with $L_{\mathrm{bol}}\geq 10^{41}\,$erg/s and $f_{\rm Edd}\geq 0.01$, respectively. 
Within the box of side length $250\,h^{-1}$Mpc, the most-massive black hole has mass $2\times 10^{11}$~\Msun, and the smallest MBH seed mass is $4\times 10^{4}$~\Msun. Hence, \astrid\ covers a BH mass range of almost 7 orders of magnitude. 
There is a bump in the black curve around $10^{9}$~\Msun, corresponding to the `critical mass' $M_{\rm crit}$ for the kinetic feedback, marking the transition from high-accretion mode to low-accretion mode. 
Below  $10^{9}$~\Msun, the AGN feedback is dominated by the high-accretion mode, allowing for fast growth for small BHs. When the BHs reach $10^{9}$~\Msun, the kinetic feedback mode turns on, and the BH growth is suppressed. 
As a result, the black holes pile up around the critical mass $10^{9}$~\Msun. 
The bump disappears when we apply the luminosity cut ($L_{\rm bol}\geq 10^{41}\ $erg/s), indicating these BHs have low accretion rates. 
To show the contribution from active BHs, we plot the mass function from the BH population with an Eddington ratio above $0.01$. Across the entire mass range, these BHs constitute $\lesssim 10\%$ of the total BH population. 
Hence, most of the BHs at $z=0$ are inactive, as the gas at the galactic center has been depleted.
We also plot the observational constraints inferred by \citet{Shankar2009} (blue solid curve) and \citet{Ueda2014}  (blue dashed curve) in Figure~\ref{fig:BHMF}.
We note there are large uncertainties in the assumptions made to arrive at these observational inferences, and discrepancies exist between different works. 
We see that \astrid\ predictions are in overall good agreement with the observational results over the mass range  $M_{\rm BH} \gtrsim 10^{7}$~\Msun. The mass function for MBHs with $L_{\mathrm{bol}}>10^{41}\ $ erg/s is closer to that from \citet{Shankar2009}, probably because those detections are biased to bright BHs. 

\subsection{AGN Luminosity Function}

The left panel of Figure~\ref{fig:Lbol_Mbh} shows the AGN bolometric luminosity, $L_{\mathrm{bol}}$, as a function of MBH mass, $M_{\mathrm{BH}}$, illustrating the correlations between them at $z=0$. 
The median $L_{\rm bol}$ is $< 10^{-3}\, L_{\rm Edd}$, consistent with other cosmological simulations including Illustris, TNG, and Horizon-AGN \citep{Habouzit2022}.
For $M_{\rm BH}\lesssim10^{8}\,M_{\odot}$, $L_{\rm bol}$ and $M_{\rm BH}$ are positively correlated.
However, the luminosity significantly decreases for MBHs of $M_{\rm BH}\geq 5\times 10^{8}$~\Msun, which reflects the impact of AGN kinetic feedback. 
The transition from high-accretion mode to low-accretion mode occurs around $M_{\rm BH}\sim5\times 10^{8}$~\Msun to $10^{9}$~\Msun, as expected from Equation~\ref{equ:chi_thr}.  
In the right panel of Figure~\ref{fig:Lbol_Mbh}, we show the relation between the galaxy stellar mass $M_{\star}$ and $L_{\mathrm{bol}}$. 
Due to the tight correlation between $M_{\rm BH}$ and $M_{\star}$, $L_{\rm bol}$ scales with $M_{\star}$  similarly to its scaling with $M_{\rm BH}$. 
The brightest AGN come from galaxies with $M_{\star}\sim 10^{11-11.5}$~\Msun. 

\begin{figure*}[htbp]
    \centering
    \includegraphics[width=1\linewidth]{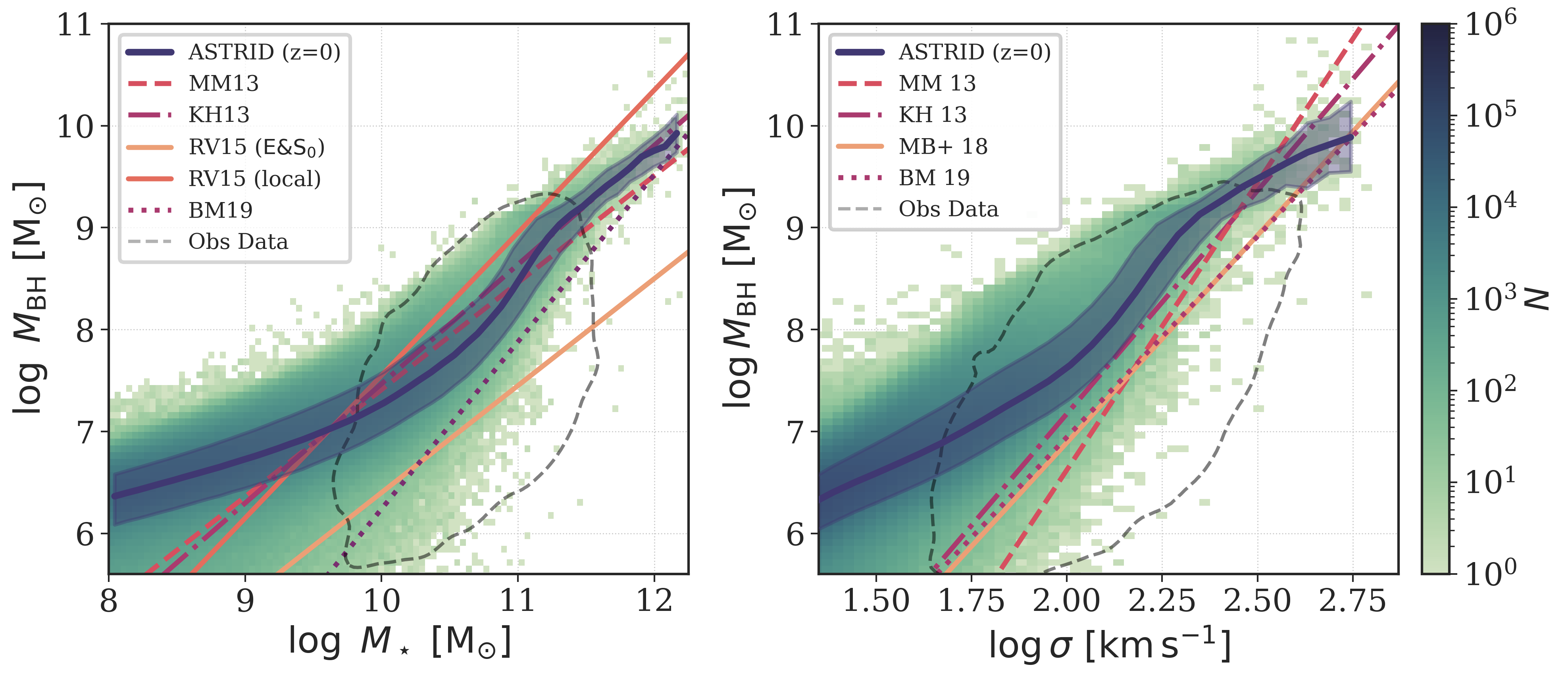}
    \caption{The correlation between MBHs and their host galaxies at $z=0$ in \astrid. We include only the most-massive BHs in each galaxy. 
    The left panel shows the relation between the MBH mass $M_{\mathrm{BH}}$ and the galaxy stellar mass $M_{\mathrm{\star}}$, and the right panel presents the correlation between galaxy central MBH mass $M_{\mathrm{BH}}$ and stellar velocity dispersion $\sigma$. $M_{\star}$ and $\sigma$ are measured based on the stars within twice the stellar half-mass-radius. 
    In both frames, the background shows the 2D distribution of galaxies in \astrid, sharing the color bar on the right. 
    The blue solid curve and the shaded area represent the median and the 16th-84th percentiles of the overall population.
    To guide the eye, we show some empirical scaling relations from the literature.
    For $M_{\rm BH}-M_{\star}$, we plot the scaling relations from 
    \citet{McConnell2013} (red shaded line), \citet{Kormendy2013} (purple dot-dashed line), \citet{Baron2019} (purple dotted line), the early type galaxy sample (orange solid line), and the local broad-line AGN sample (red solid line) in \citet{Reines2015}.
    For $M_{\rm BH}-\sigma$, we include the fitted scaling relation from 
    \citet{Kormendy2013} (orange solid line), \citet{McConnell2013} (orange dashed line), \citet{Martin-Navarro2018} (red dot-dashed line), and \citet{Baron2019} (red dotted line).
    The gray dashed contours represent the KDE estimate of the individual observed BHs (combined from \citealt{Reines2015, Bentz2018, Baron2019} for $M_{\rm BH}-M_{\star}$ and \citealt{Greene2006, Xiao2011, Baron2019} for $M_{\rm BH}-\sigma$), enclosing 97.5\% of the observational sample. They visualize the observational scatter for $M_{\rm BH}$ (not the dispersion between different fitted relations). 
    }\label{fig:BH_scaling}
\end{figure*}

\begin{figure}
    \includegraphics[width=1\linewidth]{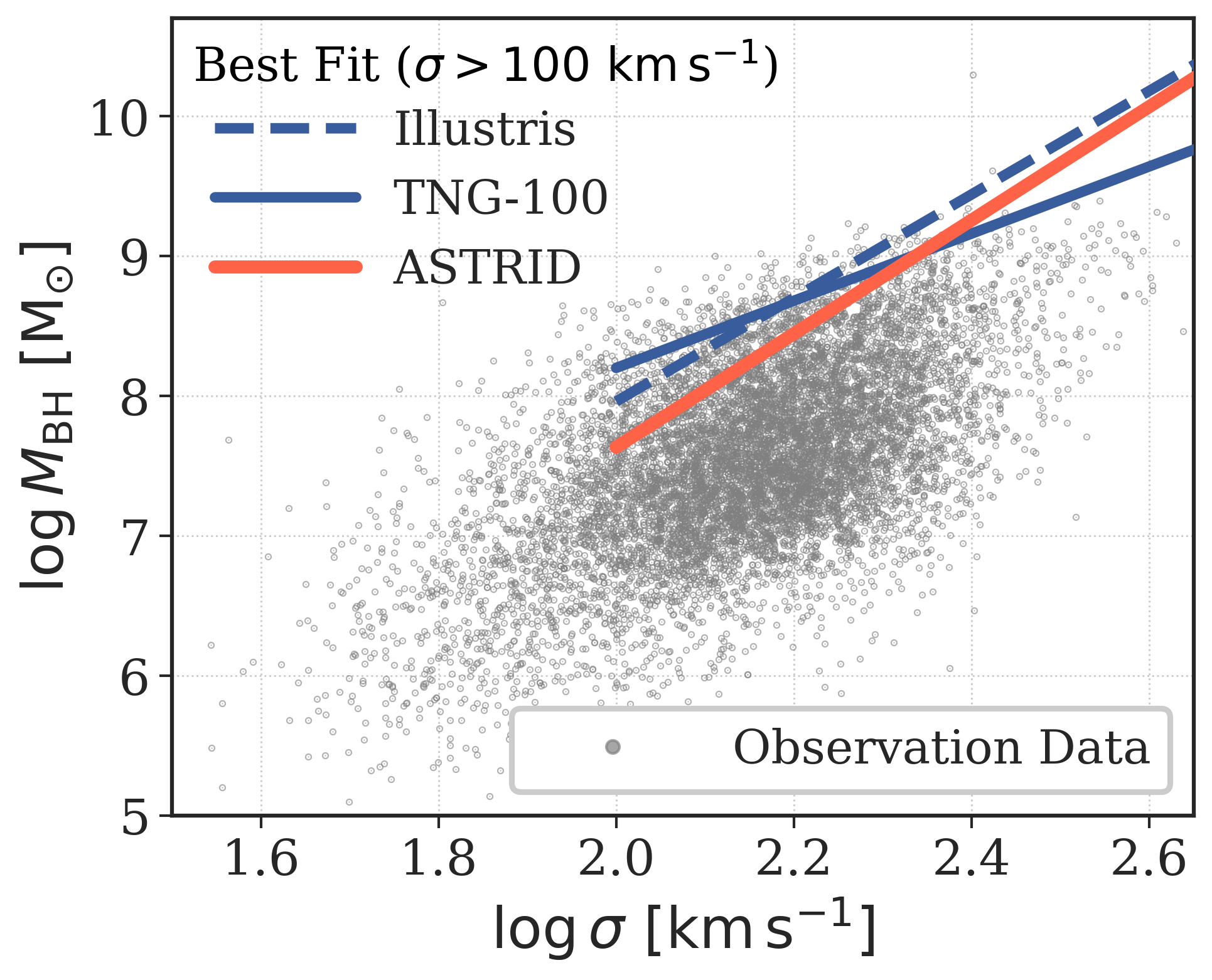}
    \caption{The fitted linear relation of $M_{\rm BH}-\sigma$ for massive galaxies with $\sigma>100\ {\rm km\,s^{-1}}$ in different simulations: Astrid (red line), Illustris (blue dashed line), and TNG-100 (blue solid line). 
    The gray dots are observed AGNs from \citet{Greene2006}, \citet{Xiao2011}, and \citet{Baron2019}.
    }\label{fig:Mbh_sigma_bestfit}
\end{figure}

Figure~\ref{fig:fedd_dist} presents the distribution of Eddington ratios $f_{\rm Edd}$ for three samples: small-mass BHs with $M_{\rm BH}<10^{7}$~\Msun\ (blue),  BHs with $M_{\rm BH}=10^{7}-10^{9}$~\Msun\ (yellow), and high-mass BHs with $M_{\rm BH} > 10^{9}$~\Msun\ (red). 
Each distribution is normalized by the total number of BHs in the corresponding sample.
Most of these MBHs have small Eddington ratios: across the entire mass range, only a very small fraction have $f_{\rm Edd} >10^{-2}$.  
For small-mass MBHs with $M_{\rm BH}<10^{7}\,M_{\odot}$, the distribution of $f_{\rm Edd}$ is relatively uniform. For BHs with $10^{7}<M_{\rm BH}\leq10^{9}\,M_{\odot}$, $f_{\rm Edd}$ peaks at $3\times 10^{-3}$.
The most-massive BHs ($M_{\rm BH}>10^{9}$~\Msun) exhibit a bimodal distribution in $f_{\rm Edd}$: besides a dominant peak at $f_{\rm Edd}=10^{-4}$, the minor, high-$f_{\rm Edd}$ peak of $f_{\rm Edd}$ is at $10^{-2}$. This reflects the transition between the two modes of AGN feedback models. 
We mark the $\chi_{\rm thr,1e9}$, the Eddington threshold for the AGN kinetic feedback (see Equation~\ref{equ:chi_thr}) evaluated at $M_{\rm BH}=10^{9}\, h^{-1}$~\Msun, using the vertical gray line. This line shows where the BHs in the red distribution shift from the high-accretion mode to the low-accretion mode. 
The minor peak coincides with $\chi_{\rm thr, 1e9}$, which means the BHs around this high-$f_{\rm Edd}$ peak are those that recently triggered kinetic feedback, while the surrounding gas has not yet been fully expelled. As more and more kinetic energy is released, the local gas density decreases significantly, moving the BHs in the $f_{\rm Edd}$ distribution to the major, low-$f_{\rm Edd}$ peak. 
The distribution of strongly accreting MBHs with $f_{\rm Edd} > 0.01$ shows no clear dependence on mass. Compared to other simulations \citep{Habouzit2022}, \astrid\ has more massive BHs with high $f_{\rm Edd}$.

The AGN luminosity function quantifies the AGN space density as a function of luminosity, and constrains a combination of BH mass and accretion rate. In Figure~\ref{fig:Lx_func} we plot the hard X-ray (2-10 keV) AGN luminosity function in \astrid\ at z=0.4 (left), z=0.2 (middle), and z=0 (right).  
In each panel, the purple/blue region shows the observational results from \citet{Shen2020}/\citet{Buchner2015}, respectively.
The $L_{\rm X}$ is calculated based on the method described in Section~\ref{sec:BHaccretion}.
As the radiative efficiency $\eta$ ranges from 10\% to 20\% in different simulations (e.g., IllustrisTNG adopts $\eta = 0.2$), we 
show two sets of AGN luminosity functions assuming $\eta=0.1$ (solid orange curves) and $\eta=0.2$ (dashed orange curves).
In \citet{Ni2024}, the authors presented  AGN luminosity functions with a higher amplitude. This is because they did not distinguish the radiatively inefficient AGN when estimating the luminosity, and used $\eta \dot{M}c^{2}$ for the entire BH population.
We find general good agreement between the \astrid\ AGN luminosity functions and those from observations for the luminosity range $10^{42}<L_{\rm X}<10^{44}\ {\rm erg}\,s^{-1}$, especially when we assume $\eta=0.2$. 
AGN feedback prevents the existence of bright AGNs with $L_{\rm X}>10^{45}\ {\rm erg}\,s^{-1}$ for $z\leq 0.4$ in \astrid\, while some bright AGNs within this luminosity range are observed.  
\citet{Habouzit2022} found that the bright end of the luminosity function is sensitive to AGN short timescale variability, which is not resolved in cosmological simulations.

\begin{figure}[htbp]
   \includegraphics[width=1\linewidth]{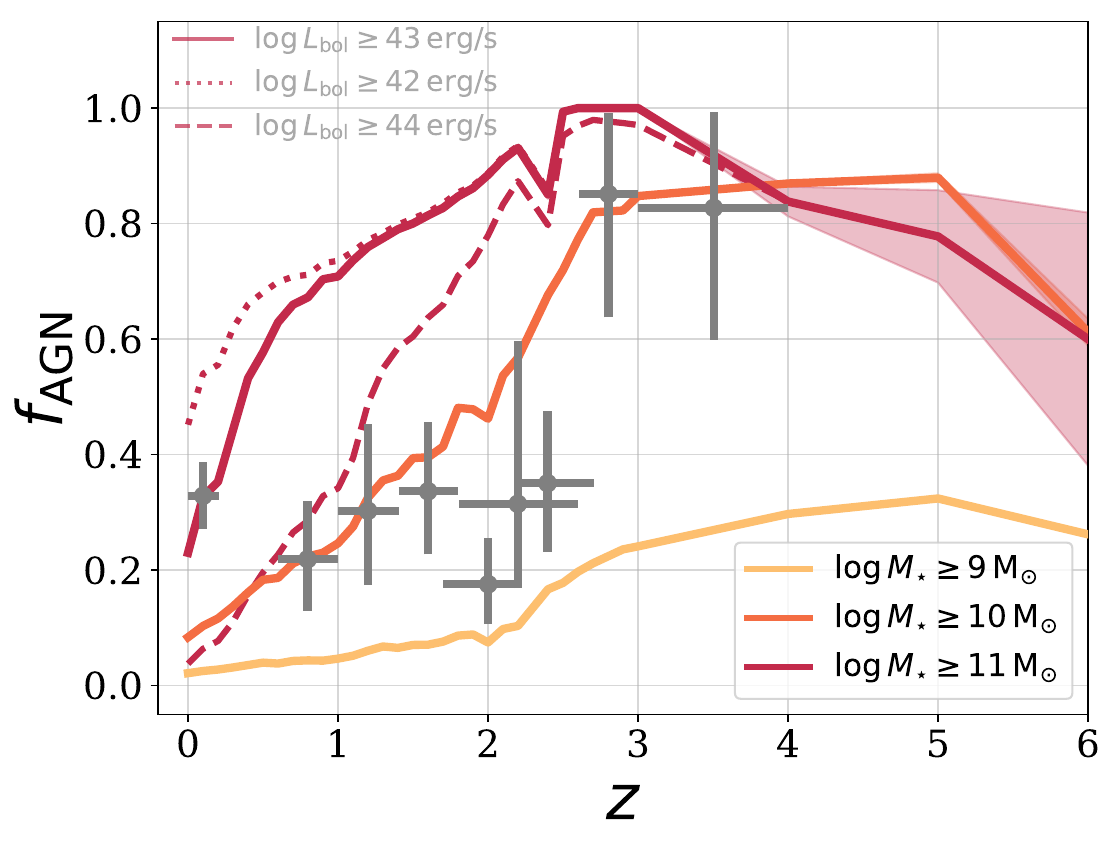}
    \caption{The evolution of AGN fraction with $L_{\mathrm{bol}}\geq 10^{43}\,$ erg/s in galaxies with different mass cuts: $M_{\star}\geq 10^{11}$~\Msun\ (yellow solid), 
    $M_{\star}\geq 10^{10}$~\Msun\ (orange solid), 
    $M_{\star}\geq 10^{9}$~\Msun\ (red solid). 
    For galaxies with $M_{\star}\geq 10^{9}$~\Msun\ (red), we plot the AGN fraction with $L_{\mathrm{bol}}\geq 10^{42}\,$ erg/s (red dotted) and $L_{\mathrm{bol}}\geq 10^{44}\,$ erg/s (red dashed). 
    The shaded areas correspond to the $1\sigma$ uncertainty from the sample variance. 
    The gray error bars show existing observational constraints compiled in \citet{Marsan2017}. 
    }\label{fig:frac_AGN}
\end{figure}

\begin{figure*}[htbp]
    \centering
    \includegraphics[width=1\linewidth]{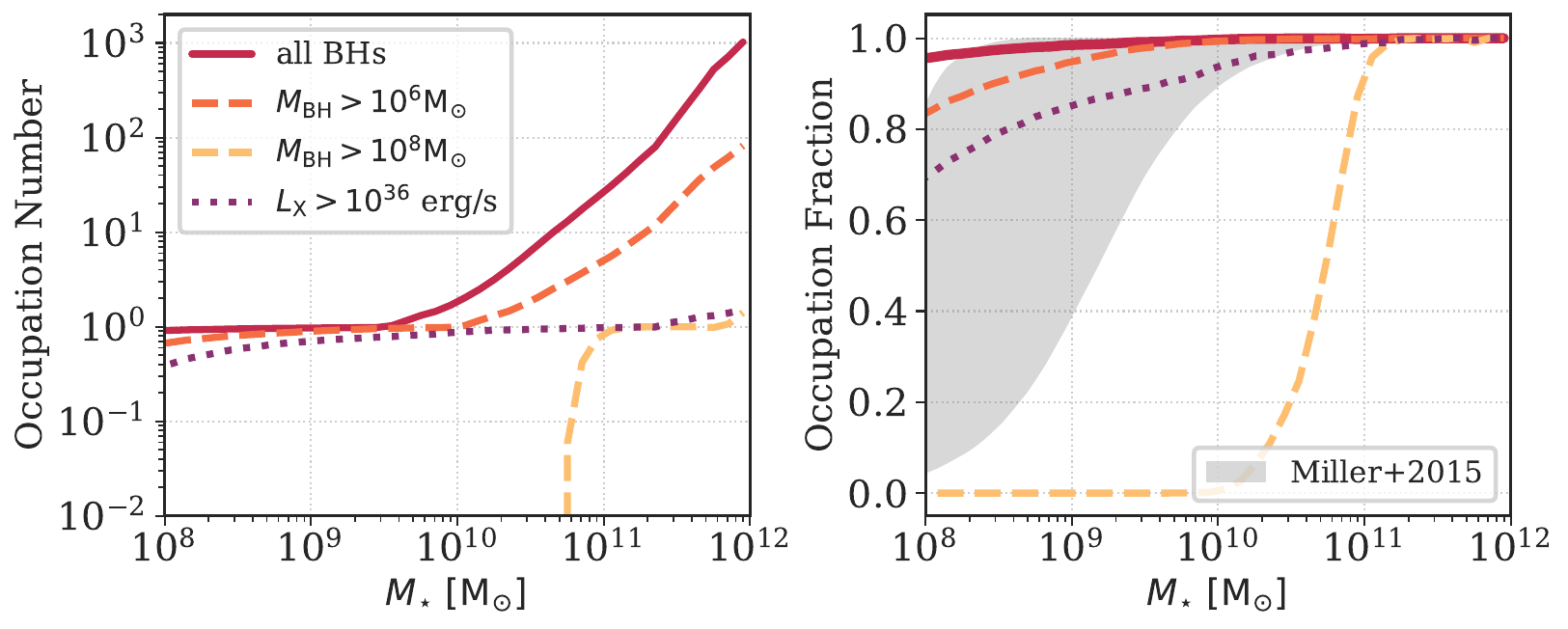}
\caption{\textit{Left:} the averaged occupation number of BHs as a function of the galaxy stellar mass.  The red lines include the entire BH population. The dashed lines represent the BHs above different mass thresholds of $M_{\rm BH}\geq 10^{6}$~\Msun (orange) and $M_{\rm BH}\geq 10^{8}$~\Msun (yellow). 
The purple dotted lines show the BH population with bolometric luminosity $L_{\rm X} \geq 10^{36}$~erg/s, corresponding to the luminosity limit of the sample in \citet{Miller2015}.
\textit{Right:} the fraction of galaxies that host at least one BH in each stellar mass bin. 
Different BH populations are shown with the same color conventions as the left panel. The gray shaded area represents the observational constraint from \citet{Miller2015}.
    }\label{fig:BH_occupy}
\end{figure*}

\subsection[MBH-M* and MBH-sigma Relation]{$M_{\mathrm{BH}}-M_{\mathrm{\star}}$ and $M_{\mathrm{BH}}-\sigma$ relation}
The strong relation between central BHs and their host galaxy properties, such as stellar mass, bulge mass, luminosity, and effective radius, points toward a close coevolution. 
In this section, we focus on the scaling relations between $M_{\rm BH}$ and the galaxy stellar mass $M_{\star}$ and the galaxy stellar 1D velocity dispersion $\sigma$. 

In the left panel of Figure~\ref{fig:BH_scaling} we present the $M_{\rm BH}-M_{\star}$ diagrams from \astrid\ at $z=0$. 
$M_{\mathrm{BH}}$ is the mass of the most-massive BHs in each galaxy, and $M_{\star}$ is measured for stars within twice the stellar half-mass-radius. 
The BH population in \astrid\ exhibits a relatively tight correlation between $M_{\rm BH}$ and $M_{\star}$, and is in good agreement with the observations at the massive end ($M_{\star}\gtrsim 10^{10}$~\Msun), with the median relation lying within the fitted scaling relation for inactive and active galaxies. 

\begin{figure*}[htbp]
    \centering
    \includegraphics[width=1\linewidth]{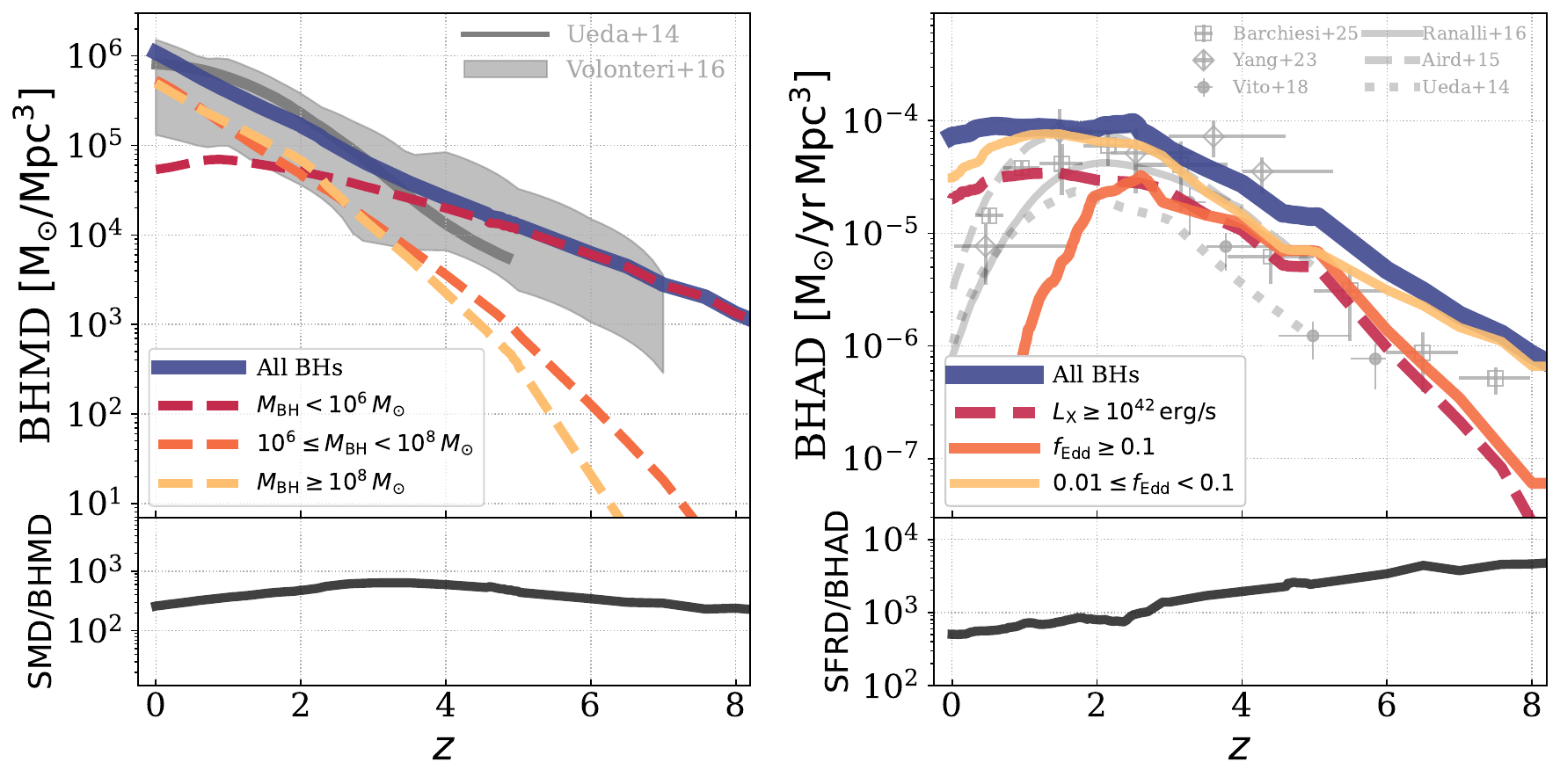}
    \caption{Evolution of MBH Mass and accretion rate densities. \textit{Left}: the top subpanel shows the evolution of the global MBH mass density (BHMD) as a function of redshift. The black curve includes all MBHs. The colored lines show the BHMD contributed by the MBHs in a given range of masses: red for $M_{\rm BH} < 10^{6}$~\Msun, orange for $10^{6}\leq M_{\rm BH}\leq 10^{8}$~\Msun, and yellow for  $M_{\rm BH}\geq 10^{8}$~\Msun. 
    We show the observational constraints from \citet{Ueda2014} (gray curve) and \citet{Volonteri2016} (gray shaded area). 
    The bottom subpanel shows the ratio between the global stellar mass density (SMD) and BHMD. 
    \textit{Right}: the top subpanel shows the evolution of the MBH accretion rate density (BHAD) as a function of redshift. The colored dashed lines correspond to different MBH populations: 
    red for MBHs with $L_{\rm X} < 10^{42}$ erg/s, orange for $L_{\rm X}\geq 10^{42}$ erg/s, and the yellow curve represents the MBHs with high Eddington ratios ($f_{\rm Edd}\geq 0.01$). 
    As indicated in the legend, the gray dots present the observational data from \citet{Barchiesi2025}, \citet{Yang2023}, and \citet{Vito2018}. We also include the fitted relation from \citet{Ranalli2016}, \citet{Aird2015}, \citet{Ueda2014}. 
    The bottom subpanel shows the ratio of the cosmic star formation rate density (SFRD) to the BHAD. 
    }\label{fig:BHAD}
\end{figure*}

Although the galaxy stellar mass is an easier quantity to measure in observations, in reality the scaling relation of $M_{\mathrm{BH}}-M_{\mathrm{\star}}$ is less tight than the relation between $M_{\rm BH}$ and the stellar velocity dispersion $\sigma$ \citep{Kormendy2013}.  
\citet{Marsden2020} suggested that $M_{\rm BH}-\sigma$ is a more fundamental relation compared to those between BH and galaxy mass, luminosity, or effective radius.  
In the right panel of Figure~\ref{fig:BH_scaling}, we plot the $M_{\rm BH}-\sigma$ relation in \astrid\ at $z=0$.  As for the left panel, we include only the central MBH (i.e., the most-massive BH) in each galaxy, and the velocity dispersion is measured within twice the stellar half-mass-radius for each galaxy. 
We found very similar results using $\sigma$ measured from all star particles within $30$~kpc 
of the galactic center. 

The BH population in \astrid\ exhibits a $M_{\rm BH}-\sigma$ correlation generally in agreement with the observations, especially 
for massive BHs ($M_{\mathrm{BH}}\gtrsim 10^{9}$~\Msun). On the other hand, at the low-mass end most BHs in \astrid\ are overmassive when compared to the empirical scaling relations.
A similar but more severe discrepancy is observed in Illustris and TNG-100, with \citet{Li2020} claiming that this is due to the $\sigma$ in simulations being smaller than that of realistic galaxies. This might come from insufficient numerical resolution, or be related to the overestimation of galaxy sizes in Illustris and TNG \citep{Xu2017, Genel2018}.
In Figure~\ref{fig:Mbh_sigma_bestfit}, we fit a linear function to  the $M_{\rm BH}-\sigma$ relation for the massive galaxies with $\sigma>100 {\rm\, km\,s^{-1}}$ in \astrid: $\log M_{\rm BH} (M_{\odot}) = 4.15\, \log \sigma (\rm {km\,s^{-1}
}) -0.73$.
We compare with the best fit for Illustris and TNG-100 made by \citet{Li2020}, as well as observations. 
\astrid\ produces an $M_{\rm BH}-\sigma$ relation more consistent with the observational constraints than either Illustris or TNG-100, although the $M_{\rm BH}$ are still slightly overmassive. 

In both BH-galaxy correlations shown in Figure~\ref{fig:BH_scaling}, we observe changes in the slope.
For $M_{\rm BH}\lesssim10^{7.5}$~\Msun and $M_{\rm BH}\gtrsim10^{9}$~\Msun, the slope for both $M_{\rm BH}-M_{\star}$ and $M_{\rm BH}-\sigma$ becomes flat. The transition at the high-mass end corresponds to the feedback critical mass $10^{9}$~\Msun. As discussed previously, BH growth is suppressed after the low-accretion mode is turned on, and lags behind the evolution of galaxies, the latter being quantified through the growth of $M_{\star}$ and $\sigma$. 
Such a transition has not been identified in the observational data, probably due to the lack of samples at the high-mass end (as the gray contours in Figure~\ref{fig:BH_scaling} show), as well as the fact that the BHs in the low-accretion mode generally have low luminosity. The latter reason is supported by the mass function plotted in Figure~\ref{fig:BHMF}: for the BH population with $L_{\rm bol}\geq 41$ erg/s no excess of BHs around $M_{\rm BH}=10^{9}$~\Msun\ appears. 
The flattening of the scaling relations for low-mass galaxies (with $\sigma < 96\ {\rm km\,s^{-1}}$ ) was also identified in observational data \citep{Martin-Navarro2018}. 
In these small galaxies, a larger fraction of BHs are not massive enough to reach the self-regulated regime. Their accretion rate is dominated by their initial mass, while less sensitive to the local environment. 
This suggests the role of AGN feedback in regulating star formation becomes subdominant in low-$\sigma$ galaxies. 
\citet{Habouzit2017} and \citet{Habouzit2021} highlighted that the strength of supernova feedback has a large impact in this mass range. 
Our results here imply that with the next generation of X-ray missions covering larger and deeper sky areas, the details of  BH-galaxy correlations will reveal important information about  AGN feedback and even the seeding mechanism. 

The scatter in  $M_{\rm BH}-\sigma$ has been studied for a long time \citep{Gebhardt2000, Gultekin2009}, and it has been shown that the scatter can be reproduced by introducing a third parameter related to the host galaxy \citep{Marconi2003, Graham2008}, which indicates that the coevolution of MBH and their host galaxies is a multi-dimensional correlation.
By plotting the $M_{\rm BH}-M_{\star}$ relations in Illustris, Horizon-AGN, EAGLE, TNG, and SIMBA, \citet{Habouzit2021} found that cosmological simulations struggle to produce the diversity of BHs observed in galaxies in the local Universe. 
By comparing the MBH in \astrid\ and the gray dashed contours plotted in Figure~\ref{fig:BH_scaling} that represent the observational data, we can see that \astrid\ reproduces most of the observed data shown by the gray contour, except for the MBHs close to the lower bound for galaxies with $M_{\star}\gtrsim 10^{11}$~\Msun. 
Notably, \astrid\ includes the small-mass MBH ($M_{\rm BH}\sim 10^{6-8}$~\Msun) in the galaxies with $M_{\star}\sim10^{10.5}-10^{11}$~\Msun. This region is not populated in many simulations \citet{Habouzit2021}. 
This can be attributed to the fact that we do not implement the repositioning algorithm, which makes some wandering black holes decay to the galactic center too fast and so undergo too much efficient accretion.
Moreover, due to the adopted small MBH seed mass and the high resolution, there are some MBHs with $M_{\rm BH}<10^{6}$~\Msun\ residing in galaxies with $M_{\star}\lesssim 10^{9}$~\Msun. These correspond to detected AGNs in local dwarf galaxies \citep{Reines2013, Baldassare2015, Mezcua2016, Han2025}. 
The failure in \astrid\ to create MBH with $M_{\rm BH}\lesssim10^{8}$~\Msun\ in galaxies with $M_{\star}=10^{11}\sim 10^{11.5}$~\Msun\ might be because of the limited simulation volume. 
The BH seeding scheme being tightly dependent on halo mass may also contribute to the smaller scatter in the $M_{\rm BH}-M_{\star}$ relation. 
In addition, the lack of inclusion of some physical processes in the simulations may also shrink the scatter of the scaling relations. For example, \citet{Bustamante2019} showed that introducing a model for  BH spin evolution and relating the AGN feedback efficiency to the BH spin would increase the scatter in the $M_{\rm BH}-M_{\star}$ relations for the massive galaxies.

\subsection{AGN Fraction and BH Occupation}
\label{sec:AGN_frac}
In this section, we investigate the AGN fraction in \astrid.
Following \citet{Habouzit2022}, we define the AGN fraction as the number of galaxies hosting at least one active BH divided by the number of galaxies hosting BHs.  
In Figure~\ref{fig:frac_AGN} we plot the evolution of the AGN fraction in galaxies with different mass cuts: $M_{\star}\geq 10^{11}$~\Msun\ (yellow solid), 
$M_{\star}\geq 10^{10}$~\Msun\ (orange solid), 
$M_{\star}\geq 10^{9}$~\Msun\ (red solid). 
We use a bolometric luminosity cut of $L_{\rm bol}\geq 10^{43}\ {\rm erg\,s^{-1}}$ to identify the AGN, which roughly corresponds to the 
AGN population that could be detected by current instruments (e.g., Chandra, XMM-Newton, and JWST).
To show the influence of applying different luminosity cuts, we also plot the AGN fraction with $L_{\mathrm{bol}}\geq 10^{42}\,$ erg/s (red dotted) and $L_{\mathrm{bol}}\geq 10^{44}\,$ erg/s (red dashed) for the galaxies with $M_{\star}\geq 10^{9}$~\Msun.
The gray points with error bars are the existing observational constraints compiled by \citet{Marsan2017}. 

Across all galaxy stellar mass bins, we find that the AGN fraction is higher at high redshifts. This trend has been confirmed by observations \citep{Aird2018}. 
For galaxies with $M_{\star}\geq 10^{11}$~\Msun, the fraction peaks at $z=3$ and reaches $1$. 
For less massive galaxies, the fraction is relatively stable for $3<z<5$: $f_{\rm AGN}\sim0.8 $ for galaxies with $M_{\star}\geq 10^{10}$~\Msun, and $f_{\rm AGN}\sim 0.3$ for galaxies with $M_{\star}\geq 10^{9}$~\Msun. 
For $z < 3$, the fraction goes through a sharp decline, especially for the bright AGNs with $L_{\rm bol}\geq 10^{44}\ {\rm erg}\,s^{-1}$. Fainter AGNs ($L_{\rm bol} \lesssim 10^{42}\ {\rm erg}\,s^{-1}$) present a smoother decline after the peak in $f_{\rm AGN}$. 
Massive galaxies in \astrid\ have a larger $f_{\rm AGN}$, a trend also found in observations \citep{Aird2012, Man2019}. 
By $z=0$, there are still $20\%$ of galaxies with $M_{\star}\geq 10^{10}$~\Msun\ that host an AGN with $L_{\rm bol} \geq 10^{43}\ {\rm erg}\,s^{-1}$. While for galaxies above $M_{\star}=10^{9}$~\Msun, this fraction is only 2\%. 

\begin{figure*}[htbp]
    \centering
    \includegraphics[width=1\linewidth]{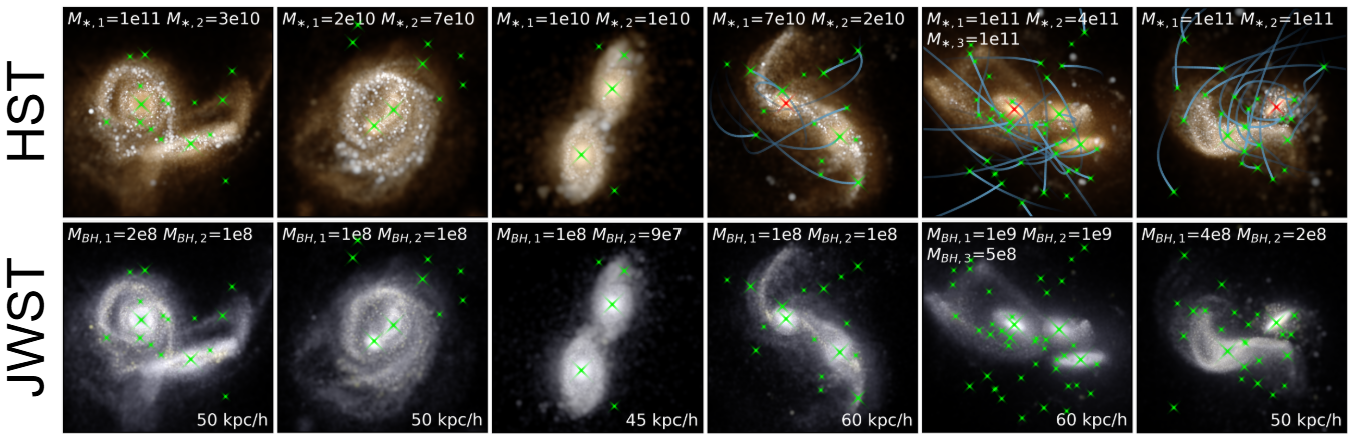}
\caption{Mock observations of 6 galaxy mergers present in the \astrid\ simulation. The top row contains HST Wide Field Camera (WFC3) mock observations with the red, green, and blue channels showing the F625W, F475W, and F390W filters, respectively. The bottom row shows James Webb Space Telescope (JWST) NIRCam mock observations with the F444W, F277W, and F115W filters as the red, green, and blue channels. 
The green crosses represent the black hole population hosted by these galaxies, whose sizes are scaled by $M_{\rm BH}$. 
The galaxy stellar mass is labeled at the top of each HST image with the unit of ~\Msun. The masses of their central MBHs are shown at the top of JWST images. 
In the right three images in the top row, we plot the black hole trajectory over the past $100$ Myr for those with $M_{\rm BH}\geq10^{5}$~\Msun\ (blue curves). These paths are plotted relative to one of the central black holes, which is indicated by a red cross.} \label{fig:mock_obs_w_BH_tracks}
\end{figure*}

There is compelling evidence that almost every massive galaxy in the local Universe contains an MBH at its center \citet{Kormendy2013}. 
In Figure~\ref{fig:BH_occupy}, we present the BH occupation number as a function of the stellar mass in the host galaxy. 
The left panel shows the occupation number of BHs, with a variety of different mass cuts, as well as a luminosity threshold of $L_{\rm  X}=10^{36}\ {\rm erg/s}$, corresponding to the luminosity limit of the AGN sample compiled in \citet{Miller2015}.
For galaxies with $M_{\star}>10^{10}$~\Msun, the occupation number increases sharply 
with stellar mass. 
This is consistent with \citet{Ricarte2021}, who found that the wandering MBH occupation number scales with galaxy mass as a power-law relation.
When we limit our analysis to the massive MBHs ($M_{\rm BH}\geq 10^{8}$~\Msun) or AGNs ($L_{\rm X}\geq 10^{46}\, {\rm erg/s}$), there is only one MBH for each galaxy, even for the most-massive galaxies. 
AGN feedback prevents individual galaxies from hosting multiple massive or luminous MBHs. Moreover, as pointed out by \citet{Ni2022_astrid} and \citet{Chen2022}, massive MBHs 
have a shorter hardening timescale and a larger merger rate compared to seed MBHs, which means that multiple massive MBHs hosted by the same galaxy would merge quickly. 
Our results indicate that a large population of smaller MBHs exists in massive galaxies. 
There are about 20 wandering MBHs in galaxies with $M_{\star}=10^{11}$~\Msun\ on average, and over 1000 wandering MBHs in galaxies with $M_{\star}=10^{12}$~\Msun. Most of these wandering MBHs are seed MBHs with $M_{\rm BH}\geq 10^{6}$~\Msun. 
When a mass cut of $M_{\rm BH}>10^{6}$~\Msun\ is applied, the averaged BH occupation number is 4 for galaxies with $M_{\star}=10^{11}$~\Msun\ and 82 for those with $M_{\star}=10^{12}$~\Msun.

The right panel plots the fraction of galaxies that host at least one MBH in each stellar mass bin. 
Almost all galaxies ($98\%$) with $M_{\star}>10^{9}$~\Msun\ host BHs, and 95\% of them host BHs with masses over $10^{6}$~\Msun. 
For galaxies at $M_{\star}=10^{8}$~\Msun, the BH occupation fraction drops to 95\% for the entire BH population, and to 83\% for those with $M_{\rm BH}>10^{6}$~\Msun. 
Multiple simulations find that MBH growth in low-mass galaxies with $M_{\star}\leq 10^{9}$~\Msun\ is regulated or stunted by supernova feedback so that low-mass galaxies in general have a lower probability of hosting an MBH \citep{Dubois2015}.
About 87\% of massive galaxies with $M_{\star}\geq10^{11}$~\Msun\ host an MBH over $M_{\rm BH}=10^{8}$~\Msun in mass.
When we consider the luminosity threshold of $L_{\rm X}>10^{36}$ erg/s, which is similar to the limit of the observed AGN sample compiled in \citet{Miller2015}, the MBH occupation fraction (purple dotted curve) is consistent with the constraints given by \citet{Miller2015} (gray shaded area).

\begin{figure*}[htbp]
    \centering
    \includegraphics[width=1\linewidth]{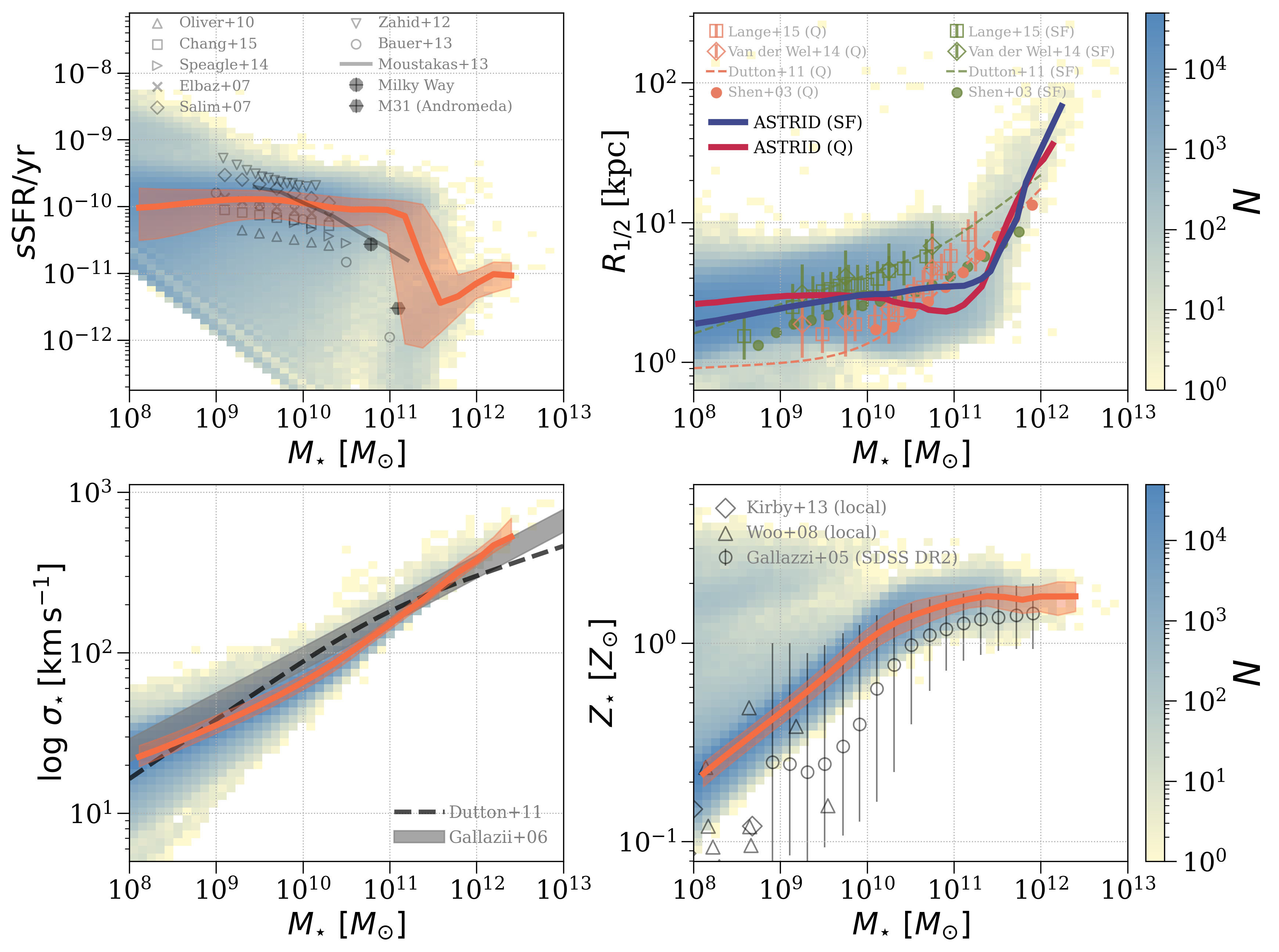}
\caption{The relation between galaxy stellar mass and specific star formation rate ($s$SFR; top left), stellar half-mass-radius (top right), stellar velocity dispersion (bottom left), and metallicity (bottom right). 
In each panel, the underlying distribution is for the galaxy population in \astrid\ at $z=0$. The x-axis represents the galaxy mass, defined as the stellar mass enclosed within twice the stellar half-mass-radius. 
Except for the bottom-right panel, we show the median value in each mass bin by the orange curve, with the shaded area indicating the 16th to 84th percentiles. 
We only plot the median for bins hosting more than 10 galaxies. 
In the $R_{\rm 1/2}$ panel, we plot the median value for the star-forming (blue) and quiescent (red) galaxy populations separately. 
The $s$SFR is calculated within 30 kpc from the galactic center and averaged over the past 200 Myr. The stellar metallicity and the velocity dispersion are estimated based on the star particles within twice the stellar half-mass-radius, and the metallicity is normalized by solar metallicity $Z_{\star}=0.0127$. 
As indicated in the legends, the simulation is compared with the observed $s$SFRs from \citet{Oliver2010, Chang2015, Speagle2014, Elbaz2007, Salim2007, Zahid2012, Bauer2013, Moustakas2013}, the stellar metallicity from \citet{Kirby2013}, \citet{Woo2008}, and \citet{Gallazzi2006}. 
We compare to the fitted $\sigma-M_{\star}$ relation from \citet{Dutton2011} and \citet{Gallazzi2006}, and the galaxy size from \citet{Lange2015}, \citet{vanderWel2014}, \citet{Dutton2011}, and \citet{Shen2003}. 
The $s$SFR and stellar masses for Milky Way and Andromeda plotted in the first panel are from \citet{Tamm2012, Rahmani2016, Licquia2015}.
    }\label{fig:galaxy_property}
\end{figure*}

\subsection{BH Mass and Accretion Rate Density}

The gas reservoir in the galaxy is believed to be the main source of both BH and galaxy mass growth.
The partitioning of gas between fueling star formation versus BH accretion determines the coevolution of MBH and their host galaxies. 
\citet{Dattathri2025} found that the interplay between the BH accretion rate density (BHAD) and global star formation rate density (SFRD) drives the evolution of the mean relation of $M_{\rm BH}-M_{\star}$. 
In this section, we focus on the global evolutionary history of the BH mass density (BHMD) and the BHAD over cosmic time, then compare them with the star mass density (SMD) and SFRD.  
In the top-left frame of Figure~\ref{fig:BHAD}, we plot the global BHMD as a function of redshift. 
The gray curve is the observational constraints from \citet{Ueda2014} based on the X-ray AGN luminosity function. The gray shaded area denotes the predicted range reported by \citet{Volonteri2016}, which is derived from the galaxy mass function of \citet{Grazian2015} and relies on assumptions about the MBH-stellar mass relation.
It can be seen that the BHMD in \astrid\ is well placed in the regime indicated by the observations. 
At high redshift, $z > 2$, most of the MBH mass is still from seed MBHs ($M_{\rm BH}<10^{6}$~\Msun). For $z < 2$, the MBH mass is dominated by larger objects.
The contributions from the BH population with $M_{\rm BH}=10^{6-8}$~\Msun\ and from those with $M_{\rm BH}\geq10^{8}$~\Msun\ are similar: in the local Universe, the former consists of 49\% of the total BH mass, and the latter consists of 47\%, about 1 order of magnitude larger than the BHMD from seed BHs. 
The bottom-left panel shows the ratio between the global SMD and BHMD. 
At high redshifts, the star formation process builds up galaxy mass faster than BH formation and accretion, the ratio between $\rm SMD$ and $\rm BHMD$ goes through a slow increase, and reaches a peak of $\sim 600$ at $z=3$, corresponding to  `Cosmic Noon' \citep{Madau2014}. At $z=0$, ${\rm SMD/BHAD} \sim 250$. 

The top-right panel of Figure~\ref{fig:BHAD} shows the redshift evolution of BHAD in \astrid. 
For $z\gtrsim 0.3$, more than half of the accretion is contributed by MBHs with moderate Eddington ratios $0.01\leq f_{\rm Edd}<0.1$. It peaks at $z\approx1.5$ where they supply $\sim 90\%$ of the total, then declines to $\sim 30\%$ by $z=0$.
The accretion from fast accretors with $f_{\rm Edd}\geq 0.1$ peaks at $z=2.5$, and its fraction among the total BHAD reaches its maximum of $45\%$ at $z=5$. While for $z< 1$, it is negligible. 
Bright AGNs with $L_{\rm X}\geq 10^{42}$ erg/s typically account for $30\sim 40$\% of the total BHAD at $z<5$.

We include the observational constraints for BHAD in Figure~\ref{fig:BHAD}: the gray dots present the data from \citet{Barchiesi2025}, \citet{Yang2023}, and \citet{Vito2018}; the gray curves are the fitted relations from \citet{Ranalli2016}, \citet{Aird2015}, \citet{Ueda2014}. 
These observational data typically cover samples above $L_{\rm X}>10^{42}$~erg/s, while suffering from incompleteness at the lower-luminosity end. The BHAD in \astrid\ contributed by MBHs with $L_{\rm X}\geq 10^{42}$~erg/s is in good agreement with the observations at $z\geq 2$, while it is an overestimate by more than 1 dex at $z=0$. 
It has been noted in previous work that the BHAD predictions made by simulations are in general higher than indicated by observations \citep{Vito2018}. However, the BHAD in \astrid\ at $z=0$ is still $\sim 0.5$ dex larger compared to other simulations such as Horizon-AGN \citep{Volonteri2016} and Illustris \citet{Sijacki2015}. This implies that the AGN feedback in \astrid\ is not sufficiently efficient. 

The bottom panel shows the ratio of the cosmic SFRD to the global BHAD. The ratio goes through a slow decrease across the redshift: changing from $\approx 4500$ at $z=8$ to $\approx 500$ at $z=0$. \citet{Dattathri2025} found that \astrid\ has much minor redshift evolution in this ratio compared to TNG300. 
The SFRD/BHAD in \astrid\ is small compared to results from previous works:
\citet{Kormendy2013} found $5000\times {\rm BHAD}$ fits well the observed SFRD contained by \citet{Hopkins2004}; \citet{Vito2018} adopted a ratio of 3000 to scale the ${\rm SFRD}$ in \citet{Bouwens2015}, getting a density aligned with the observed BHAD.

\begin{figure}[htbp]
    \centering
    \includegraphics[width=1\linewidth]{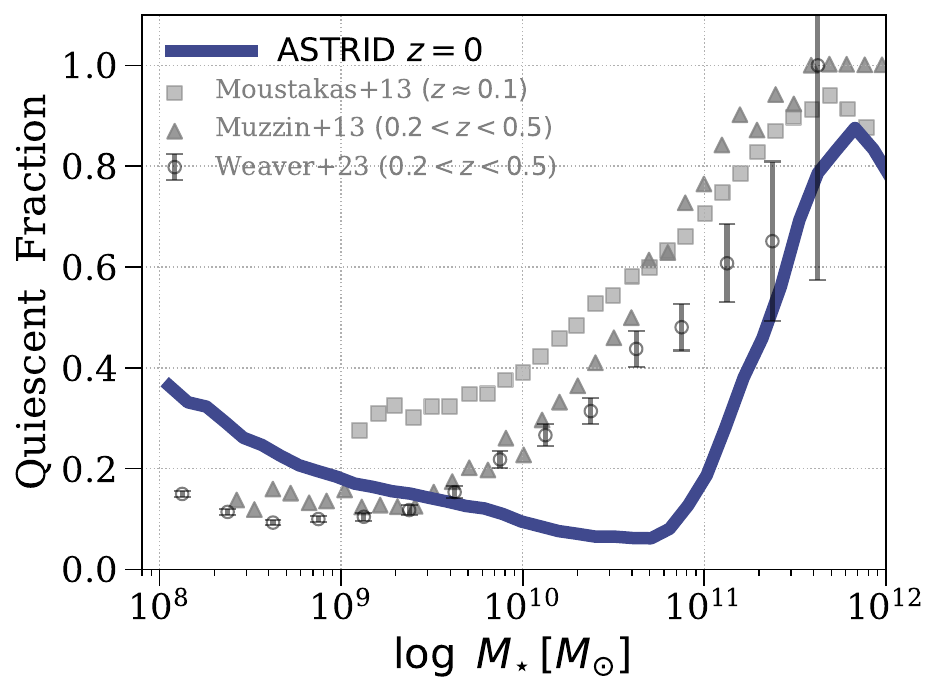}
    \caption{Quiescent fraction of galaxies as a function of the galaxy stellar mass. The blue curve shows the fraction of the quiescent galaxies in \astrid\ at z=0. The quiescent galaxies are defined as those with the $s{\rm SFR}< 10^{-11}/{\rm yr}$. The data points show the observational results compiled in 
    \citet{Moustakas2013}, \citet{Muzzin2013}, and \citet{Weaver2023}.
    }\label{fig:QuiescentFrac}
\end{figure}

\section{Galaxies}
\label{sec:galaxy}

In this section, we analyze the $z=0$ galaxy population whose stellar mass within twice the stellar half-mass-radius is larger than $10^{8}\, h^{-1}$~\Msun\ in \astrid.  
With its relatively low particle mass, \astrid\ is able to well resolve these galaxies as each of them contains at least a few hundred stellar particles. Mock observations of a variety of these galaxies can be seen in both Figures~\ref{fig:large_scale} and \ref{fig:mock_obs_w_BH_tracks}. The process of creating these mock observations is detailed in Section~\ref{sec:GalaxyProperty}.

\subsection{Galaxy Properties}
\label{sec:GalaxyProperty}

In Figure~\ref{fig:galaxy_property}, we show the relation between galaxy stellar mass and  specific star formation rate ($s$SFR; top left),
stellar half-mass-radius (top right),
stellar velocity dispersion (bottom left), and
metallicity (bottom right). 
The galaxy mass is defined as the stellar mass enclosed within twice the stellar half-mass-radius. 
In each panel, the underlying 2D distribution is for the galaxy population in \astrid\ at $z=0$. 
Except for the top-right panel, we show the median value in each mass bin using an orange curve, with the shaded area indicating the 16th to 84th percentiles. 
We only plot the median for bins hosting more than 10 galaxies.

Although many simulations report galaxies' SFR based on the star-forming gas particles, this instantaneous SFR is not an observable. 
From an observational point of view, the galaxy's SFR is typically inferred from the luminosity in specific bands. Widely used star formation tracers span the full electromagnetic spectrum, including those from X-ray, ultraviolet (UV), and radio wavelengths. They are sensitive to star formation history over different characteristic timescales, ranging from 10 Myr (H$\alpha$-based) to 200 Myr (UV-based) \citep{Kennicutt2012}. 
To make our results more comparable to observational values, we derive the galaxy's SFR from the amount of stellar mass formed over the last 200 Myr. 
We choose 200 Myr, which is close to the upper limit of the timescale the observational SFR indicators can trace, to mitigate the resolution effects in the simulation \citep[see][Appendix~A]{Donnari2019}.
Moreover, to bracket the typical galaxy apertures used in observational surveys, we apply a fixed 3D aperture of 30 kpc when measuring the SFRs. 
We have tested a range of timescales (100 Myr, 50 Myr, and 10 Myr) and aperture cuts (20 kpc and twice the half-mass-radius) to define the galaxy's SFR, and found that our results are robust to the specific choice of these parameters. 
In the first panel of Figure~\ref{fig:galaxy_property}, 
we compare our simulated $s$SFR with observational constraints from prior literature.
It can be seen that various observational data differ by up to 1 dex for a given galaxy mass bin. 
For galaxies with $M_{\star} \lesssim 10^{10.5}$~\Msun, \astrid\ predicts a $s$SFR in good agreement with observations.
However, while observations show a rapid decline in $s$SFR around $M_{\star}=10^{10.5}$~\Msun, marking the transition to quiescence, \astrid\ exhibits this drop only beyond $M_{\star}=10^{11}$~\Msun, where the AGN kinetic feedback takes effect. 
This discrepancy may indicate that the adopted threshold for AGN kinetic feedback, i.e., the critical mass $M_{\rm circ}$ in Equation~\ref{equ:chi_thr}, was set too high. 
Adopting a lower threshold $M_{\rm circ}=10^{8}$~\Msun, TNG-100 generates a $s$SFR turnover at $10^{10.5}$~\Msun\ \citep{Donnari2019}.
Another noticeable feature in the \astrid\ $s$SFR-$M_{\star}$ relation appears at the high-mass end. For massive galaxies with $M_{\star}\geq 10^{12}$~\Msun, where observational data are lacking, the $s$SFR stops falling and begins to rise.
This trend can be attributed to the frequent mergers experienced by these massive systems, which can trigger significant star formation activities. 
This interpretation is supported by \citet{Zhou2025_PTA_CW}, which shows $s$SFR-$M_{\star}$ for galaxies hosting recent MBH mergers in \astrid. The $s$SFR of these merger-host galaxies is higher than the median $s$SFR for the overall galaxies we present here.

\begin{figure}[htbp]
    \centering
    \includegraphics[width=1\linewidth]{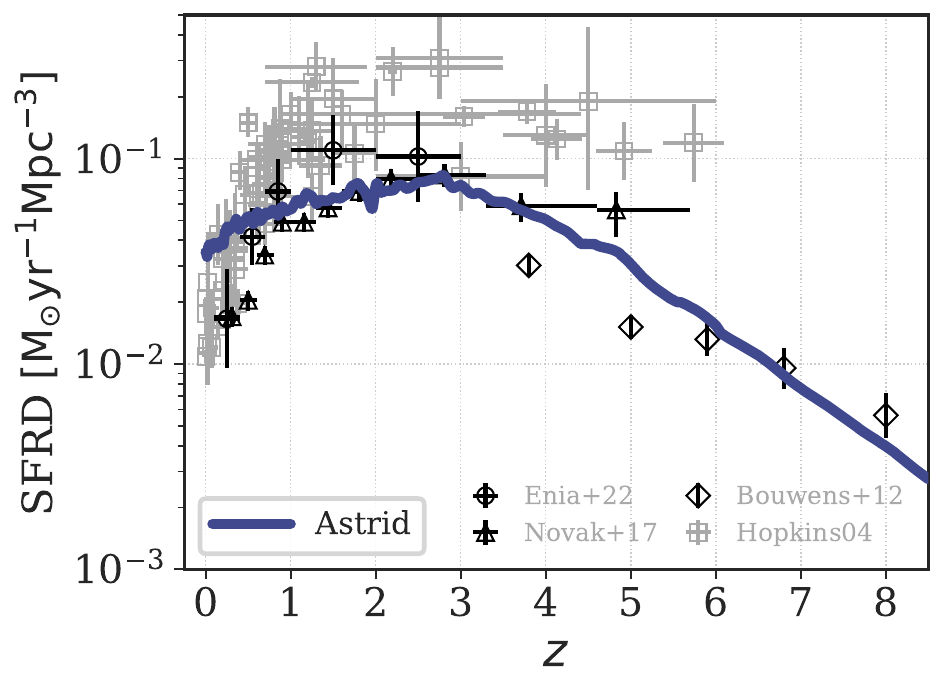}
    \caption{The evolution of cosmic star formation rate density (SFRD) as a function of redshift.
    We compare the SFRD in \astrid\ (blue curve) with the observational constraints compiled from \citet{Enia2022}, \citet{Novak2017}, \citet{Bouwens2015}, and \citet{Hopkins2004}. 
    }\label{fig:SFRD}
\end{figure}

\begin{figure*}[htbp]
    \centering
    \includegraphics[width=0.95\linewidth]{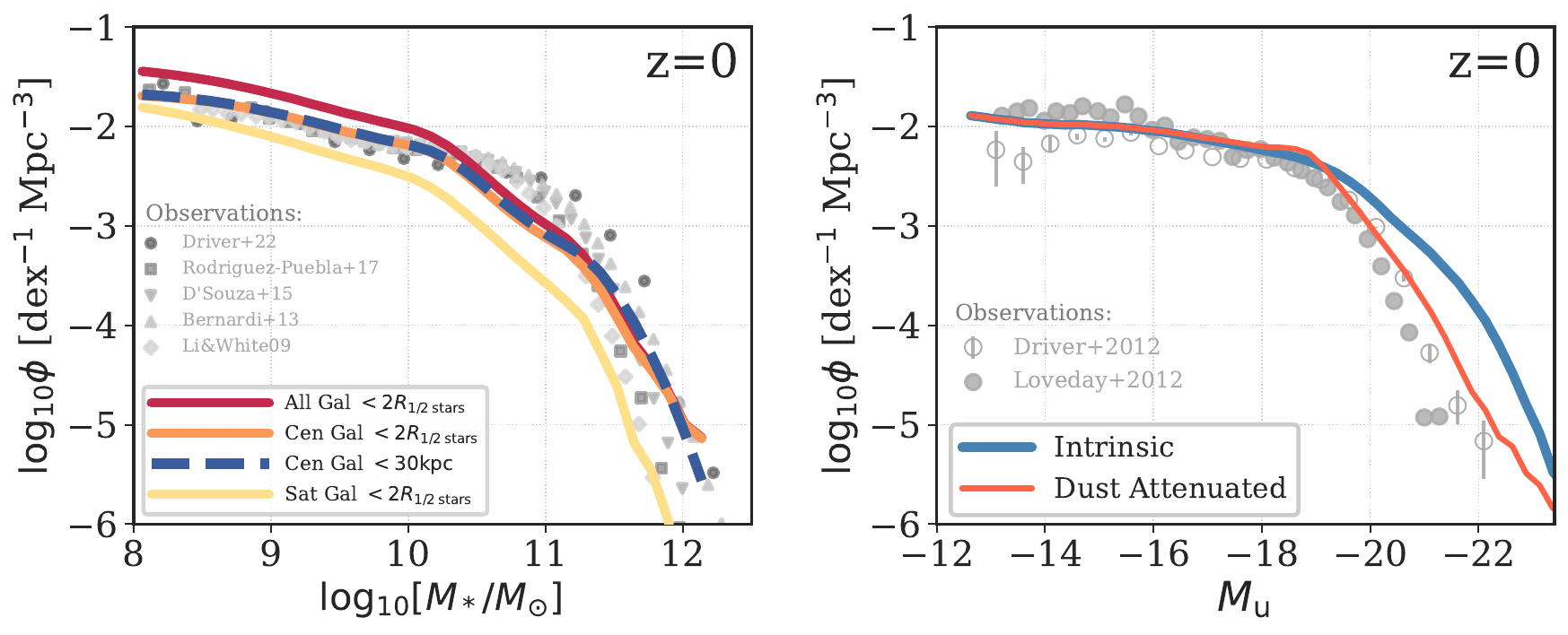}
    \caption{ \textit{Left:} the $z=0$ galaxy stellar mass function in \astrid. The red, orange, and yellow solid curves account for the stellar mass within twice the stellar half-mass-radius for all the galaxies, central galaxies, and satellite galaxies, respectively.
    The blue dashed line uses the stellar mass within 30 kpc for the entire galaxy population. We also show the observational constraints from \citet{Driver2022}, \citet{Rodriguez-Puebla2017}, \citet{D'Souza2015}, \citet{Bernardi2013}, and \citet{Li2009} using the gray dots.
    \textit{Right:} the galaxy luminosity function from $z=0$ \astrid\ observed at SDSS u band. The blue curve represents the intrinsic magnitude, and the red curve incorporates the dust attenuation. Observations from \citet{loveday2012galaxy} and \citet{driver2012galaxy} are marked by gray circles. 
    }\label{fig:gal_mass_func}
\end{figure*}

We further explore the quiescent fraction of galaxies. 
Following the definition adopted \citet{Schaye2015, Schaye2023}, we classify galaxies at $z=0$ as quiescent if their $s$SFR is smaller than $10^{-11}/{\rm yr}$; galaxies above this threshold are considered star-forming. 
In observational studies, galaxies are divided into star-forming (late type) or quiescent (early type or passive) populations using a variety of diagnostics: 
the color-color diagram \cite[e.g.][]{Muzzin2013}, S\'ersic index \cite[e.g.][]{Lange2015}, or the concentration index \cite[e.g.][]{Shen2003}. 
Replicating these diagnostics in simulations would require a more sophisticated dust attenuation model tailored to specific color bands, which is beyond the scope of this work. 
In Figure~\ref{fig:QuiescentFrac}, we plot the quiescent fraction predicted by \astrid\ as a function of galaxy stellar mass at $z=0$. 
Overall, the quiescent fraction increases with galaxy mass, consistent with the general observational trend. 
The upturn at $M_{\star}
\lesssim 10^{10}$~\Msun\ is a resolution effect: as we rely on the newly formed stars over the past 200 Myr to estimate the galaxy's SFR, the finite particle mass imposes a lower limit on the resolvable SFR. 
This limitation also appears in the $s$SFR panel of Figure~\ref{fig:galaxy_property}: no galaxies populate the lower-left region of the $s$SFR-$M_{\star}$ plane. 
In the intermediate mass range $M_{\star}=10^{10-11}$~\Msun, the quiescent fraction in \astrid\ is lower than observations. 
The discrepancy is particularly pronounced around $5\times 10^{10}$~\Msun, where \astrid\ predicts approximately 10\% quiescent fraction, while observational constraint is over 40\% \citep{Weaver2023}.  
This mass range corresponds to the regime in the top-left panel of Figure~\ref{fig:galaxy_property} where \astrid\ shows high $s$SFR compared to observations. 
At the high-mass end ($M_{\star}>10^{11.5}$~\Msun), the quiescent fraction begins to decline, a trend that is validated by one of the observational works \citet{Moustakas2013}. As mentioned before, this may be attributed to the frequent mergers experienced by these massive galaxies. 

The top-right panel of Figure~\ref{fig:galaxy_property} presents the correlation between the half-mass-radius $R_{1/2}$ and the galaxy stellar mass. 
We plot the median value for the star-forming (blue) and quiescent (red) galaxy populations separately. 
For small galaxies with $M_{\star}<5\times10^{11}$~\Msun, the median $R_{1/2}$ over the galaxy mass is flat for both star-forming and quiescent galaxies, varying between $2$ and $3$ kpc. For $M_{\star}>5\times 10^{11}$~\Msun, $R_{\rm 1/2}$ increases steeply, with $\log R_{\rm 1/2} \propto 1.1 \log M_{\star}$.
At $M_{\star}\sim10^{10-11}$~\Msun, we observed a clear drop in $R_{\rm 1/2}$ for quenched galaxies.
\cite{Ni2024} found similar results and demonstrated that the central black holes in these galaxies lie above the $M_{\rm BH}-M_{\star}$ relations. These massive MBHs are fueled by intense gas inflows, resulting in a compact galaxy morphology, and their AGN feedback simultaneously suppresses star formation.  
We also plot some results from observations: \citet{Lange2015}, \citet{vanderWel2014}, \citet{Dutton2011}, and \citet{Shen2003}. 
Among them, \citet{Shen2003} classified star-forming/quiescent galaxies according to their concentration factor, \citet{Lange2015} based their
classification on the S\'ersic index, and \citet{vanderWel2014} and \citet{Dutton2011} used  color-color diagrams. 
The differences in these identification methods and the criteria adopted in this work (i.e., solely based on the $s$SFR) are likely to cause some of the deviations seen in the plot.  
For a detailed discussion of various systematic factors involved in the comparison of $R_{\rm 1/2}-M_{\star}$ relations from observations to simulations, we refer readers to \citet{Genel2018}.

In the bottom panels of Figure~\ref{fig:galaxy_property}, both the stellar metallicity and the velocity dispersion are estimated based on the star particles within twice the stellar half-mass-radius. 
For the $\sigma-M_{\star}$ correlation, we plot the fitted relation from \citet{Dutton2011} and \citet{Gallazzi2006}.
\astrid\ generates a $\sigma-M_{\star}$ correlation in remarkable agreement with the observations.

\begin{figure}[htbp]
    \centering
    \includegraphics[width=1\linewidth]{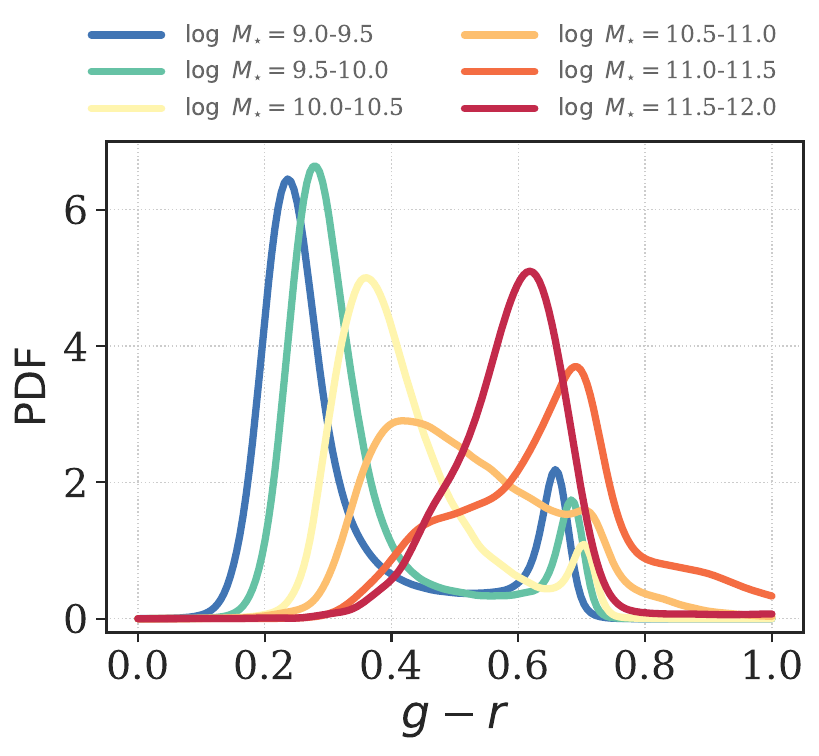}
    \caption{The distribution of simulated SDSS $g-r$ colors for galaxies in \astrid\ at $z=0$. Dust attenuation is implemented. 
    We show the galaxies in different mass bins separately, as indicated by the legend. 
    }\label{fig:galaxy_color}
\end{figure}

\begin{figure*}[htbp]
    \centering
    \includegraphics[width=0.9\linewidth]{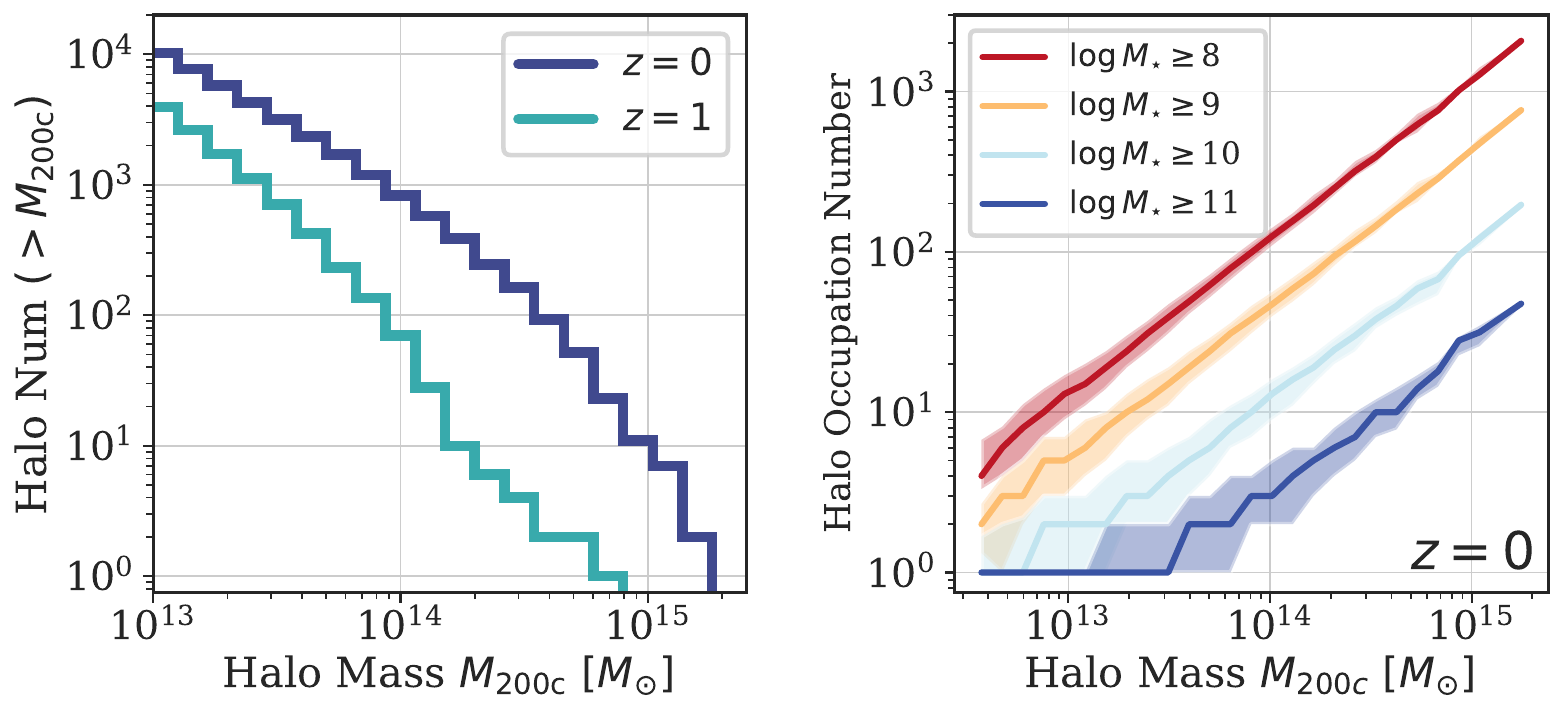}
    \caption{Galaxy cluster and groups in \astrid, i.e., halos with $M_{\rm 200c}\geq 10^{13}$~\Msun. 
    \textit{Left}: cumulative halo number above the given halo mass $M_{\rm 200c}$ at $z=0$ (blue) and $z=1$ (cyan). 
    \textit{Right}: halo occupation number in \astrid, i.e., average number of member galaxies within $R_{\rm 200c}$ as a function of halo mass. 
    Different mass cuts are applied for the member galaxies: $M_{\star}\geq 10^{8}$~\Msun\ (red), $M_{\star}\geq 10^{9}$~\Msun\ (yellow), $M_{\star}\geq 10^{10}$~\Msun\ (cyan), and $M_{\star}\geq 10^{11}$~\Msun\ (blue).  
    The shaded area covers the 16th-84th percentile. 
    }\label{fig:halo_mass_func}
\end{figure*}

The stellar metallicity $Z_{\star}$ shown in the bottom-right panel is normalized by the solar metallicity $Z_{\star}=0.0127$. 
$Z_{\star}$ increases with the stellar mass and flattens off for $M_{\star}>10^{11}$~\Msun, as a result of the suppressed star formation. 
Compared to the observational results from the Sloan Digital Sky Survey (SDSS) DR2 \citep{Gallazzi2006}, the galaxies in \astrid\ have a larger metallicity, while still residing in the regime of statistical error of the observational constraints. 

\subsection{Galaxy Stellar Mass Function}
\label{sec:GSMF}

Figure~\ref{fig:SFRD} plots the global SFRD over cosmic time. 
In addition to the evolution of SFRD in \astrid\ (blue curve), we also include the observations compiled in \citet{Enia2022}, \citet{Novak2017}, \citet{Bouwens2015}, and \citet{Hopkins2004}. 
At redshifts above $z=2$, the SFRD in \astrid\ is in good agreement with the observational constraints. However, at low redshift, the SFRD in the real Universe is observed to drop rapidly. Compared to the peak of $0.1\,$~\Msun${\rm yr}^{-1}{\rm Mpc}^{-3}$ at $z=1$, the SFRD at the local Universe is 1 order of magnitude lower. In \astrid, the SFRD drops relatively slowly. At $z=0$, SRFD is about $3\times 10^{-2}$~\Msun${\rm yr}^{-1}{\rm Mpc}^{-3}$. Such behavior is similar to the evolution of BHAD shown in Figure~\ref{fig:BHAD} as the black hole accretion rate closely tracks the cosmic SFRD. Both of them can be explained by the inefficient AGN feedback at low redshifts. 

In the left panel of Figure~\ref{fig:gal_mass_func} we show the galaxy stellar mass function (GSMF) at $z=0$.
The blue solid curve, red dashed line, and the orange dashed line account for the stellar mass within twice the stellar half-mass-radius for all the galaxies, central galaxies, and satellite galaxies, respectively.
As mentioned in \citet{Ni2024}, some massive galaxies have a diffuse component caused by tidal stripping, so that the mass within $R_{1/2\ \rm star}$ cannot properly account for the galaxy mass. 
Hence, to more closely mimic observational data, we also present the results using a 3D spherical aperture of $30$~ckpc/h applied to the central galaxies (yellow dot-dashed curve).
The choice of aperture cut only affects galaxies at the high-mass end ($M_{\star}
\gtrsim 10^{12}$~\Msun).
We compare to  a compendium of observational data from \citet{Driver2022}, \citet{Rodriguez-Puebla2017}, \citet{D'Souza2015}, \citet{Bernardi2013}, and \citet{Li2009}, which is plotted using  gray dots.
Overall, \astrid\ satisfies the available observational constraints, especially the mass function based on the central galaxy within twice the stellar half-mass-radius. 
The galaxy abundance at the low-mass end ($M_{\star}<10^{9}$~\Msun) is regulated mainly by the stellar feedback. Hence, the consistency demonstrated here implies that the SN wind feedback performs well in \astrid. 
The observational samples exhibit a more pronounced `knee' around $M_{\star}\sim 10^{11}$, and \astrid\ predicts a mass function up to 0.5 dex lower compared to the Galaxy and Mass Assembly (GAMA) survey \citet{Driver2022}.
Similar deficiencies appear as early as $z=2$ in \astrid\ \citep{Ni2024}, and also generated by FABLE simulation suite \citep{Henden2018_fable, Bigwood2025_xfable}.
This can be attributed to the high SFR of the galaxies in this mass range, as shown in the top-left frame of Figure~\ref{fig:galaxy_property}. The larger SFR makes the galaxies grow faster, thus fewer galaxies within this mass range are observed, while more massive galaxies are produced.

\subsection{Galaxy Luminosity Function}


The galaxy luminosity function and galaxy colors are 
key observables that constrain models of star formation and stellar evolution \citep{Kauffmann2003, Baldry2006}. 
In this section, we present the $z=0$ luminosity function and  SDSS $g-r$ colors of galaxies in \astrid, facilitating direct comparison with observations. 
We model each star particle as a simple stellar population (SSP) with its birth time, metallicity, and the mass extracted from the simulation.
We use the \textsc{FSPS} stellar population synthesis code \citep{Conroy2009, Conroy2010} with the PARSEC isochrones \citep{Bressan2012_parsec}, MILES stellar library \citep{Sanchez-Blazquez2006_miles}, assuming a Chabrier initial mass function \citep{Chabrier2003}. 
We attenuate the stellar light based on the metals along each star's line of sight, with a wavelength dependence following a simple power-law relation with a slope of $\gamma=-1.0$ \citep{lachance2024, Wilkins2017}. This is steeper than the Starburst curve \citep{Calzetti2000}, but flatter than the Small Magellanic Cloud curve \citep{Pei1992}. The metal surface density ($\Sigma(x,y,z)$) is calculated by creating a 3D metal density map from the gas particles in the galaxy and summing the column in front of each star particle. This is used to determine the optical depth at a given wavelength via the equation
\begin{equation}
\tau_{\mathrm{ISM}}(\lambda) = -\kappa_{\mathrm{ISM}} \, \Sigma(x,y,z)\left(\frac{\lambda}{0.55\,\mu m}\right)^{\gamma},
\end{equation}
$\kappa_{\mathrm{ISM}}$ is a calibration parameter that we set to $10^{3.0}$ by calibrating against observed galaxy luminosity functions in the SDSS u,g, and r bands \citep{driver2012galaxy, loveday2012galaxy}. 
The luminosity for an individual galaxy is computed by summing the emission of all of the star particles contained in this galaxy.

The stellar luminosities calculated during this process are also used in the creation of the mock observations shown in Figure~\ref{fig:large_scale} and Figure~\ref{fig:mock_obs_w_BH_tracks} following the process described in \citet{lachance2024}. 
They are created by adding each star particle's luminosity in a given filter to the mock observation based on its position, and smoothing it with the SPH kernel according to its smoothing length. For the JWST images, we use a pixel resolution of 0.030" with the NIRCam F444W, F277W, and F115W filters being the red, green, and blue channels, respectively. For the HST images, the WFC3 F625W, F475W, and F390W filters are used for the three color channels, with a pixel resolution of 0.040". All mock observations are made as though the galaxies are at a distance of z=0.05. 

The impact of dust attenuation is most pronounced in the u band, and we show the galaxy luminosity function in this band in the right panel of Figure~\ref{fig:gal_mass_func}. 
After applying the dust attenuation, \astrid\ produces a luminosity function in good agreement with the observational constraints. 
The distribution of dust-attenuated galaxy color (SDSS $g-r$) is presented in Figure~\ref{fig:galaxy_color}, and galaxies within different mass bins are plotted separately, as indicated by the legend. 
It can be seen that \astrid\ successfully reproduces the galaxy color bimodality. 
At the low-mass end ($\log M_{\star}<10$), the color peaks around $g-r=0.3$. 
With increasing stellar mass, galaxies become redder, and the peak of the distribution is located at $g-r=0.65$ for galaxies with $11.5<\log \ M_{\star}<12.0$~\Msun. 
A sharp transition occurs at a mass scale $M_{\star}\approx10^{11}$~\Msun. 
This transition shifts to a higher mass compared to IllustrisTNG (at $M_{\star}\approx 10^{10}$~\Msun) \citep{Nelson2018} and MillenniumTNG (at $M_{\star}\approx 10^{10.5}$~\Msun) \citep{Pakmor2023}. 
We associate this with differences in the efficiency of the kinetic AGN feedback.

\section{Groups and clusters of galaxies} 
\label{sec:groups}

Due to its large simulation volume, \astrid\ hosts a population of massive galaxy clusters or groups ($M_{\rm 200c}\geq 10^{13}$~\Msun). In this section, we present their demographics and provide a census of their stellar mass content.

\begin{figure}[htbp]
    \centering
    \includegraphics[width=1\linewidth]{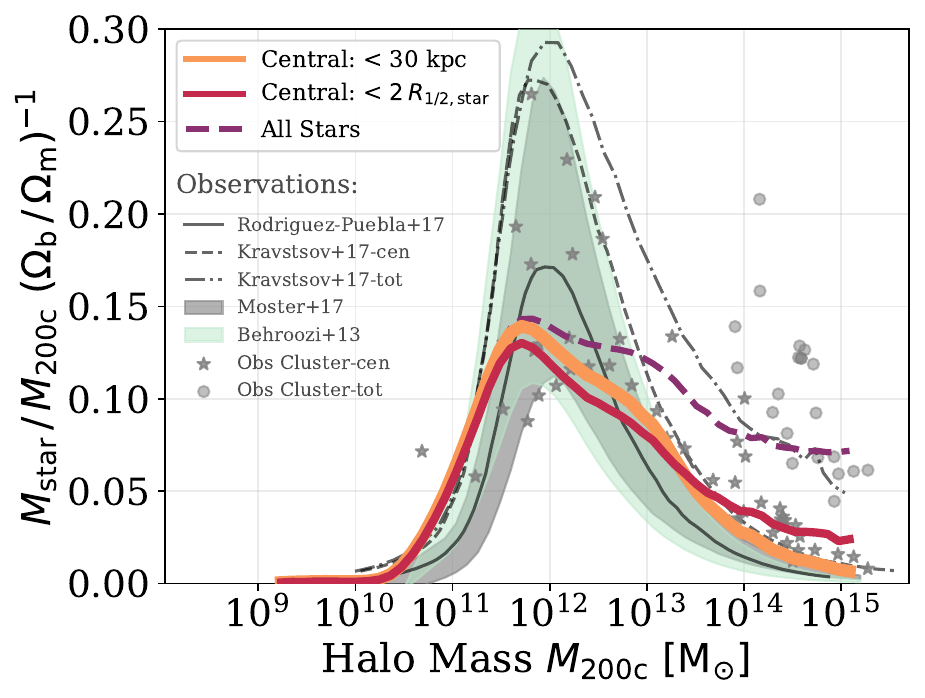}
    \caption{ The stellar mass-halo mass (SMHM) relation for halos in \astrid\ at $z=0$. The yellow and red curves are the median values based on the central galaxies' stellar mass within fixed spherical apertures of 30 kpc, and twice the stellar half-mass-radius, respectively. 
    The purple dashed line is the median value using all the stars in the halos. 
    The black solid / dash / dash-solid curve represents results from the abundance matching or semi-empirical models from \citet{Rodriguez-Gomez2015_illu_mergingrate} / central galaxies in \citet{Kravtsov2018} / total star in \citet{Kravtsov2018}. 
    The black/green shaded areas are the $1\sigma$ region for \citet{Moster2018} and \citet{Behroozi2013}. The gray stars and circles are observational data from \citet{Kravtsov2018}.
    }\label{fig:SMHM}
\end{figure}

\begin{table*}[t]
\centering
\caption{Best-fitting Parameters to the Relation between Stellar Components and Halo Mass for Groups and Clusters in \astrid\ at $z=0$. 
}\label{table:cluster_fitting}
\begin{tabular*}{0.85\textwidth}{c@{\hspace*{26pt}}c@{\hspace*{24pt}}c@{\hspace*{35pt}}
c@{\hspace*{25pt}}c@{\hspace*{26pt}}c@{\hspace*{25pt}}}
\specialrule{0.1em}{0.05pt}{2pt}     
\specialrule{0.08em}{0.05pt}{2pt}     
      $Y$
      & $X$
      & 3D apertures
      & $a$
      & $b$
      & scatter
      \\ \specialrule{0.08em}{4pt}{4pt}
 $M_{\rm star,all}$ &  $ M_{\rm 500c}$ & $<R_{\rm 500c}$ & $0.793\pm0.002$  &  $1.010\pm0.033$ & $0.079$ \\
 $M_{\rm star,cen}$ & $ M_{\rm 500c}$ & $>30$ kpc & $0.437\pm0.004$  & $5.538\pm0.053$ & $0.127$ \\
 $M_{\rm star,cen}$ & $ M_{\rm 500c}$ & $>100$ kpc & $0.527\pm0.004$  & $4.390\pm0.051$ & $0.123$ \\
 $M_{\rm star,cen}$ &  $ M_{\rm 500c}$ & $>2R_{\rm1/2\, stars}$ & $0.678\pm0.003$  & $2.348\pm0.044$ & $0.106$ \\
 $M_{\rm star,sat}$ & $ M_{\rm 500c}$ & $30\,{\rm kpc} - R_{\rm 500c}$  & $1.273\pm0.010$  & $-6.089\pm0.133$ & $0.319$ \\
 $M_{\rm star,sat}$ & $ M_{\rm 500c}$ &  $100\,{\rm kpc} - R_{\rm 500c}$ & $1.362\pm0.011$  & $-7.362\pm0.143$ & $0.343$ \\
 $M_{\rm star,ICL}$ & $ M_{\rm 200c}$ &  $30\,{\rm kpc} - R_{\rm 200c}$ & $1.276\pm0.005$  & $-6.435\pm0.068$ & $0.162$ \\
 $M_{\rm star,ICL}$ & $ M_{\rm 200c}$ &  $100\,{\rm kpc} - R_{\rm 200c}$ & $1.632\pm0.005$  & $-11.729\pm0.068$ & $0.163$ \\
\bottomrule
\end{tabular*}
\begin{minipage}{0.85\textwidth}
\footnotesize
\noindent\textbf{Note.} The adopted fitting function is
$\log_{10}Y = a\,\log_{10}X + b$. The third column is the 3D aperture
applied on the stellar components (the first column).
The last column is the standard deviation of the residuals.
\end{minipage}

\end{table*}

\subsection{Halo Mass Function}
The left panel of Figure~\ref{fig:halo_mass_func} gives the cumulative histogram of halo number above a given mass at $z=0$ (blue) as well as $z=1$ (cyan). 
At $z=0$, the most-massive halo has $M_{\rm 200c} = 1.8\times 10^{15}$~\Msun\ ($M_{\rm 500c}=1.5\times 10^{15}$~\Msun), which is a Coma-like cluster, i.e., $M_{\rm 200c} \sim 1.3-2 \times 10^{15}$~\Msun\ at $z=0.023$ \citep{Weinmann2011}. Its central galaxy has a stellar mass of $M_{\star}=10^{12.8}$~\Msun\ measured within twice the stellar half-mass-radius. 
\citet{Zhou2025_PTA_CW} found that the most-massive halo in \astrid\ hosts a triple merger event involving two consecutive MBH mergers occurring within a 500 Myr duration, and both of these two mergers are potential PTA CW sources with high detectability. This provides insights into the environments where CW sources will be observed, offering valuable guidance for the future search for their EM counterpart. 
At $z=0$, there are 7 massive galaxy clusters, halos with $M_{\rm 200c} > 10^{15}$~\Msun. There are 9709 galaxy groups, halos with $M_{\rm 200c} \geq 10^{13}$~\Msun. 
At $z=1$, the most-massive halo is $M_{\rm 200c}=7\times 10^{14}$~\Msun, and there are 3715 halos with $M_{\rm 200c} > 10^{13}$~\Msun. 
The number of halos with $M_{\rm 200c}\geq 10^{14}$~\Msun\ increases by 1 order of magnitude from $z=1$ to $z=0$. 
Based on this, \astrid\ provides a large sample of galaxy clusters or groups, maintaining good statistics at the high-mass end.

In the right panel of Figure~\ref{fig:halo_mass_func}, we plot the richness, or halo occupation number, i.e., the number of galaxies above a given stellar mass threshold in an individual halo. 
Only galaxies within $R_{\rm 200c}$ are counted. 
The galaxy richness is a steep function of the total halo mass, whose slope is insensitive to the adopted galaxy mass threshold. 
Galaxies clusters with $M_{\rm 200c}\geq 10^{15}$~\Msun\ host about 30 massive galaxies ($M_{\star}\geq 10^{11}$~\Msun), and over 1000 galaxies with $M_{\star}\geq 10^{8}$~\Msun. Low-mass halos ($M_{\rm 200c}\sim 10^{13}$~\Msun) have fewer than 20 member galaxies. 
Galaxies with $M_{\star}\geq 10^{11}$~\Msun\ can only be found in halos of $M_{\rm 200c} > 10^{13}$~\Msun.

\begin{figure*}[!htbp]
    \centering
    \includegraphics[width=1\linewidth]{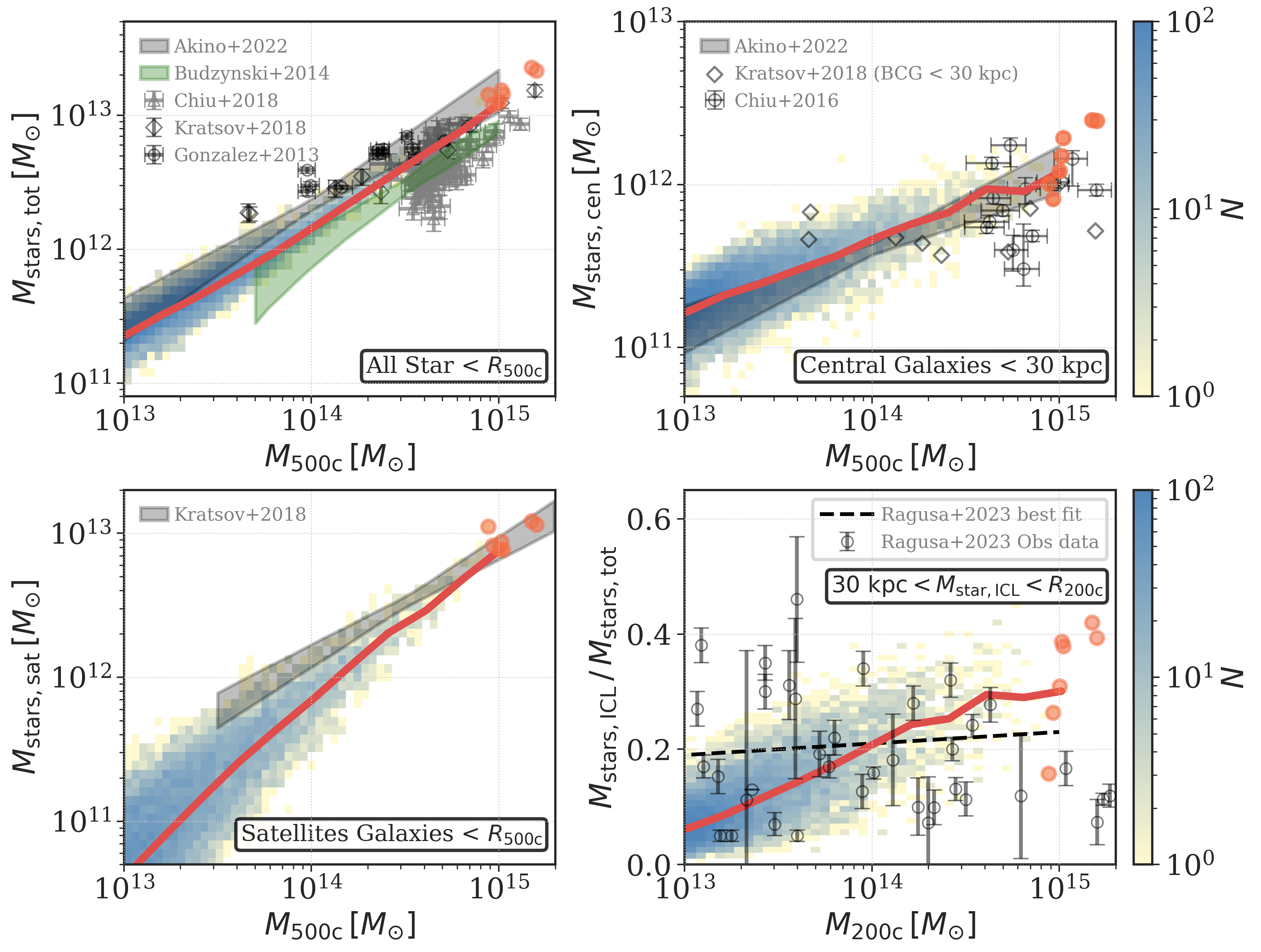}
    \caption{ The stellar mass budget of galaxy groups and clusters. 
    In each panel, the underlying distribution corresponds to the number density of the halos in \astrid\ at $z=0$. The four panels share the same color bars, which are shown on the right. 
    The red curve represents the median value for the halos in the given mass bin.  
    The seven most-massive halos with $M_{\rm 200c}\geq 10^15$~\Msun\ are highlighted by the red dots. 
    \textit{Top left:} the total stellar mass in the halo (including central galaxies, satellite galaxies, and ICL) within $R_{\rm 500c}$ as a function of the halo mass $M_{\rm 500c}$. 
    As indicated by the legend, we include the observational constraints from \citet{Akino2022}, \citet{Budzynski2014}, \citet{Chiu2018}, \citet{Kravtsov2018}, and \citet{Gonzalez2013}.
    \textit{Top right:} the star mass in the central galaxies versus the $M_{\rm 500c}$. The central stellar mass is measured within $30$~kpc from the galactic center. To guide the eye, we show the observational data from \citet{Akino2022}, \citet{Kravtsov2018}, and \citet{Chiu2018}. 
    \textit{Bottom left}: the cumulative stellar mass in satellites within $R_{\rm 500c}$. The gray shaded area is the observational result from \citet{Kravtsov2018}. 
    \textit{Bottom right:} the stellar mass fraction in the ICL as a function of $M_{\rm 200c}$. Note we use $M_{\rm 200c}$ for this panel rather than $M_{\rm 500c}$ like the other three subplots to enable a direct comparison with the observational constraints compiled in \citet{Ragusa2023} (the gray circles), which includes the observed galaxy clusters or groups with $z<0.05$ from \citet{Pildis1995, DaRocha2005, DaRocha2008, Mihos2017, Jimenez-Teja2019, Ragusa2021, Kluge2021, Poliakov2021, Montes2021}). The black dashed curve is the best fit provided in \citet{Ragusa2023} based on this dataset.     
    }\label{fig:cluster_mass_budget}
\end{figure*}

\subsection{The SMHM Relation}


In Figure~\ref{fig:SMHM}, we show the stellar mass-halo mass (SMHM) relation for halos in \astrid\ at $z=0$, parameterized by $M_{\rm star}/M_{\rm 200c}(\Omega_{\rm b}/\Omega_{\rm m})^{-1}$.
The yellow and red curves use the central galaxies' stellar mass within a fixed spherical aperture of 30 kpc and twice the stellar half-mass-radius, respectively. 
We use different spherical apertures to account for the galaxy stellar mass to give a better comparison to the observational data. 
The purple dashed line includes all the stars in the halos. 
We also show a variety of observational surveys. There is limited consensus among observational results, as the measurement of $M_{\rm star}/M_{\rm 200c}$ for a halo is challenging, and these observations suffer from a lack of common calibration and definition of halo mass.

\astrid\ produces a strongly peaked SMHM relation, in agreement with the observational constraints. with
It has long been recognized that the efficiency with which halos convert baryons to stars peaks around $M_{\rm 200c}\sim 10^{12}$~\Msun\ and is reduced for higher-mass halos \citep{Conroy2009a,Leauthaud2012}.   
In \astrid, the peak is at a slightly lower mass: $M_{\rm 200c}=10^{11.8}$~\Msun, with a conversion efficiency of 14\%. 
The location of the SMHM peak is sensitive to details of the stellar and AGN feedback prescriptions, including the adopted BH feedback efficiency, whether galactic winds are implemented isotropically / with metallicity dependence, or whether a minimum wind velocity is imposed \citep{Pillepich2018_tngmodel, Ni2023_camels}.
Toward the low-mass end, $M_{\rm star}/M_{\rm 200c}$ declines steeply: the star-forming efficiency halves at $M_{\rm 200c}=10^{11}$~\Msun. 
In the range $10^{10.5}\lesssim M_{\rm 200c}\lesssim 10^{11.5}$, \astrid\ predicts a higher stellar conversion efficiency than typically inferred from observations.
At the high-mass end, the drop is more gradual, halving at $M_{\rm 200c}\approx 10^{13}$~\Msun.
For the most-massive galaxy clusters with $M_{\rm 200c}>10^{14}$~\Msun, \astrid\ yields a stellar mass fraction in good agreement with most observational constraints. 
For central galaxies, applying a fixed $30$~kpc aperture produces measurements that are more consistent with observations than those based on twice the stellar half-mass-radius.

\subsection{Stellar Mass Budget}
\label{sec:stellar_mass_budget}

In this section, we present the stellar mass census of the cluster components: central galaxies, satellite galaxies, and intracluster light (ICL). 
Following \citet{Pillepich2019}, we define the ICL as the stellar mass beyond a fixed aperture from the center of the halo to a maximum boundary set by the virial radius (e.g., $R_{\rm 200c}$ or $R_{\rm 500c}$), with the satellites and gravitationally unbound stars removed. 
In this work, we adopt the fixed aperture boundary of 30 kpc. 

We focus on the FOF halos in \astrid\ at $z=0$ with $M_{\rm 500c} \geq 10^{13}$~\Msun, which is similar to the observed galaxy groups and clusters, and present their stellar mass budget in Figure~\ref{fig:cluster_mass_budget}.
In the top-left frame, we show the mass of all stars within $R_{\rm 500c}$ as a function of the halo mass $M_{\rm 500c}$. 
In the top-right panel, we plot the mass of the central galaxy in each halo, and compare it to observational constraints. 
As can be seen in the plot, both $M_{\rm star,tot}-M_{\rm 500c}$ and $M_{\rm star,cen}-M_{\rm 500c}$ correlations generated by \astrid\ are in good agreement with the observations, especially at the high-mass end, where most of the observational data are located.

We show the total stellar mass of satellite galaxies within $R_{\rm 500c}$ in the bottom-left panel of Figure~\ref{fig:cluster_mass_budget}, compared to X-ray observations \citep{Kravtsov2018}. 
The stellar mass in satellites is a much steeper function of the halo mass than in central galaxies.
For $10^{13} < M_{\rm 500c}< 10^{15}$~\Msun, the increase in the central galaxy stellar mass is  $\approx 1$ order of magnitude, while for the satellite galaxies, their total stellar mass is boosted by more than 2 orders of magnitude. 
For $M_{\rm 500c} \sim 10^{13}$~\Msun, almost $80\%$ of the total stellar mass within an aperture of 30 kpc is contributed by the central galaxies. At the high-mass end ($M_{\rm 500c}\gtrsim10^{15}$~\Msun), the fraction of stellar mass from central galaxies drops below 20\%, and 60\% of the stellar mass is in the satellite galaxies. 
This indicates that in the massive galaxy clusters of \astrid, the central galaxies are surrounded by many massive satellite galaxies.
The less-massive galaxy groups are relatively isolated, with the central galaxy dominating the total stellar mass. 

In the bottom-right panel, we present the fraction of the total stellar mass that contributes to the ICL as a function of $M_{\rm 200c}$. 
We use $M_{\rm 200c}$ for the ICL fraction rather than $M_{\rm 500c}$ like the other three panels to enable a direct comparison with the observational constraints compiled in \citet{Ragusa2023} (the gray circles in the figure). 
As can be seen, the ICL fraction for \astrid\ in the investigated mass range does not exhibit a significant evolution, and there is only a minor increase of a factor of 2 from $M_{\rm 200c}=10^{13}$~\Msun\ to $10^{15}$~\Msun.
The evolution of the ICL fraction in the observational data is even less obvious, as the best fit from  \citet{Ragusa2023} changes by $<0.05$ over this mass range.

The correlations of all components (central galaxies, satellite galaxies, ICL, and total stellar mass) with the halo mass are well described by power-law relations. 
To facilitate future comparisons, we fit the relation between different stellar components with $M_{\rm 500c}$ with $\log M = A \, \log M_{\rm 500c}\ +\ B$, and provide the optimal parameter set ${A, \ B}$ and the corresponding scatter in Table~\ref{table:cluster_fitting}.
Based on the slopes of these power laws, we can see that the stellar mass of central galaxies evolves relatively slowly as a function of $M_{\rm 500c}$, while the stellar mass bound to satellites and the ICL mass is a steeper function of the halo mass.

\begin{figure*}[!htbp]
    \centering
    \includegraphics[width=0.9\linewidth]{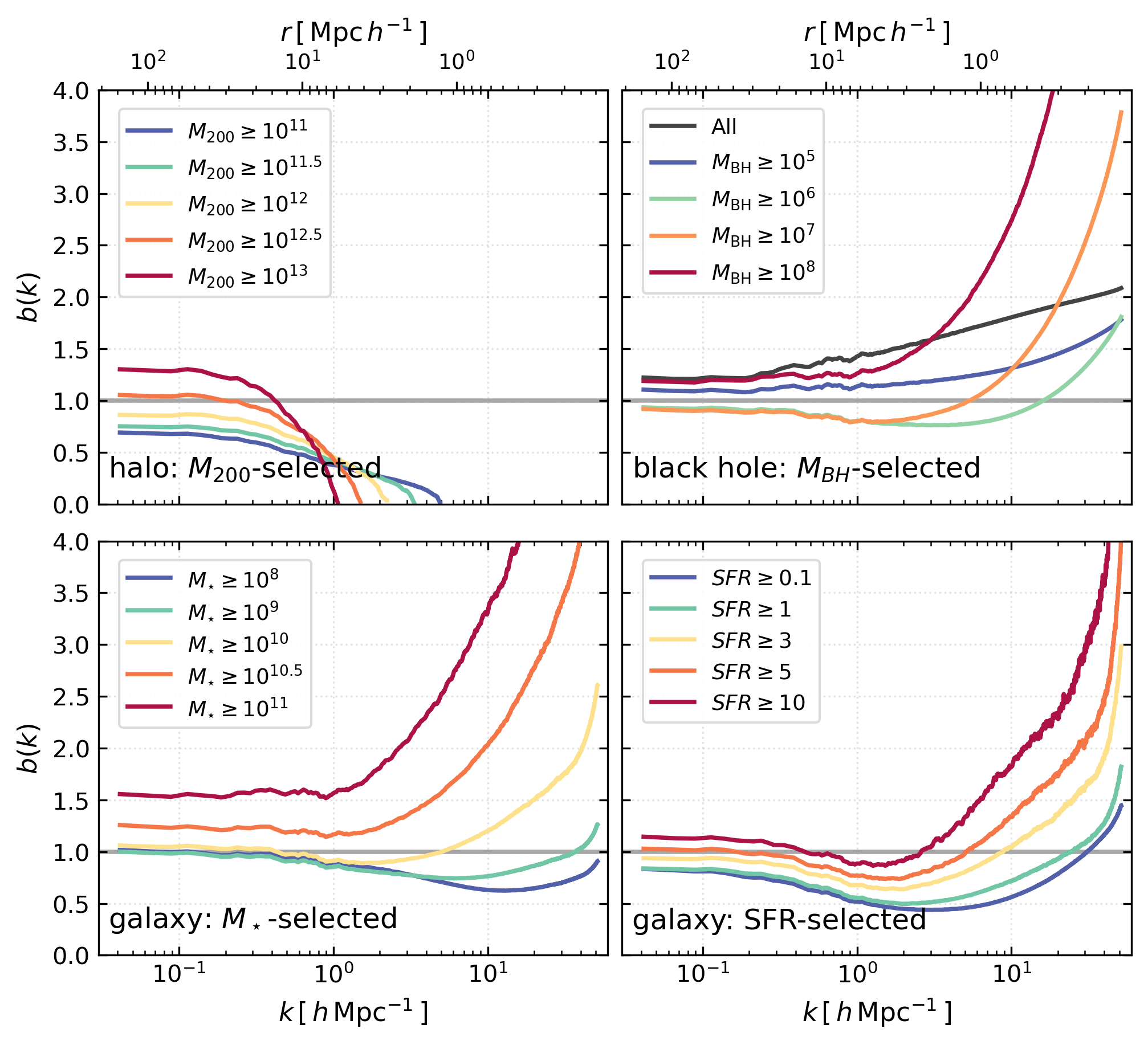}
    \caption{Scale-dependent bias for mass and SFR-selected tracers. \textit{Top left}: halos selected by $M_{200}$; \textit{top right}: black holes by $M_{\rm BH}$; \textit{bottom-left}: galaxies by stellar mass $M_\star$ and \textit{bottom-right}: galaxies by SFR. The top $x$-axis indicates real space distance $r=2\pi/k$ and the horizontal gray line shows $b(k) = 1$.}\label{fig:PS_masscut}
\end{figure*}

\section{Matter and galaxy clustering} 
\label{sec:clustering}



With a large simulation box of $250\,h^{-1}{\rm Mpc}$ per side, 
\astrid\ has sufficient volume to make a precise prediction for clustering on cosmologically relevant scales. 
In this section, we investigate the matter clustering and bias of different tracers in \astrid.
\revise{We will further study the baryonic effects on matter clustering in future work, using a dark-matter-only counterpart to ASTRID.}

\subsection{Scale-Dependent Bias}

In Figure~\ref{fig:PS_masscut}, we present the clustering bias $b(k)=[P_{\rm tracer}(k)/P_{\rm matter}(k)]^{1/2}$ as a function of wavenumber $k$. 
Observers rely on different observational tracers to study the large-scale structure of our Universe. 
These tracers are identified in different surveys according to various criteria. 
To examine the influence of sample selection criteria, we investigate four types of tracers: halos selected by $M_{\rm 200}$ (top left), MBHs selected by their mass (top right), galaxies selected by $M_{\star}$ (bottom left), and galaxies selected by SFR (bottom right), here used as a proxy for rest-frame UV selection. 
In each panel, the curves with colors from blue to red correspond to tracers with a lower mass/SFR cut (higher abundance) to a higher mass/SFR cut (lower abundance). 
For the halo samples, the bias decreases sharply with $k$ on small scales ($k>0.3\, h\,{\rm Mpc}^{-1}$), especially for the massive halos. 
The bias of BHs increases monotonically with $k$ across the entire mass range. This trend is also observed for the $M_{\star}$-selected galaxy samples, except for the small-mass population with $M_{\star}<10^{10}$~\Msun. 
For the SFR-selected galaxies, there is a suppression of bias at intermediate scales ($k\approx 1 \ h\,{\rm Mpc}^{-1}$), followed by a strong rise toward small scales.

\begin{figure}[htbp]
    \centering
    \includegraphics[width=1\linewidth]{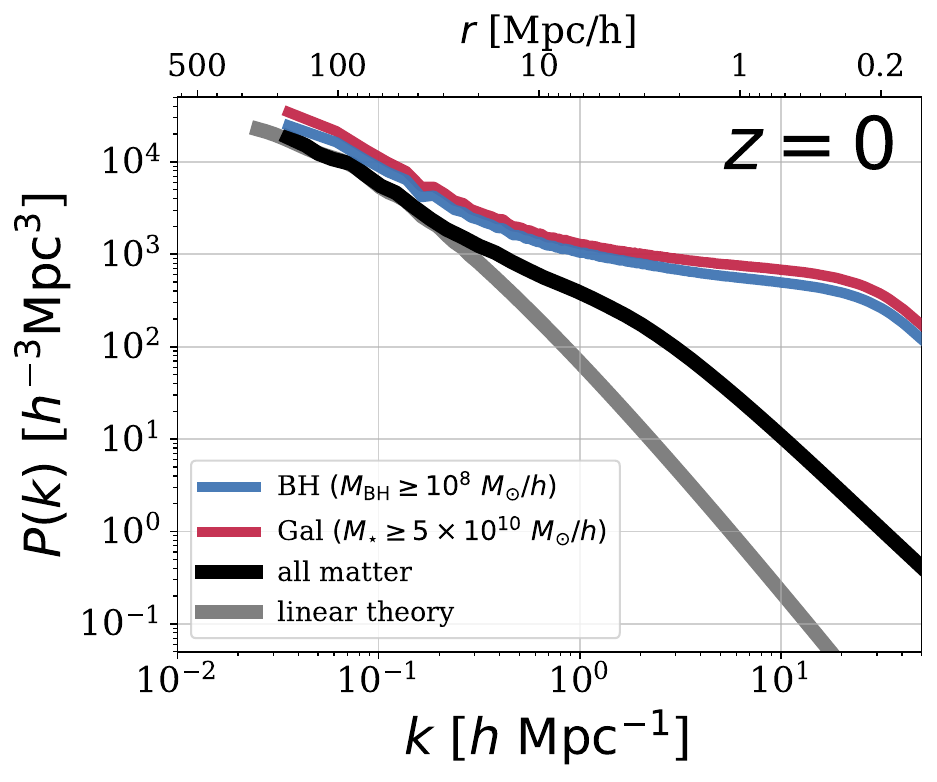}
    \caption{Matter power spectra for BHs with $M_{\rm BH}\geq 10^{8}\,h^{-1}$\Msun (blue), galaxies with $M_{\star}\geq 5\times 10^{10}\,h^{-1}$\Msun (red), and all matter (black) in \astrid\ at $z=0$.
    The gray line shows the $z=0$ matter power spectrum predicted by linear perturbation theory. 
    }\label{fig:tot_PS}
\end{figure}

The bias of BHs with $M_{\rm BH}\geq10^{8}$~\Msun\ and galaxies with $M_{\star}\geq10^{10.5}$~\Msun, stays above 1 and is quite stable for $k<1\ h\,{\rm Mpc}^{-1}$, which indicates they are  good tracers of  large-scale structure. 
In Figure~\ref{fig:tot_PS}, we plot their power spectra.
We apply the mass cut of $M_{\rm BH}\geq 10^{8}$~\Msun$h^{-1}$ for BHs and $M_{\star}\geq 5\times 10^{10}$~\Msun$h^{-1}$ for galaxies. 
For both BHs and galaxies, this corresponds to a number density of $\approx 2\times 10^{-3}\ h^{3}\,{\rm Mpc}^{3}$. 
As a reference, the typical galaxy space number densities observed by existing surveys vary between $3\times 10^{-4}$ to $5\times 10^{-3}\ h^{3}\,{\rm Mpc}^{-3}$ \citep{Zhou2023,Yuan2021,Liivamagi2012, Vakili2023}. 
For comparison, 
we also plot the total matter power spectrum in \astrid\ (black) and that predicted by linear perturbation theory (gray).
For the all-matter spectrum, the slope changes around $k=2\, h\,{\rm Mpc}^{-1}$. On smaller scales, the spectrum is dominated by the particle pairs within the same halo (`one-halo term'), and for larger scales (smaller $k$), the clustering mainly comes from the pairs in different halos (`two-halo term') \citep{Cooray2002}.  
The effects of non-linear evolution of the power spectrum are clearly visible for $k > 0.1\ h\,{\rm Mpc}^{-1}$. 
With the adopted mass cut, the BH and galaxy populations have similar power spectra across all scales, staying higher than the all-matter spectrum. 
As we found in Figure~\ref{fig:PS_masscut}, their bias is stable for the `two-halo' regime ($k < 2\, h\,{\rm Mpc}^{-1}$), while increasing significantly on smaller scales.  

\begin{figure}[htbp]
    \centering
    \includegraphics[width=1\linewidth]{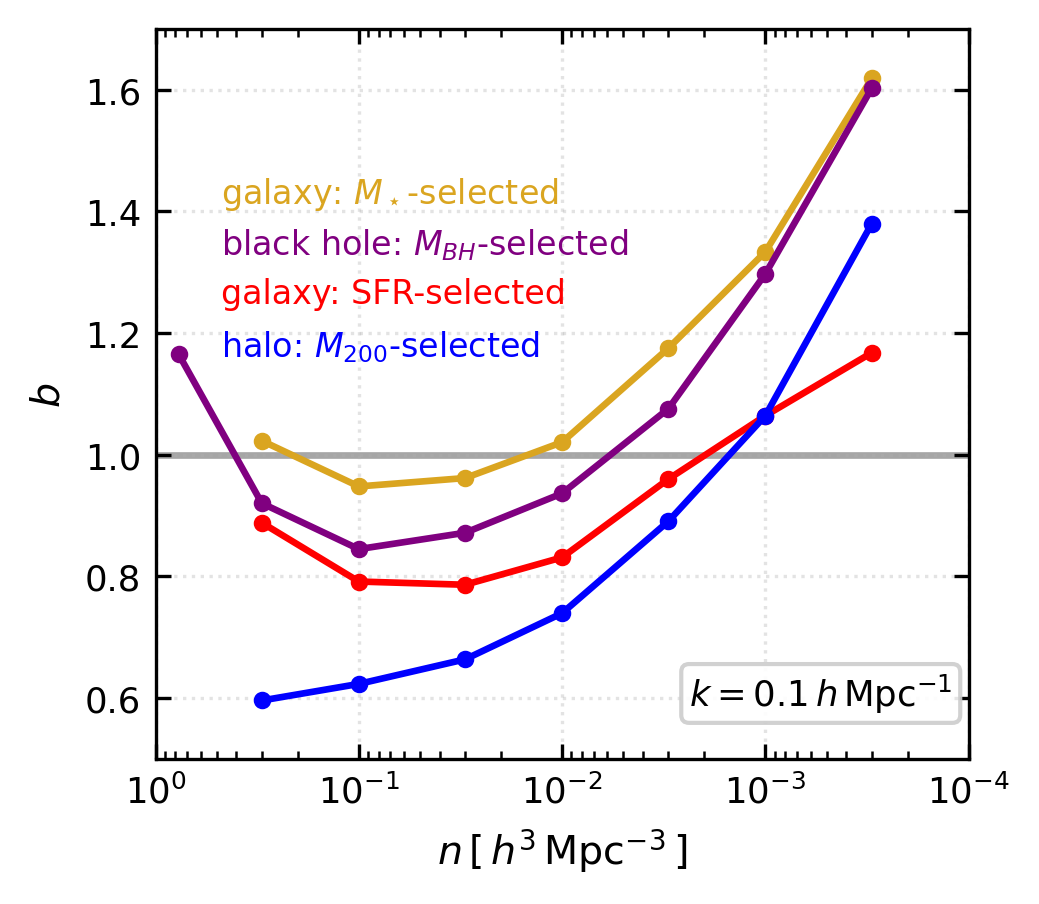}
    \caption{Large-scale (constant or linear) bias measured at $k= 0.1\ h^{-1}{\rm Mpc}$ as a function of tracer space number density $n$. We show the bias of galaxies selected by stellar mass (yellow), black holes selected by black hole mass (purple), galaxies selected by SFR (red), and halos selected by $M_{200}$ (blue). 
    The BH sample with the largest abundance (the leftmost purple point) corresponds to the entire population. 
    The horizontal gray line indicates the unit level. Note that we invert the x-axis: rarer samples are to the right. 
    }
    \label{fig:bias_vs_n}
\end{figure}

In Figure~\ref{fig:bias_vs_n}, we present the large-scale bias measured at $k=0.1\ h{\rm Mpc}^{-1}$ for different tracers. 
To allow for a direct comparison between these objects, we consider samples with different space densities $n$.
It can be seen in Figure~\ref{fig:bias_vs_n} that tracers with the same number density but of different types can exhibit substantially different biases. 
For the $n$ range achieved by current surveys ($n\lesssim 5\times 10^{3}\ h^{3}{\rm Mpc}^{-3}$), galaxies selected by $M_{\star}$ have the highest clustering, followed by BHs selected by $M_{\rm BH}$.
The highest bias comes from the rarest samples: $n=3\times 10^{-4}\,h^{3}\,{\rm Mpc}^{-3}$ and $b=1.6$. 
The galaxy population selected by SFR has a much lower bias, and crosses $b=1$ at $n\approx2\times 10^{-3}\,h^{3}\,{\rm Mpc}^{-3}$. 
For $M_{\star}$-selected galaxies and BHs, the bias becomes negative at higher $n$: $n=0.02\,h^{3}\,{\rm Mpc}^{-3}$ and $n=0.006\,h^{3}\,{\rm Mpc}^{-3}$, respectively. 
Among these tracers, halos and black holes have the largest dependence on the number density. The galaxies selected using SFR show the weakest variation of bias with number density. 
Similar results were also found in \citet{Springel2018}.
This comes from the fact that high-SFR galaxies do not tend to populate the most-massive halos, as most of the massive halos are quenched by $z=0$. 

\begin{figure*}[htbp]
    \centering
    \includegraphics[width=1\linewidth]{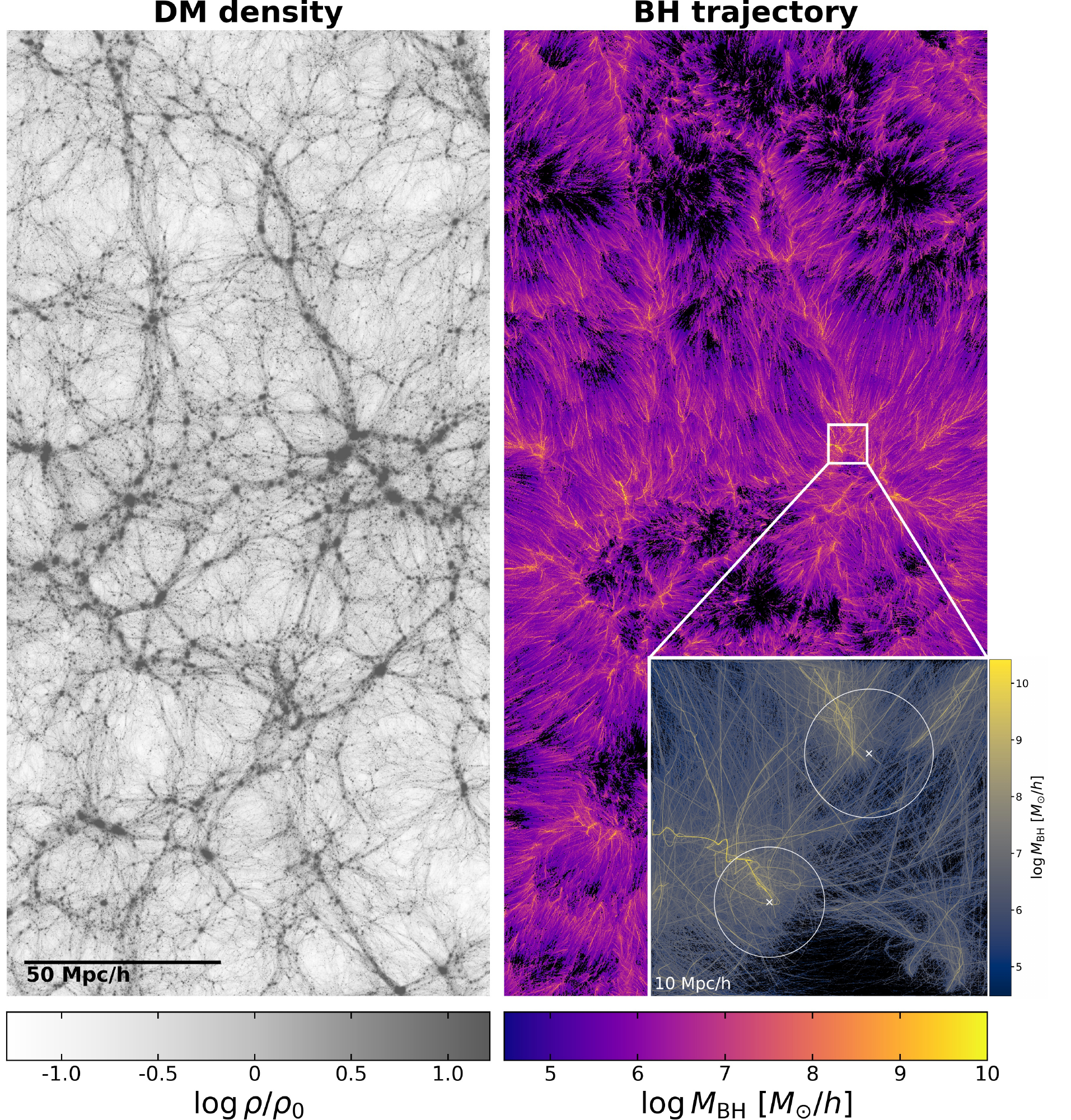}
    \caption{Black hole tracks in \astrid\ compared with the large-scale structures. 
    \textit{Left:} the DM density field in a slice of $125\times 250\times 25\, h^{-3}{\rm Mpc}^{3}$ at $z=0$. 
    \textit{Right:} the trajectories of all black holes in the same slice plotted on the right. 
    We trace the MBHs from $z=15$ to $z=0$. 
    The orbits are colored by MBH mass, with the yellow regions corresponding to the massive MBHs.
    The small inset shows the MBH trajectory in a zoomed-in $10$ Mpc/h box. We change the color bar to highlight the smaller MBHs. The two circles mark the $R_{\rm 200}$ of the most-massive halos in this region, which have $\log M_{\rm 200}$ of $14.7$ and $14.9$, respectively. The white crosses are the positions of the halo centers. }
    \label{fig:BH_tracks}
\end{figure*}

\begin{figure}[htbp]
    \centering
    \includegraphics[width=1\linewidth]{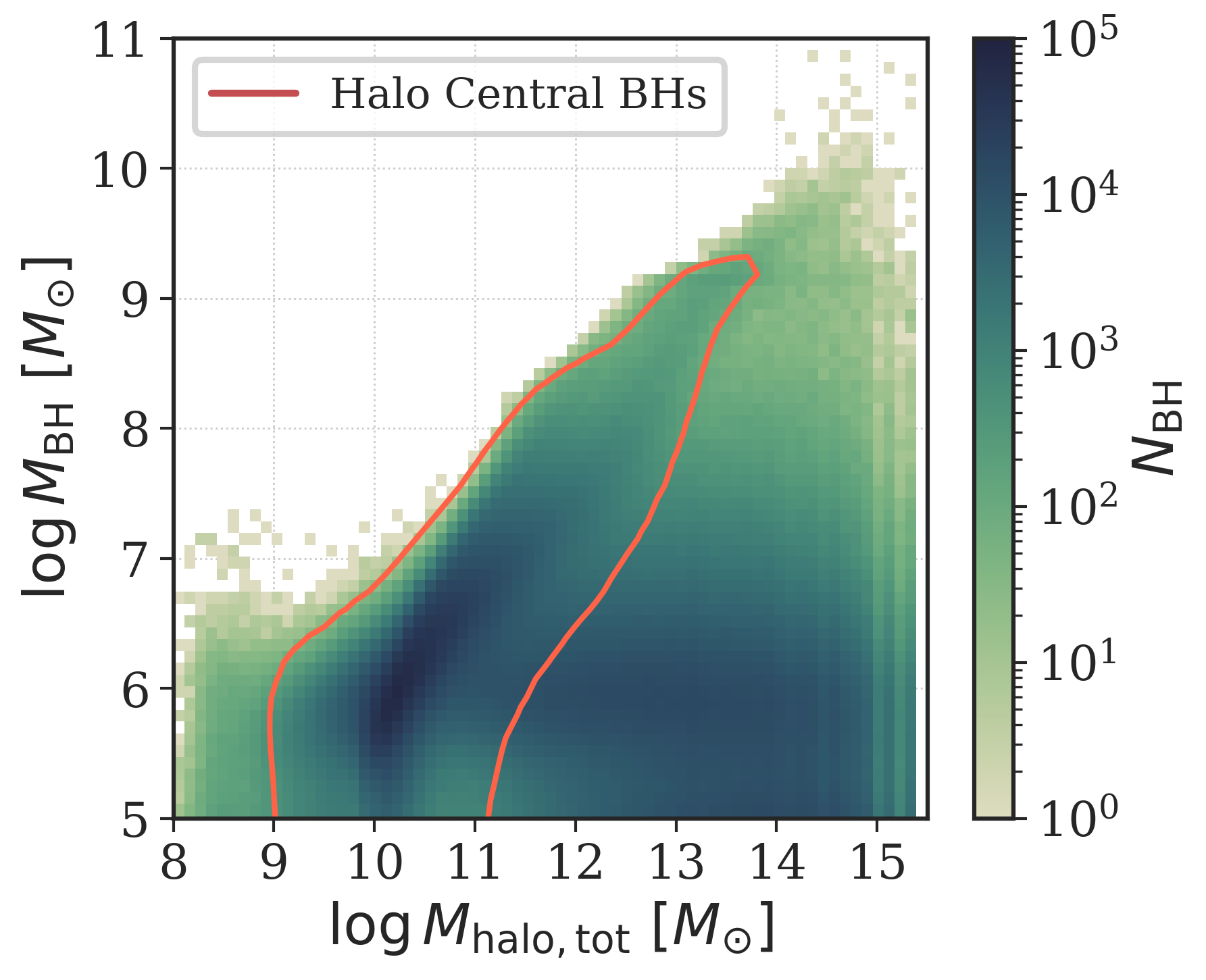}
    \caption{The relation between the black hole mass $M_{\rm BH}$ and the mass of their host halos. The red contour represents the 3$\sigma$ regions of halo central BHs. 
    }\label{fig:Mbh_Mhalo}
\end{figure}

\subsection{The clustering of wandering MBH}
We further explore the extent to which MBHs trace large-scale structure based on the large MBH population in \astrid.
In the right panel of Figure~\ref{fig:BH_tracks}, we present the trajectory of all MBHs in a slice of $125\times 250\times 25\, h^{-3}{\rm Mpc}^{3}$ at $z=0$. We trace the MBHs from $z=15$ to $z=0$. 
The DM density field of the same slice is plotted in the left panel. 
The comparison between the MBH trajectory and the DM density shows that MBHs, especially those with high mass, trace the large-scale structure closely. 
In the inset of the right panel, we zoom into a 10 Mpc/h cube to show the trajectories in more detail, allowing us to view those of smaller MBHs more easily. 
The zoomed-in region includes two massive halos, which have a mass of $M_{\rm 200}$ of $10^{14.9}$ (upper) and $10^{14.9}$~\Msun\ (lower), respectively. 
We mark the halo center positions using white crosses, and show their $R_{\rm 200}$ using white circles.
The largest MBHs move along the dense filament and end up close to the halo center. 
The tracks of the smaller MBHs are generally more diffuse and space-filling, appearing throughout the low-density region (even the lower central region of the small inset). We notice that a dense cloud of trajectories from low-mass MBHs surrounds the few most-massive black holes in these two halos.


A remarkable feature presented in Figure~\ref{fig:bias_vs_n} is that the relation between the large-scale bias and the tracer density is not monotonic for BHs.
Below $n=0.1\,h^{3}\,{\rm Mpc}^{-3}$, the bias rises steadily toward rarer samples. 
While for $n>0.1\,h^{3}\,{\rm Mpc}^{-3}$, the bias for high-density tracers increases. 
The BH bias is below 1 at $0.5<n<0.01\ \,h^{3}\,{\rm Mpc}^{-3}$. 
However, when we include the entire population (the leftmost purple point), the bias goes up to 1.2. 
This indicates the seed MBHs reflect the large-scale structure to some degree. 
To explain this, in Figure~\ref{fig:Mbh_Mhalo}, we investigate the BH population hosted by halos with different masses. 
We plot the $M_{\rm BH}$ and the mass of their host halo for individual BHs, colored by the BH number. 
The red contour shows the regime of the central BHs of halos, which demonstrates that the mass of central BHs increases with the mass of their host halos.  
Outside the regime occupied by these central BHs, there is a large group of wandering BHs in the massive halos, which have a mass of $M_{\rm BH}\lesssim 10^{6}$~\Msun\ located in the halo of $M_{\rm halo}>10^{12}$~\Msun.
These wandering BHs are populated in the massive halos, boosting the BH bias. 
Similarly, compared to low-mass halos, the massive halos host a large population of small-mass satellite galaxies, which is supported by Figure~\ref{fig:halo_mass_func}. This explains why the galaxy bias increases at the high-abundance end (low-mass cut) in Figure~\ref{fig:bias_vs_n}.
A caveat is that our results presented here are partly determined by our BH seeding prescription, for which we seed BHs in halos above $M_{\rm halo, FOF}>5\times 10^{9}\,h^{-1}M_{\odot}$.  This also implies the possibility that the clustering of wandering BHs can be used to test the BH seeding model.

Current and upcoming GW detectors, such as LISA and PTA, will detect catalogs of MBH binaries. The clustering of these sources, even if small in number, can be inferred from their cross-correlation with the galaxy population in large surveys
and used to constrain the population of MBHs and cosmology over a wide redshift range (e.g., \citealt{bosi23}). The relatively high bias of the wandering MBH population we have seen in Astrid will be highly relevant to these future analyses.


\section{Conclusion}
\label{sec:conclusion}
With a simulation box of $250\,h^{-1}\ {\rm Mpc}$ per side and $2\times 5500^{3}$ particles,
\astrid\ is one of the largest cosmological hydrodynamic simulations in terms of particle numbers that have yet reached $z=0$. 
The unique combination of large volume and high resolution enables detailed studies of galaxy and MBH evolution, while capturing rare, high-mass systems in a statistically representative volume. 
\astrid\ includes MBHs in a mass range spanning nearly 7 orders of magnitude, and records over 3 million MBH mergers, encompassing the GW sources targeted by LISA and PTA experiments. 
The inclusion of a dynamical friction subgrid model provides an improved description of MBH dynamics than earlier large simulations, and provides realistic MBH velocities and trajectories, improving the predictions for merger environments and event rates. 
All of these advantages make \astrid\ a powerful tool for investigating  MBHs and their mergers in a fully cosmological context.  

In this work, we present $z=0$ results of \astrid\ simulation, spanning the MBH population, galaxy, galaxy groups and clusters, and matter clustering on large scales. Our main results are summarized as follows:

\begin{enumerate}

\item The black hole mass function in \astrid\ at $z=0$ matches the observational data for $M_{\rm BH}\geq 10^{7}$~\Msun. 
There is an excess of small-mass black holes with $M_{\rm BH}<10^{7}$~\Msun, which could be due to an overabundance of seed BHs. 
The AGN hard X-ray luminosity function (2-10 keV) is generally consistent with observational constraints for $L_{\rm X}= 10^{42-45}$ erg/s when we assume a radiative efficiency ranging from $0.1 < \eta <0.2$.   

\item \astrid\ successfully captures the coevolution of MBHs and their host galaxies, generating  $M_{\rm BH}$-$M_{\star}$ and $M_{\rm BH}-\sigma$ relations in good agreement with observational constraints. 
Remarkably, the scatter in both scaling relations is comparable to that in the observed population at $M_{\star}< 10^{11}$~\Msun. In particular, \astrid\ reproduces the small-mass black holes ($M_{\rm BH}=10^{6-8}$~\Msun) in galaxies with $M_{\star}=10^{10.5-11}$~\Msun. This regime is not well reproduced in many simulations, which indicates \astrid\ generates better MBH diversity. 
This can be attributed to the fact that we do not implement a black hole repositioning algorithm, but instead allow black holes to merge more naturally with dynamical friction.

\item The galaxy stellar mass function at $z=0$ generally aligns well with observational data, especially when we only include the central galaxies with a fixed 3D aperture of 30 kpc. However, near the `knee' $M_{\star}\sim10^{11}$~\Msun, \astrid\ underpredicts the galaxy number density by up to 0.5 dex relative to the GAMA survey.
This deficiency likely reflects an overly high critical mass adopted for the kinetic AGN feedback 
as galaxies in the same mass range have a high $s$SFR compared to the observations. 
After applying dust attenuation, the $z=0$ galaxy luminosity function from \astrid\ agrees well with observational constraints, and the SDSS $g-r$ color distribution exhibits the red-blue bimodality.
We also present the distribution of galaxy $s$SFR, stellar half-mass-radius, stellar velocity dispersion, and stellar metallicity. All of them exhibit a correlation with galaxy mass broadly consistent with available observations.  

\item Due to its large simulation volume, \astrid\ hosts a large population of massive galaxy groups and clusters. 
At $z=0$, the most-massive halo is a Coma-like cluster, having a mass of  $M_{\rm 200c}=1.8\times 10^{15}$~\Msun\ ($M_{\rm 500c}=1.5\times 10^{15}$~\Msun). There are 9709 halos with $M_{\rm 200c}>10^{13}$~\Msun\ and seven of them have a mass over $M_{\rm 200c}=10^{15}$~\Msun. 
We quantify the stellar mass content in these halos, including the central galaxies, satellite galaxies, and ICL, and find that their correlations with halo mass are in good agreement with observational constraints.  
To facilitate future comparisons, we provide the fitted power-law relations between different stellar components and the halo mass.  

\item We present the 
power spectra for the black hole and the galaxy spatial distributions at $z=0$, as well as their scale-dependent bias. 
We then examine the influence of sample selection criteria by studying the bias for four 
types of tracers: halos selected by $M_{\rm 200}$, BH selected by $M_{\rm BH}$, galaxies selected by $M_{\star}$, and galaxies selected by SFR (used as a proxy for rest-frame UV). 
Among the tracers we investigate, $M_{\star}$-selected galaxies and MBHs have the largest bias at a given number density. 
SFR-selected galaxies are less clustered compared to the $M_{\star}$-selected galaxies. 
We find that MBHs with $M_{\rm BH}\geq10^{8}$~\Msun\ and galaxies with $M_{\star}\geq10^{10.5}$~\Msun\ are good tracers of large-scale structure, as they both have a large bias, and the bias remains scale-independent for $k<2\ h\,{\rm Mpc}^{-1}$. 

\end{enumerate}


\section*{Acknowledgements}
YZ acknowledges helpful discussions with TianQing Zhang, Lei Hu, and Zhiwei Shao. 
YZ and TDM acknowledge the support from the NASA FINESST grant 80NSSC25K0318. 
TDM acknowledges funding from NASA ATP 80NSSC20K0519, NSF PHY-2020295, NASA ATP NNX17AK56G, and NASA ATP 80NSSC18K101, NASA Theory grant 80NSSC22K072.
NC acknowledges support from the Schmidt Futures Fund. 
YN acknowledges support from the ITC Postdoctoral Fellowship.
SB acknowledges funding from NASA ATP 80NSSC22K1897 and NSF AST-2509639.
\astrid~was run on the Frontera facility at the Texas Advanced Computing Center.

\section*{Data Availability}
The \astrid\ simulations are publicly available and accessible at \url{https://astrid.psc.edu}, where we share the full particle-in-group data, the FOFGroups and SubFind subgroup catalogs, and the MBH merger catalog.

\bibliography{reference}{}

@ARTICLE{bosi23,
       author = {{Bosi}, M. and {Bellomo}, N. and {Raccanelli}, A.},
        title = "{Constraining extended cosmologies with GW{\texttimes}LSS cross-correlations}",
      journal = {\jcap},
         year = 2023,
        month = nov,
       volume = {2023},
       number = {11},
          eid = {086},
        pages = {086},
          doi = {10.1088/1475-7516/2023/11/086},
archivePrefix = {arXiv},
       eprint = {2306.03031},
 primaryClass = {astro-ph.CO},
       adsurl = {https://ui.adsabs.harvard.edu/abs/2023JCAP...11..086B}
}

@misc{jeon202,
      title={The Emerging Black Hole Mass Function in the High-Redshift Universe}, 
      author={Junehyoung Jeon and Boyuan Liu and Anthony J. Taylor and Vasily Kokorev and John Chisholm and Dale D. Kocevski and Steven L. Finkelstein and Volker Bromm},
      year={2025},
      eprint={2503.14703},
      archivePrefix={arXiv},
      primaryClass={astro-ph.GA},
      url={https://arxiv.org/abs/2503.14703}, 
}

@ARTICLE{DiMatteo2005_BH_model,
       author = {{Di Matteo}, Tiziana and {Springel}, Volker and {Hernquist}, Lars},
        title = "{Energy input from quasars regulates the growth and activity of black holes and their host galaxies}",
      journal = {\nat},
         year = 2005,
        month = feb,
       volume = {433},
       number = {7026},
        pages = {604-607},
          doi = {10.1038/nature03335},
archivePrefix = {arXiv},
       eprint = {astro-ph/0502199},
 primaryClass = {astro-ph},
       adsurl = {https://ui.adsabs.harvard.edu/abs/2005Natur.433..604D}
}

@ARTICLE{Shakura1973_BH,
       author = {{Shakura}, N.~I. and {Sunyaev}, R.~A.},
        title = "{Black holes in binary systems. Observational appearance.}",
      journal = {\aap},
         year = 1973,
        month = jan,
       volume = {24},
        pages = {337-355},
       adsurl = {https://ui.adsabs.harvard.edu/abs/1973A&A....24..337S}
}

@ARTICLE{Ni2022_astrid,
       author = {{Ni}, Yueying and {Di Matteo}, Tiziana and {Bird}, Simeon and {Croft}, Rupert and {Feng}, Yu and {Chen}, Nianyi and {Tremmel}, Michael and {DeGraf}, Colin and {Li}, Yin},
        title = "{The ASTRID simulation: the evolution of supermassive black holes}",
      journal = {\mnras},
         year = 2022,
        month = jun,
       volume = {513},
       number = {1},
        pages = {670-692},
          doi = {10.1093/mnras/stac351},
archivePrefix = {arXiv},
       eprint = {2110.14154},
 primaryClass = {astro-ph.GA},
       adsurl = {https://ui.adsabs.harvard.edu/abs/2022MNRAS.513..670N}
}

@ARTICLE{Tremaine2002,
       author = {{Tremaine}, Scott and {Gebhardt}, Karl and {Bender}, Ralf and {Bower}, Gary and {Dressler}, Alan and {Faber}, S.~M. and {Filippenko}, Alexei V. and {Green}, Richard and {Grillmair}, Carl and {Ho}, Luis C. and {Kormendy}, John and {Lauer}, Tod R. and {Magorrian}, John and {Pinkney}, Jason and {Richstone}, Douglas},
        title = "{The Slope of the Black Hole Mass versus Velocity Dispersion Correlation}",
      journal = {\apj},
         year = 2002,
        month = aug,
       volume = {574},
       number = {2},
        pages = {740-753},
          doi = {10.1086/341002},
archivePrefix = {arXiv},
       eprint = {astro-ph/0203468},
 primaryClass = {astro-ph},
       adsurl = {https://ui.adsabs.harvard.edu/abs/2002ApJ...574..740T}
}

@ARTICLE{Dattathri2025,
       author = {{Dattathri}, Shashank and {Natarajan}, Priyamvada and {Porras-Valverde}, Antonio J. and {Burke}, Colin J. and {Chen}, Nianyi and {Di Matteo}, Tiziana and {Ni}, Yueying},
        title = "{The Redshift Evolution of the M$_{BH}${\textendash}M$_{*}$ Scaling Relation: New Insights from Cosmological Simulations and Semianalytic Models}",
      journal = {\apj},
         year = 2025,
        month = may,
       volume = {984},
       number = {2},
          eid = {122},
        pages = {122},
          doi = {10.3847/1538-4357/adbeef},
archivePrefix = {arXiv},
       eprint = {2410.13958},
 primaryClass = {astro-ph.GA},
       adsurl = {https://ui.adsabs.harvard.edu/abs/2025ApJ...984..122D}
}

@ARTICLE{Ni2023_camels,
       author = {{Ni}, Yueying and {Genel}, Shy and {Angl{\'e}s-Alc{\'a}zar}, Daniel and {Villaescusa-Navarro}, Francisco and {Jo}, Yongseok and {Bird}, Simeon and {Di Matteo}, Tiziana and {Croft}, Rupert and {Chen}, Nianyi and {de Santi}, Natal{\'\i} S.~M. and {Gebhardt}, Matthew and {Shao}, Helen and {Pandey}, Shivam and {Hernquist}, Lars and {Dave}, Romeel},
        title = "{The CAMELS Project: Expanding the Galaxy Formation Model Space with New ASTRID and 28-parameter TNG and SIMBA Suites}",
      journal = {\apj},
         year = 2023,
        month = dec,
       volume = {959},
       number = {2},
          eid = {136},
        pages = {136},
          doi = {10.3847/1538-4357/ad022a},
archivePrefix = {arXiv},
       eprint = {2304.02096},
 primaryClass = {astro-ph.CO},
       adsurl = {https://ui.adsabs.harvard.edu/abs/2023ApJ...959..136N}
}

@ARTICLE{Ni2024,
       author = {{Ni}, Yueying and {Chen}, Nianyi and {Zhou}, Yihao and {Park}, Minjung and {Yang}, Yanhui and {DiMatteo}, Tiziana and {Bird}, Simeon and {Croft}, Rupert},
        title = "{The Astrid Simulation: Evolution of black holes and galaxies to z=0.5 and different evolution pathways for galaxy quenching}",
      journal = {arXiv e-prints},
         year = 2024,
        month = sep,
          eid = {arXiv:2409.10666},
        pages = {arXiv:2409.10666},
          doi = {10.48550/arXiv.2409.10666},
archivePrefix = {arXiv},
       eprint = {2409.10666},
 primaryClass = {astro-ph.GA},
       adsurl = {https://ui.adsabs.harvard.edu/abs/2024arXiv240910666N}
}

@ARTICLE{Zhou2025,
       author = {{Zhou}, Yihao and {Mukherjee}, Diptajyoti and {Chen}, Nianyi and {Di Matteo}, Tiziana and {Johansson}, Peter H. and {Rantala}, Antti and {Partmann}, Christian and {Di Carlo}, Ugo Niccol{\`o} and {Bird}, Simeon and {Ni}, Yueying},
        title = "{MAGICS. II. Seed Black Holes Stripped of Their Surrounding Stars Do Not Sink}",
      journal = {\apj},
         year = 2025,
        month = feb,
       volume = {980},
       number = {1},
          eid = {79},
        pages = {79},
          doi = {10.3847/1538-4357/ada283},
archivePrefix = {arXiv},
       eprint = {2409.19914},
 primaryClass = {astro-ph.GA},
       adsurl = {https://ui.adsabs.harvard.edu/abs/2025ApJ...980...79Z}
}

@ARTICLE{Gultekin2009,
       author = {{G{\"u}ltekin}, Kayhan and {Richstone}, Douglas O. and {Gebhardt}, Karl and {Lauer}, Tod R. and {Tremaine}, Scott and {Aller}, M.~C. and {Bender}, Ralf and {Dressler}, Alan and {Faber}, S.~M. and {Filippenko}, Alexei V. and {Green}, Richard and {Ho}, Luis C. and {Kormendy}, John and {Magorrian}, John and {Pinkney}, Jason and {Siopis}, Christos},
        title = "{The M-{\ensuremath{\sigma}} and M-L Relations in Galactic Bulges, and Determinations of Their Intrinsic Scatter}",
      journal = {\apj},
         year = 2009,
        month = jun,
       volume = {698},
       number = {1},
        pages = {198-221},
          doi = {10.1088/0004-637X/698/1/198},
archivePrefix = {arXiv},
       eprint = {0903.4897},
 primaryClass = {astro-ph.GA},
       adsurl = {https://ui.adsabs.harvard.edu/abs/2009ApJ...698..198G}
}

@ARTICLE{Bustamante2019,
       author = {{Bustamante}, Sebastian and {Springel}, Volker},
        title = "{Spin evolution and feedback of supermassive black holes in cosmological simulations}",
      journal = {\mnras},
         year = 2019,
        month = dec,
       volume = {490},
       number = {3},
        pages = {4133-4153},
          doi = {10.1093/mnras/stz2836},
archivePrefix = {arXiv},
       eprint = {1902.04651},
 primaryClass = {astro-ph.GA},
       adsurl = {https://ui.adsabs.harvard.edu/abs/2019MNRAS.490.4133B}
}

@ARTICLE{Kennicutt2012,
       author = {{Kennicutt}, Robert C. and {Evans}, Neal J.},
        title = "{Star Formation in the Milky Way and Nearby Galaxies}",
      journal = {\araa},
         year = 2012,
        month = sep,
       volume = {50},
        pages = {531-608},
          doi = {10.1146/annurev-astro-081811-125610},
archivePrefix = {arXiv},
       eprint = {1204.3552},
 primaryClass = {astro-ph.GA},
       adsurl = {https://ui.adsabs.harvard.edu/abs/2012ARA&A..50..531K}
}

@ARTICLE{Zahid2012,
       author = {{Zahid}, H.~J. and {Dima}, G.~I. and {Kewley}, L.~J. and {Erb}, D.~K. and {Dav{\'e}}, R.},
        title = "{A Census of Oxygen in Star-forming Galaxies: An Empirical Model Linking Metallicities, Star Formation Rates, and Outflows}",
      journal = {\apj},
         year = 2012,
        month = sep,
       volume = {757},
       number = {1},
          eid = {54},
        pages = {54},
          doi = {10.1088/0004-637X/757/1/54},
archivePrefix = {arXiv},
       eprint = {1207.5509},
 primaryClass = {astro-ph.CO},
       adsurl = {https://ui.adsabs.harvard.edu/abs/2012ApJ...757...54Z}
}

@ARTICLE{Rahmani2016,
       author = {{Rahmani}, S. and {Lianou}, S. and {Barmby}, P.},
        title = "{Star formation laws in the Andromeda galaxy: gas, stars, metals and the surface density of star formation}",
      journal = {\mnras},
         year = 2016,
        month = mar,
       volume = {456},
       number = {4},
        pages = {4128-4144},
          doi = {10.1093/mnras/stv2951},
archivePrefix = {arXiv},
       eprint = {1512.06675},
 primaryClass = {astro-ph.GA},
       adsurl = {https://ui.adsabs.harvard.edu/abs/2016MNRAS.456.4128R}
}

@ARTICLE{Conroy2009,
       author = {{Conroy}, Charlie and {Gunn}, James E. and {White}, Martin},
        title = "{The Propagation of Uncertainties in Stellar Population Synthesis Modeling. I. The Relevance of Uncertain Aspects of Stellar Evolution and the Initial Mass Function to the Derived Physical Properties of Galaxies}",
      journal = {\apj},
         year = 2009,
        month = jul,
       volume = {699},
       number = {1},
        pages = {486-506},
          doi = {10.1088/0004-637X/699/1/486},
archivePrefix = {arXiv},
       eprint = {0809.4261},
 primaryClass = {astro-ph},
       adsurl = {https://ui.adsabs.harvard.edu/abs/2009ApJ...699..486C}
}

@ARTICLE{Chen2022,
       author = {{Chen}, Nianyi and {Ni}, Yueying and {Holgado}, A. Miguel and {Di Matteo}, Tiziana and {Tremmel}, Michael and {DeGraf}, Colin and {Bird}, Simeon and {Croft}, Rupert and {Feng}, Yu},
        title = "{Massive black hole mergers with orbital information: predictions from the ASTRID simulation}",
      journal = {\mnras},
         year = 2022,
        month = aug,
       volume = {514},
       number = {2},
        pages = {2220-2238},
          doi = {10.1093/mnras/stac1432},
archivePrefix = {arXiv},
       eprint = {2112.08555},
 primaryClass = {astro-ph.GA},
       adsurl = {https://ui.adsabs.harvard.edu/abs/2022MNRAS.514.2220C}
}

@ARTICLE{Pillepich2018_tngmodel,
       author = {{Pillepich}, Annalisa and {Springel}, Volker and {Nelson}, Dylan and {Genel}, Shy and {Naiman}, Jill and {Pakmor}, R{\"u}diger and {Hernquist}, Lars and {Torrey}, Paul and {Vogelsberger}, Mark and {Weinberger}, Rainer and {Marinacci}, Federico},
        title = "{Simulating galaxy formation with the IllustrisTNG model}",
      journal = {\mnras},
         year = 2018,
        month = jan,
       volume = {473},
       number = {3},
        pages = {4077-4106},
          doi = {10.1093/mnras/stx2656},
archivePrefix = {arXiv},
       eprint = {1703.02970},
 primaryClass = {astro-ph.GA},
       adsurl = {https://ui.adsabs.harvard.edu/abs/2018MNRAS.473.4077P}
}

@ARTICLE{Crain2015,
       author = {{Crain}, Robert A. and {Schaye}, Joop and {Bower}, Richard G. and {Furlong}, Michelle and {Schaller}, Matthieu and {Theuns}, Tom and {Dalla Vecchia}, Claudio and {Frenk}, Carlos S. and {McCarthy}, Ian G. and {Helly}, John C. and {Jenkins}, Adrian and {Rosas-Guevara}, Yetli M. and {White}, Simon D.~M. and {Trayford}, James W.},
        title = "{The EAGLE simulations of galaxy formation: calibration of subgrid physics and model variations}",
      journal = {\mnras},
         year = 2015,
        month = jun,
       volume = {450},
       number = {2},
        pages = {1937-1961},
          doi = {10.1093/mnras/stv725},
archivePrefix = {arXiv},
       eprint = {1501.01311},
 primaryClass = {astro-ph.GA},
       adsurl = {https://ui.adsabs.harvard.edu/abs/2015MNRAS.450.1937C}
}

@ARTICLE{Vakili2023,
       author = {{Vakili}, Mohammadjavad and {Hoekstra}, Henk and {Bilicki}, Maciej and {Fortuna}, Maria Cristina and {Kuijken}, Konrad and {Wright}, Angus H. and {Asgari}, Marika and {Brown}, Michael and {Dombrovskij}, Elisabeth and {Erben}, Thomas and {Giblin}, Benjamin and {Heymans}, Catherine and {Hildebrandt}, Hendrik and {Johnston}, Harry and {Joudaki}, Shahab and {Kannawadi}, Arun},
        title = "{Clustering of red sequence galaxies in the fourth data release of the Kilo-Degree Survey}",
      journal = {\aap},
         year = 2023,
        month = jul,
       volume = {675},
          eid = {A202},
        pages = {A202},
          doi = {10.1051/0004-6361/202039293},
archivePrefix = {arXiv},
       eprint = {2008.13154},
 primaryClass = {astro-ph.CO},
       adsurl = {https://ui.adsabs.harvard.edu/abs/2023A&A...675A.202V}
}

@ARTICLE{Liivamagi2012,
       author = {{Liivam{\"a}gi}, L.~J. and {Tempel}, E. and {Saar}, E.},
        title = "{SDSS DR7 superclusters. The catalogues}",
      journal = {\aap},
         year = 2012,
        month = mar,
       volume = {539},
          eid = {A80},
        pages = {A80},
          doi = {10.1051/0004-6361/201016288},
archivePrefix = {arXiv},
       eprint = {1012.1989},
 primaryClass = {astro-ph.CO},
       adsurl = {https://ui.adsabs.harvard.edu/abs/2012A&A...539A..80L}
}

@ARTICLE{Yuan2021,
       author = {{Yuan}, Sihan and {Hadzhiyska}, Boryana and {Bose}, Sownak and {Eisenstein}, Daniel J. and {Guo}, Hong},
        title = "{Evidence for galaxy assembly bias in BOSS CMASS redshift-space galaxy correlation function}",
      journal = {\mnras},
         year = 2021,
        month = apr,
       volume = {502},
       number = {3},
        pages = {3582-3598},
          doi = {10.1093/mnras/stab235},
archivePrefix = {arXiv},
       eprint = {2010.04182},
 primaryClass = {astro-ph.CO},
       adsurl = {https://ui.adsabs.harvard.edu/abs/2021MNRAS.502.3582Y}
}

@ARTICLE{Zhou2023,
       author = {{Zhou}, Rongpu and {Dey}, Biprateep and {Newman}, Jeffrey A. and {Eisenstein}, Daniel J. and {Dawson}, K. and {Bailey}, S. and {Berti}, A. and {Guy}, J. and {Lan}, Ting-Wen and {Zou}, H. and {Aguilar}, J. and {Ahlen}, S. and {Alam}, Shadab and {Brooks}, D. and {de la Macorra}, A. and {Dey}, A. and {Dhungana}, G. and {Fanning}, K. and {Font-Ribera}, A. and {Gontcho}, S. Gontcho A. and {Honscheid}, K. and {Ishak}, Mustapha and {Kisner}, T. and {Kov{\'a}cs}, A. and {Kremin}, A. and {Landriau}, M. and {Levi}, Michael E. and {Magneville}, C. and {Manera}, Marc and {Martini}, P. and {Meisner}, Aaron M. and {Miquel}, R. and {Moustakas}, J. and {Myers}, Adam D. and {Nie}, Jundan and {Palanque-Delabrouille}, N. and {Percival}, W.~J. and {Poppett}, C. and {Prada}, F. and {Raichoor}, A. and {Ross}, A.~J. and {Schlafly}, E. and {Schlegel}, D. and {Schubnell}, M. and {Tarl{\'e}}, Gregory and {Weaver}, B.~A. and {Wechsler}, R.~H. and {Y{\'e}che}, Christophe and {Zhou}, Zhimin},
        title = "{Target Selection and Validation of DESI Luminous Red Galaxies}",
      journal = {\aj},
         year = 2023,
        month = feb,
       volume = {165},
       number = {2},
          eid = {58},
        pages = {58},
          doi = {10.3847/1538-3881/aca5fb},
archivePrefix = {arXiv},
       eprint = {2208.08515},
 primaryClass = {astro-ph.CO},
       adsurl = {https://ui.adsabs.harvard.edu/abs/2023AJ....165...58Z}
}

@ARTICLE{Weinberger2018,
       author = {{Weinberger}, Rainer and {Springel}, Volker and {Pakmor}, R{\"u}diger and {Nelson}, Dylan and {Genel}, Shy and {Pillepich}, Annalisa and {Vogelsberger}, Mark and {Marinacci}, Federico and {Naiman}, Jill and {Torrey}, Paul and {Hernquist}, Lars},
        title = "{Supermassive black holes and their feedback effects in the IllustrisTNG simulation}",
      journal = {\mnras},
         year = 2018,
        month = sep,
       volume = {479},
       number = {3},
        pages = {4056-4072},
          doi = {10.1093/mnras/sty1733},
archivePrefix = {arXiv},
       eprint = {1710.04659},
 primaryClass = {astro-ph.GA},
       adsurl = {https://ui.adsabs.harvard.edu/abs/2018MNRAS.479.4056W}
}

@ARTICLE{Ricarte2021,
       author = {{Ricarte}, Angelo and {Tremmel}, Michael and {Natarajan}, Priyamvada and {Zimmer}, Charlotte and {Quinn}, Thomas},
        title = "{Origins and demographics of wandering black holes}",
      journal = {\mnras},
         year = 2021,
        month = jun,
       volume = {503},
       number = {4},
        pages = {6098-6111},
          doi = {10.1093/mnras/stab866},
archivePrefix = {arXiv},
       eprint = {2103.12124},
 primaryClass = {astro-ph.GA},
       adsurl = {https://ui.adsabs.harvard.edu/abs/2021MNRAS.503.6098R}
}

@ARTICLE{Dubois2015,
       author = {{Dubois}, Yohan and {Volonteri}, Marta and {Silk}, Joseph and {Devriendt}, Julien and {Slyz}, Adrianne and {Teyssier}, Romain},
        title = "{Black hole evolution - I. Supernova-regulated black hole growth}",
      journal = {\mnras},
         year = 2015,
        month = sep,
       volume = {452},
       number = {2},
        pages = {1502-1518},
          doi = {10.1093/mnras/stv1416},
archivePrefix = {arXiv},
       eprint = {1504.00018},
 primaryClass = {astro-ph.GA},
       adsurl = {https://ui.adsabs.harvard.edu/abs/2015MNRAS.452.1502D}
}

@ARTICLE{Conroy2010,
       author = {{Conroy}, Charlie and {Gunn}, James E.},
        title = "{The Propagation of Uncertainties in Stellar Population Synthesis Modeling. III. Model Calibration, Comparison, and Evaluation}",
      journal = {\apj},
         year = 2010,
        month = apr,
       volume = {712},
       number = {2},
        pages = {833-857},
          doi = {10.1088/0004-637X/712/2/833},
archivePrefix = {arXiv},
       eprint = {0911.3151},
 primaryClass = {astro-ph.CO},
       adsurl = {https://ui.adsabs.harvard.edu/abs/2010ApJ...712..833C}
}

@ARTICLE{Licquia2015,
       author = {{Licquia}, Timothy C. and {Newman}, Jeffrey A.},
        title = "{Improved Estimates of the Milky Way's Stellar Mass and Star Formation Rate from Hierarchical Bayesian Meta-Analysis}",
      journal = {\apj},
         year = 2015,
        month = jun,
       volume = {806},
       number = {1},
          eid = {96},
        pages = {96},
          doi = {10.1088/0004-637X/806/1/96},
archivePrefix = {arXiv},
       eprint = {1407.1078},
 primaryClass = {astro-ph.GA},
       adsurl = {https://ui.adsabs.harvard.edu/abs/2015ApJ...806...96L}
}

@ARTICLE{Tamm2012,
       author = {{Tamm}, A. and {Tempel}, E. and {Tenjes}, P. and {Tihhonova}, O. and {Tuvikene}, T.},
        title = "{Stellar mass map and dark matter distribution in M 31}",
      journal = {\aap},
         year = 2012,
        month = oct,
       volume = {546},
          eid = {A4},
        pages = {A4},
          doi = {10.1051/0004-6361/201220065},
archivePrefix = {arXiv},
       eprint = {1208.5712},
 primaryClass = {astro-ph.CO},
       adsurl = {https://ui.adsabs.harvard.edu/abs/2012A&A...546A...4T}
}

@ARTICLE{Salim2007,
       author = {{Salim}, Samir and {Rich}, R. Michael and {Charlot}, St{\'e}phane and {Brinchmann}, Jarle and {Johnson}, Benjamin D. and {Schiminovich}, David and {Seibert}, Mark and {Mallery}, Ryan and {Heckman}, Timothy M. and {Forster}, Karl and {Friedman}, Peter G. and {Martin}, D. Christopher and {Morrissey}, Patrick and {Neff}, Susan G. and {Small}, Todd and {Wyder}, Ted K. and {Bianchi}, Luciana and {Donas}, Jos{\'e} and {Lee}, Young-Wook and {Madore}, Barry F. and {Milliard}, Bruno and {Szalay}, Alex S. and {Welsh}, Barry Y. and {Yi}, Sukyoung K.},
        title = "{UV Star Formation Rates in the Local Universe}",
      journal = {\apjs},
         year = 2007,
        month = dec,
       volume = {173},
       number = {2},
        pages = {267-292},
          doi = {10.1086/519218},
archivePrefix = {arXiv},
       eprint = {0704.3611},
 primaryClass = {astro-ph},
       adsurl = {https://ui.adsabs.harvard.edu/abs/2007ApJS..173..267S}
}

@ARTICLE{Elbaz2007,
       author = {{Elbaz}, D. and {Daddi}, E. and {Le Borgne}, D. and {Dickinson}, M. and {Alexander}, D.~M. and {Chary}, R. -R. and {Starck}, J. -L. and {Brandt}, W.~N. and {Kitzbichler}, M. and {MacDonald}, E. and {Nonino}, M. and {Popesso}, P. and {Stern}, D. and {Vanzella}, E.},
        title = "{The reversal of the star formation-density relation in the distant universe}",
      journal = {\aap},
         year = 2007,
        month = jun,
       volume = {468},
       number = {1},
        pages = {33-48},
          doi = {10.1051/0004-6361:20077525},
archivePrefix = {arXiv},
       eprint = {astro-ph/0703653},
 primaryClass = {astro-ph},
       adsurl = {https://ui.adsabs.harvard.edu/abs/2007A&A...468...33E}
}

@ARTICLE{Speagle2014,
       author = {{Speagle}, J.~S. and {Steinhardt}, C.~L. and {Capak}, P.~L. and {Silverman}, J.~D.},
        title = "{A Highly Consistent Framework for the Evolution of the Star-Forming ``Main Sequence'' from z \raisebox{-0.5ex}\textasciitilde 0-6}",
      journal = {\apjs},
         year = 2014,
        month = oct,
       volume = {214},
       number = {2},
          eid = {15},
        pages = {15},
          doi = {10.1088/0067-0049/214/2/15},
archivePrefix = {arXiv},
       eprint = {1405.2041},
 primaryClass = {astro-ph.GA},
       adsurl = {https://ui.adsabs.harvard.edu/abs/2014ApJS..214...15S}
}

@ARTICLE{Chang2015,
       author = {{Chang}, Yu-Yen and {van der Wel}, Arjen and {da Cunha}, Elisabete and {Rix}, Hans-Walter},
        title = "{Stellar Masses and Star Formation Rates for 1M Galaxies from SDSS+WISE}",
      journal = {\apjs},
         year = 2015,
        month = jul,
       volume = {219},
       number = {1},
          eid = {8},
        pages = {8},
          doi = {10.1088/0067-0049/219/1/8},
archivePrefix = {arXiv},
       eprint = {1506.00648},
 primaryClass = {astro-ph.GA},
       adsurl = {https://ui.adsabs.harvard.edu/abs/2015ApJS..219....8C}
}

@ARTICLE{Oliver2010,
       author = {{Oliver}, Seb and {Frost}, M. and {Farrah}, D. and {Gonzalez-Solares}, E. and {Shupe}, D.~L. and {Henriques}, B. and {Roseboom}, I. and {Alfonso-Luis}, A. and {Babbedge}, T.~S.~R. and {Frayer}, D. and {Lencz}, C. and {Lonsdale}, C.~J. and {Masci}, F. and {Padgett}, D. and {Polletta}, M. and {Rowan-Robinson}, M. and {Siana}, B. and {Smith}, H.~E. and {Surace}, J.~A. and {Vaccari}, M.},
        title = "{Specific star formation and the relation to stellar mass from 0 < z < 2 as seen in the far-infrared at 70 and 160 {\ensuremath{\mu}}m}",
      journal = {\mnras},
         year = 2010,
        month = jul,
       volume = {405},
       number = {4},
        pages = {2279-2294},
          doi = {10.1111/j.1365-2966.2010.16643.x},
archivePrefix = {arXiv},
       eprint = {1003.2446},
 primaryClass = {astro-ph.CO},
       adsurl = {https://ui.adsabs.harvard.edu/abs/2010MNRAS.405.2279O}
}

@ARTICLE{Donnari2019,
       author = {{Donnari}, Martina and {Pillepich}, Annalisa and {Nelson}, Dylan and {Vogelsberger}, Mark and {Genel}, Shy and {Weinberger}, Rainer and {Marinacci}, Federico and {Springel}, Volker and {Hernquist}, Lars},
        title = "{The star formation activity of IllustrisTNG galaxies: main sequence, UVJ diagram, quenched fractions, and systematics}",
      journal = {\mnras},
         year = 2019,
        month = jun,
       volume = {485},
       number = {4},
        pages = {4817-4840},
          doi = {10.1093/mnras/stz712},
archivePrefix = {arXiv},
       eprint = {1812.07584},
 primaryClass = {astro-ph.GA},
       adsurl = {https://ui.adsabs.harvard.edu/abs/2019MNRAS.485.4817D}
}

@ARTICLE{Habouzit2017,
       author = {{Habouzit}, M{\'e}lanie and {Volonteri}, Marta and {Dubois}, Yohan},
        title = "{Blossoms from black hole seeds: properties and early growth regulated by supernova feedback}",
      journal = {\mnras},
         year = 2017,
        month = jul,
       volume = {468},
       number = {4},
        pages = {3935-3948},
          doi = {10.1093/mnras/stx666},
archivePrefix = {arXiv},
       eprint = {1605.09394},
 primaryClass = {astro-ph.GA},
       adsurl = {https://ui.adsabs.harvard.edu/abs/2017MNRAS.468.3935H}
}

@ARTICLE{Marconi2003,
       author = {{Marconi}, Alessandro and {Hunt}, Leslie K.},
        title = "{The Relation between Black Hole Mass, Bulge Mass, and Near-Infrared Luminosity}",
      journal = {\apjl},
         year = 2003,
        month = may,
       volume = {589},
       number = {1},
        pages = {L21-L24},
          doi = {10.1086/375804},
archivePrefix = {arXiv},
       eprint = {astro-ph/0304274},
 primaryClass = {astro-ph},
       adsurl = {https://ui.adsabs.harvard.edu/abs/2003ApJ...589L..21M}
}

@ARTICLE{Graham2008,
       author = {{Graham}, Alister W.},
        title = "{Fundamental Planes and the Barless M$_{BH}$-{\ensuremath{\sigma}} Relation for Supermassive Black Holes}",
      journal = {\apj},
         year = 2008,
        month = jun,
       volume = {680},
       number = {1},
        pages = {143-153},
          doi = {10.1086/587473},
archivePrefix = {arXiv},
       eprint = {0801.1548},
 primaryClass = {astro-ph},
       adsurl = {https://ui.adsabs.harvard.edu/abs/2008ApJ...680..143G}
}

@ARTICLE{Gebhardt2000,
       author = {{Gebhardt}, Karl and {Bender}, Ralf and {Bower}, Gary and {Dressler}, Alan and {Faber}, S.~M. and {Filippenko}, Alexei V. and {Green}, Richard and {Grillmair}, Carl and {Ho}, Luis C. and {Kormendy}, John and {Lauer}, Tod R. and {Magorrian}, John and {Pinkney}, Jason and {Richstone}, Douglas and {Tremaine}, Scott},
        title = "{A Relationship between Nuclear Black Hole Mass and Galaxy Velocity Dispersion}",
      journal = {\apjl},
         year = 2000,
        month = aug,
       volume = {539},
       number = {1},
        pages = {L13-L16},
          doi = {10.1086/312840},
archivePrefix = {arXiv},
       eprint = {astro-ph/0006289},
 primaryClass = {astro-ph},
       adsurl = {https://ui.adsabs.harvard.edu/abs/2000ApJ...539L..13G}
}

@ARTICLE{Aird2012,
       author = {{Aird}, James and {Coil}, Alison L. and {Moustakas}, John and {Blanton}, Michael R. and {Burles}, Scott M. and {Cool}, Richard J. and {Eisenstein}, Daniel J. and {Smith}, M. Stephen M. and {Wong}, Kenneth C. and {Zhu}, Guangtun},
        title = "{PRIMUS: The Dependence of AGN Accretion on Host Stellar Mass and Color}",
      journal = {\apj},
         year = 2012,
        month = feb,
       volume = {746},
       number = {1},
          eid = {90},
        pages = {90},
          doi = {10.1088/0004-637X/746/1/90},
archivePrefix = {arXiv},
       eprint = {1107.4368},
 primaryClass = {astro-ph.CO},
       adsurl = {https://ui.adsabs.harvard.edu/abs/2012ApJ...746...90A}
}

@ARTICLE{Man2019,
       author = {{Man}, Zhong-yi and {Peng}, Ying-jie and {Kong}, Xu and {Guo}, Ke-xin and {Zhang}, Cheng-peng and {Dou}, Jing},
        title = "{The dependence of AGN activity on environment in SDSS}",
      journal = {\mnras},
         year = 2019,
        month = sep,
       volume = {488},
       number = {1},
        pages = {89-98},
          doi = {10.1093/mnras/stz1706},
archivePrefix = {arXiv},
       eprint = {1907.01563},
 primaryClass = {astro-ph.GA},
       adsurl = {https://ui.adsabs.harvard.edu/abs/2019MNRAS.488...89M}
}

@ARTICLE{Aird2018,
       author = {{Aird}, J. and {Coil}, A.~L. and {Georgakakis}, A.},
        title = "{X-rays across the galaxy population - II. The distribution of AGN accretion rates as a function of stellar mass and redshift.}",
      journal = {\mnras},
         year = 2018,
        month = jan,
       volume = {474},
        pages = {1225-1249},
          doi = {10.1093/mnras/stx2700},
archivePrefix = {arXiv},
       eprint = {1705.01132},
 primaryClass = {astro-ph.HE},
       adsurl = {https://ui.adsabs.harvard.edu/abs/2018MNRAS.474.1225A}
}

@ARTICLE{Genina2024,
       author = {{Genina}, Anna and {Springel}, Volker and {Rantala}, Antti},
        title = "{A calibrated model for N-body dynamical friction acting on supermassive black holes}",
      journal = {\mnras},
         year = 2024,
        month = oct,
       volume = {534},
       number = {1},
        pages = {957-977},
          doi = {10.1093/mnras/stae2144},
archivePrefix = {arXiv},
       eprint = {2405.08870},
 primaryClass = {astro-ph.GA},
       adsurl = {https://ui.adsabs.harvard.edu/abs/2024MNRAS.534..957G}
}

@ARTICLE{Chen2023_dualAGN,
       author = {{Chen}, Nianyi and {Di Matteo}, Tiziana and {Ni}, Yueying and {Tremmel}, Michael and {DeGraf}, Colin and {Shen}, Yue and {Holgado}, A. Miguel and {Bird}, Simeon and {Croft}, Rupert and {Feng}, Yu},
        title = "{Properties and evolution of dual and offset AGN in the ASTRID simulation at z   2}",
      journal = {\mnras},
         year = 2023,
        month = jun,
       volume = {522},
       number = {2},
        pages = {1895-1913},
          doi = {10.1093/mnras/stad834},
archivePrefix = {arXiv},
       eprint = {2208.04970},
 primaryClass = {astro-ph.GA},
       adsurl = {https://ui.adsabs.harvard.edu/abs/2023MNRAS.522.1895C}
}

@ARTICLE{Rodriguez-Gomez2015_illu_mergingrate,
       author = {{Rodriguez-Gomez}, Vicente and {Genel}, Shy and {Vogelsberger}, Mark and {Sijacki}, Debora and {Pillepich}, Annalisa and {Sales}, Laura V. and {Torrey}, Paul and {Snyder}, Greg and {Nelson}, Dylan and {Springel}, Volker and {Ma}, Chung-Pei and {Hernquist}, Lars},
        title = "{The merger rate of galaxies in the Illustris simulation: a comparison with observations and semi-empirical models}",
      journal = {\mnras},
         year = 2015,
        month = may,
       volume = {449},
       number = {1},
        pages = {49-64},
          doi = {10.1093/mnras/stv264},
archivePrefix = {arXiv},
       eprint = {1502.01339},
 primaryClass = {astro-ph.GA},
       adsurl = {https://ui.adsabs.harvard.edu/abs/2015MNRAS.449...49R}
}

@ARTICLE{Vogelsberger2014,
       author = {{Vogelsberger}, M. and {Genel}, S. and {Springel}, V. and {Torrey}, P. and {Sijacki}, D. and {Xu}, D. and {Snyder}, G. and {Bird}, S. and {Nelson}, D. and {Hernquist}, L.},
        title = "{Properties of galaxies reproduced by a hydrodynamic simulation}",
      journal = {\nat},
         year = 2014,
        month = may,
       volume = {509},
       number = {7499},
        pages = {177-182},
          doi = {10.1038/nature13316},
archivePrefix = {arXiv},
       eprint = {1405.1418},
 primaryClass = {astro-ph.CO},
       adsurl = {https://ui.adsabs.harvard.edu/abs/2014Natur.509..177V}
}

@ARTICLE{Vogelsberger2014a,
       author = {{Vogelsberger}, Mark and {Genel}, Shy and {Springel}, Volker and {Torrey}, Paul and {Sijacki}, Debora and {Xu}, Dandan and {Snyder}, Greg and {Nelson}, Dylan and {Hernquist}, Lars},
        title = "{Introducing the Illustris Project: simulating the coevolution of dark and visible matter in the Universe}",
      journal = {\mnras},
         year = 2014,
        month = oct,
       volume = {444},
       number = {2},
        pages = {1518-1547},
          doi = {10.1093/mnras/stu1536},
archivePrefix = {arXiv},
       eprint = {1405.2921},
 primaryClass = {astro-ph.CO},
       adsurl = {https://ui.adsabs.harvard.edu/abs/2014MNRAS.444.1518V}
}

@ARTICLE{Schaye2015,
       author = {{Schaye}, Joop and {Crain}, Robert A. and {Bower}, Richard G. and {Furlong}, Michelle and {Schaller}, Matthieu and {Theuns}, Tom and {Dalla Vecchia}, Claudio and {Frenk}, Carlos S. and {McCarthy}, I.~G. and {Helly}, John C. and {Jenkins}, Adrian and {Rosas-Guevara}, Y.~M. and {White}, Simon D.~M. and {Baes}, Maarten and {Booth}, C.~M. and {Camps}, Peter and {Navarro}, Julio F. and {Qu}, Yan and {Rahmati}, Alireza and {Sawala}, Till and {Thomas}, Peter A. and {Trayford}, James},
        title = "{The EAGLE project: simulating the evolution and assembly of galaxies and their environments}",
      journal = {\mnras},
         year = 2015,
        month = jan,
       volume = {446},
       number = {1},
        pages = {521-554},
          doi = {10.1093/mnras/stu2058},
archivePrefix = {arXiv},
       eprint = {1407.7040},
 primaryClass = {astro-ph.GA},
       adsurl = {https://ui.adsabs.harvard.edu/abs/2015MNRAS.446..521S}
}

@ARTICLE{Dave2019,
       author = {{Dav{\'e}}, Romeel and {Angl{\'e}s-Alc{\'a}zar}, Daniel and {Narayanan}, Desika and {Li}, Qi and {Rafieferantsoa}, Mika H. and {Appleby}, Sarah},
        title = "{SIMBA: Cosmological simulations with black hole growth and feedback}",
      journal = {\mnras},
         year = 2019,
        month = jun,
       volume = {486},
       number = {2},
        pages = {2827-2849},
          doi = {10.1093/mnras/stz937},
archivePrefix = {arXiv},
       eprint = {1901.10203},
 primaryClass = {astro-ph.GA},
       adsurl = {https://ui.adsabs.harvard.edu/abs/2019MNRAS.486.2827D}
}

@ARTICLE{Artale2017,
       author = {{Artale}, M. Celeste and {Pedrosa}, Susana E. and {Trayford}, James W. and {Theuns}, Tom and {Farrow}, Daniel J. and {Norberg}, Peder and {Zehavi}, Idit and {Bower}, Richard G. and {Schaller}, Matthieu},
        title = "{Small-scale galaxy clustering in the eagle simulation}",
      journal = {\mnras},
         year = 2017,
        month = sep,
       volume = {470},
       number = {2},
        pages = {1771-1787},
          doi = {10.1093/mnras/stx1263},
archivePrefix = {arXiv},
       eprint = {1611.05064},
 primaryClass = {astro-ph.GA},
       adsurl = {https://ui.adsabs.harvard.edu/abs/2017MNRAS.470.1771A}
}

@ARTICLE{Pillepich2019,
       author = {{Pillepich}, Annalisa and {Nelson}, Dylan and {Springel}, Volker and {Pakmor}, R{\"u}diger and {Torrey}, Paul and {Weinberger}, Rainer and {Vogelsberger}, Mark and {Marinacci}, Federico and {Genel}, Shy and {van der Wel}, Arjen and {Hernquist}, Lars},
        title = "{First results from the TNG50 simulation: the evolution of stellar and gaseous discs across cosmic time}",
      journal = {\mnras},
         year = 2019,
        month = dec,
       volume = {490},
       number = {3},
        pages = {3196-3233},
          doi = {10.1093/mnras/stz2338},
archivePrefix = {arXiv},
       eprint = {1902.05553},
 primaryClass = {astro-ph.GA},
       adsurl = {https://ui.adsabs.harvard.edu/abs/2019MNRAS.490.3196P}
}

@ARTICLE{Eisert2023,
       author = {{Eisert}, Lukas and {Pillepich}, Annalisa and {Nelson}, Dylan and {Klessen}, Ralf S. and {Huertas-Company}, Marc and {Rodriguez-Gomez}, Vicente},
        title = "{ERGO-ML I: inferring the assembly histories of IllustrisTNG galaxies from integral observable properties via invertible neural networks}",
      journal = {\mnras},
         year = 2023,
        month = feb,
       volume = {519},
       number = {2},
        pages = {2199-2223},
          doi = {10.1093/mnras/stac3295},
archivePrefix = {arXiv},
       eprint = {2202.06967},
 primaryClass = {astro-ph.GA},
       adsurl = {https://ui.adsabs.harvard.edu/abs/2023MNRAS.519.2199E}
}

@ARTICLE{Nandra2013,
       author = {{Nandra}, Kirpal and {Barret}, Didier and {Barcons}, Xavier and {Fabian}, Andy and {den Herder}, Jan-Willem and {Piro}, Luigi and {Watson}, Mike and {Adami}, Christophe and {Aird}, James and {Afonso}, Jose Manuel and {Alexander}, Dave and {Argiroffi}, Costanza and {Amati}, Lorenzo and {Arnaud}, Monique and {Atteia}, Jean-Luc and {Audard}, Marc and {Badenes}, Carles and {Ballet}, Jean and {Ballo}, Lucia and {Bamba}, Aya and {Bhardwaj}, Anil and {Stefano Battistelli}, Elia and {Becker}, Werner and {De Becker}, Micha{\"e}l and {Behar}, Ehud and {Bianchi}, Stefano and {Biffi}, Veronica and {B{\^\i}rzan}, Laura and {Bocchino}, Fabrizio and {Bogdanov}, Slavko and {Boirin}, Laurence and {Boller}, Thomas and {Borgani}, Stefano and {Borm}, Katharina and {Bouch{\'e}}, Nicolas and {Bourdin}, Herv{\'e} and {Bower}, Richard and {Braito}, Valentina and {Branchini}, Enzo and {Branduardi-Raymont}, Graziella and {Bregman}, Joel and {Brenneman}, Laura and {Brightman}, Murray and {Br{\"u}ggen}, Marcus and {Buchner}, Johannes and {Bulbul}, Esra and {Brusa}, Marcella and {Bursa}, Michal and {Caccianiga}, Alessandro and {Cackett}, Ed and {Campana}, Sergio and {Cappelluti}, Nico and {Cappi}, Massimo and {Carrera}, Francisco and {Ceballos}, Maite and {Christensen}, Finn and {Chu}, You-Hua and {Churazov}, Eugene and {Clerc}, Nicolas and {Corbel}, Stephane and {Corral}, Amalia and {Comastri}, Andrea and {Costantini}, Elisa and {Croston}, Judith and {Dadina}, Mauro and {D'Ai}, Antonino and {Decourchelle}, Anne and {Della Ceca}, Roberto and {Dennerl}, Konrad and {Dolag}, Klaus and {Done}, Chris and {Dovciak}, Michal and {Drake}, Jeremy and {Eckert}, Dominique and {Edge}, Alastair and {Ettori}, Stefano and {Ezoe}, Yuichiro and {Feigelson}, Eric and {Fender}, Rob and {Feruglio}, Chiara and {Finoguenov}, Alexis and {Fiore}, Fabrizio and {Galeazzi}, Massimiliano and {Gallagher}, Sarah and {Gandhi}, Poshak and {Gaspari}, Massimo and {Gastaldello}, Fabio and {Georgakakis}, Antonis and {Georgantopoulos}, Ioannis and {Gilfanov}, Marat and {Gitti}, Myriam and {Gladstone}, Randy and {Goosmann}, Rene and {Gosset}, Eric and {Grosso}, Nicolas and {Guedel}, Manuel and {Guerrero}, Martin and {Haberl}, Frank and {Hardcastle}, Martin and {Heinz}, Sebastian and {Alonso Herrero}, Almudena and {Herv{\'e}}, Anthony and {Holmstrom}, Mats and {Iwasawa}, Kazushi and {Jonker}, Peter and {Kaastra}, Jelle and {Kara}, Erin and {Karas}, Vladimir and {Kastner}, Joel and {King}, Andrew and {Kosenko}, Daria and {Koutroumpa}, Dimita and {Kraft}, Ralph and {Kreykenbohm}, Ingo and {Lallement}, Rosine and {Lanzuisi}, Giorgio and {Lee}, J. and {Lemoine-Goumard}, Marianne and {Lobban}, Andrew and {Lodato}, Giuseppe and {Lovisari}, Lorenzo and {Lotti}, Simone and {McCharthy}, Ian and {McNamara}, Brian and {Maggio}, Antonio and {Maiolino}, Roberto and {De Marco}, Barbara and {de Martino}, Domitilla and {Mateos}, Silvia and {Matt}, Giorgio and {Maughan}, Ben and {Mazzotta}, Pasquale and {Mendez}, Mariano and {Merloni}, Andrea and {Micela}, Giuseppina and {Miceli}, Marco and {Mignani}, Robert and {Miller}, Jon and {Miniutti}, Giovanni and {Molendi}, Silvano and {Montez}, Rodolfo and {Moretti}, Alberto and {Motch}, Christian and {Naz{\'e}}, Ya{\"e}l and {Nevalainen}, Jukka and {Nicastro}, Fabrizio and {Nulsen}, Paul and {Ohashi}, Takaya and {O'Brien}, Paul and {Osborne}, Julian and {Oskinova}, Lida and {Pacaud}, Florian and {Paerels}, Frederik and {Page}, Mat and {Papadakis}, Iossif and {Pareschi}, Giovanni and {Petre}, Robert and {Petrucci}, Pierre-Olivier and {Piconcelli}, Enrico and {Pillitteri}, Ignazio and {Pinto}, C. and {de Plaa}, Jelle and {Pointecouteau}, Etienne and {Ponman}, Trevor and {Ponti}, Gabriele and {Porquet}, Delphine and {Pounds}, Ken and {Pratt}, Gabriel and {Predehl}, Peter and {Proga}, Daniel and {Psaltis}, Dimitrios and {Rafferty}, David and {Ramos-Ceja}, Miriam and {Ranalli}, Piero and {Rasia}, Elena and {Rau}, Arne and {Rauw}, Gregor and {Rea}, Nanda and {Read}, Andy and {Reeves}, James and {Reiprich}, Thomas and {Renaud}, Matthieu and {Reynolds}, Chris and {Risaliti}, Guido and {Rodriguez}, Jerome and {Rodriguez Hidalgo}, Paola and {Roncarelli}, Mauro and {Rosario}, David and {Rossetti}, Mariachiara and {Rozanska}, Agata and {Rovilos}, Emmanouil and {Salvaterra}, Ruben and {Salvato}, Mara and {Di Salvo}, Tiziana and {Sanders}, Jeremy and {Sanz-Forcada}, Jorge and {Schawinski}, Kevin and {Schaye}, Joop and {Schwope}, Axel and {Sciortino}, Salvatore},
        title = "{The Hot and Energetic Universe: A White Paper presenting the science theme motivating the Athena+ mission}",
      journal = {arXiv e-prints},
         year = 2013,
        month = jun,
          eid = {arXiv:1306.2307},
        pages = {arXiv:1306.2307},
          doi = {10.48550/arXiv.1306.2307},
archivePrefix = {arXiv},
       eprint = {1306.2307},
 primaryClass = {astro-ph.HE},
       adsurl = {https://ui.adsabs.harvard.edu/abs/2013arXiv1306.2307N}
}

@ARTICLE{Ranalli2016,
       author = {{Ranalli}, P. and {Koulouridis}, E. and {Georgantopoulos}, I. and {Fotopoulou}, S. and {Hsu}, L. -T. and {Salvato}, M. and {Comastri}, A. and {Pierre}, M. and {Cappelluti}, N. and {Carrera}, F.~J. and {Chiappetti}, L. and {Clerc}, N. and {Gilli}, R. and {Iwasawa}, K. and {Pacaud}, F. and {Paltani}, S. and {Plionis}, E. and {Vignali}, C.},
        title = "{The 2-10 keV unabsorbed luminosity function of AGN from the LSS, CDFS, and COSMOS surveys}",
      journal = {\aap},
         year = 2016,
        month = may,
       volume = {590},
          eid = {A80},
        pages = {A80},
          doi = {10.1051/0004-6361/201527013},
archivePrefix = {arXiv},
       eprint = {1512.05563},
 primaryClass = {astro-ph.HE},
       adsurl = {https://ui.adsabs.harvard.edu/abs/2016A&A...590A..80R}
}

@ARTICLE{Ivezic2019,
       author = {{Ivezi{\'c}}, {\v{Z}}eljko and {Kahn}, Steven M. and {Tyson}, J. Anthony and {Abel}, Bob and {Acosta}, Emily and {Allsman}, Robyn and {Alonso}, David and {AlSayyad}, Yusra and {Anderson}, Scott F. and {Andrew}, John and {Angel}, James Roger P. and {Angeli}, George Z. and {Ansari}, Reza and {Antilogus}, Pierre and {Araujo}, Constanza and {Armstrong}, Robert and {Arndt}, Kirk T. and {Astier}, Pierre and {Aubourg}, {\'E}ric and {Auza}, Nicole and {Axelrod}, Tim S. and {Bard}, Deborah J. and {Barr}, Jeff D. and {Barrau}, Aurelian and {Bartlett}, James G. and {Bauer}, Amanda E. and {Bauman}, Brian J. and {Baumont}, Sylvain and {Bechtol}, Ellen and {Bechtol}, Keith and {Becker}, Andrew C. and {Becla}, Jacek and {Beldica}, Cristina and {Bellavia}, Steve and {Bianco}, Federica B. and {Biswas}, Rahul and {Blanc}, Guillaume and {Blazek}, Jonathan and {Blandford}, Roger D. and {Bloom}, Josh S. and {Bogart}, Joanne and {Bond}, Tim W. and {Booth}, Michael T. and {Borgland}, Anders W. and {Borne}, Kirk and {Bosch}, James F. and {Boutigny}, Dominique and {Brackett}, Craig A. and {Bradshaw}, Andrew and {Brandt}, William Nielsen and {Brown}, Michael E. and {Bullock}, James S. and {Burchat}, Patricia and {Burke}, David L. and {Cagnoli}, Gianpietro and {Calabrese}, Daniel and {Callahan}, Shawn and {Callen}, Alice L. and {Carlin}, Jeffrey L. and {Carlson}, Erin L. and {Chandrasekharan}, Srinivasan and {Charles-Emerson}, Glenaver and {Chesley}, Steve and {Cheu}, Elliott C. and {Chiang}, Hsin-Fang and {Chiang}, James and {Chirino}, Carol and {Chow}, Derek and {Ciardi}, David R. and {Claver}, Charles F. and {Cohen-Tanugi}, Johann and {Cockrum}, Joseph J. and {Coles}, Rebecca and {Connolly}, Andrew J. and {Cook}, Kem H. and {Cooray}, Asantha and {Covey}, Kevin R. and {Cribbs}, Chris and {Cui}, Wei and {Cutri}, Roc and {Daly}, Philip N. and {Daniel}, Scott F. and {Daruich}, Felipe and {Daubard}, Guillaume and {Daues}, Greg and {Dawson}, William and {Delgado}, Francisco and {Dellapenna}, Alfred and {de Peyster}, Robert and {de Val-Borro}, Miguel and {Digel}, Seth W. and {Doherty}, Peter and {Dubois}, Richard and {Dubois-Felsmann}, Gregory P. and {Durech}, Josef and {Economou}, Frossie and {Eifler}, Tim and {Eracleous}, Michael and {Emmons}, Benjamin L. and {Fausti Neto}, Angelo and {Ferguson}, Henry and {Figueroa}, Enrique and {Fisher-Levine}, Merlin and {Focke}, Warren and {Foss}, Michael D. and {Frank}, James and {Freemon}, Michael D. and {Gangler}, Emmanuel and {Gawiser}, Eric and {Geary}, John C. and {Gee}, Perry and {Geha}, Marla and {Gessner}, Charles J.~B. and {Gibson}, Robert R. and {Gilmore}, D. Kirk and {Glanzman}, Thomas and {Glick}, William and {Goldina}, Tatiana and {Goldstein}, Daniel A. and {Goodenow}, Iain and {Graham}, Melissa L. and {Gressler}, William J. and {Gris}, Philippe and {Guy}, Leanne P. and {Guyonnet}, Augustin and {Haller}, Gunther and {Harris}, Ron and {Hascall}, Patrick A. and {Haupt}, Justine and {Hernandez}, Fabio and {Herrmann}, Sven and {Hileman}, Edward and {Hoblitt}, Joshua and {Hodgson}, John A. and {Hogan}, Craig and {Howard}, James D. and {Huang}, Dajun and {Huffer}, Michael E. and {Ingraham}, Patrick and {Innes}, Walter R. and {Jacoby}, Suzanne H. and {Jain}, Bhuvnesh and {Jammes}, Fabrice and {Jee}, M. James and {Jenness}, Tim and {Jernigan}, Garrett and {Jevremovi{\'c}}, Darko and {Johns}, Kenneth and {Johnson}, Anthony S. and {Johnson}, Margaret W.~G. and {Jones}, R. Lynne and {Juramy-Gilles}, Claire and {Juri{\'c}}, Mario and {Kalirai}, Jason S. and {Kallivayalil}, Nitya J. and {Kalmbach}, Bryce and {Kantor}, Jeffrey P. and {Karst}, Pierre and {Kasliwal}, Mansi M. and {Kelly}, Heather and {Kessler}, Richard and {Kinnison}, Veronica and {Kirkby}, David and {Knox}, Lloyd and {Kotov}, Ivan V. and {Krabbendam}, Victor L. and {Krughoff}, K. Simon and {Kub{\'a}nek}, Petr and {Kuczewski}, John and {Kulkarni}, Shri and {Ku}, John and {Kurita}, Nadine R. and {Lage}, Craig S. and {Lambert}, Ron and {Lange}, Travis and {Langton}, J. Brian and {Le Guillou}, Laurent and {Levine}, Deborah and {Liang}, Ming and {Lim}, Kian-Tat and {Lintott}, Chris J. and {Long}, Kevin E. and {Lopez}, Margaux and {Lotz}, Paul J. and {Lupton}, Robert H. and {Lust}, Nate B. and {MacArthur}, Lauren A. and {Mahabal}, Ashish and {Mandelbaum}, Rachel and {Markiewicz}, Thomas W. and {Marsh}, Darren S. and {Marshall}, Philip J. and {Marshall}, Stuart and {May}, Morgan and {McKercher}, Robert and {McQueen}, Michelle and {Meyers}, Joshua and {Migliore}, Myriam and {Miller}, Michelle and {Mills}, David J.},
        title = "{LSST: From Science Drivers to Reference Design and Anticipated Data Products}",
      journal = {\apj},
         year = 2019,
        month = mar,
       volume = {873},
       number = {2},
          eid = {111},
        pages = {111},
          doi = {10.3847/1538-4357/ab042c},
archivePrefix = {arXiv},
       eprint = {0805.2366},
 primaryClass = {astro-ph},
       adsurl = {https://ui.adsabs.harvard.edu/abs/2019ApJ...873..111I}
}

@ARTICLE{Wang2022_HLS,
       author = {{Wang}, Yun and {Zhai}, Zhongxu and {Alavi}, Anahita and {Massara}, Elena and {Pisani}, Alice and {Benson}, Andrew and {Hirata}, Christopher M. and {Samushia}, Lado and {Weinberg}, David H. and {Colbert}, James and {Dor{\'e}}, Olivier and {Eifler}, Tim and {Heinrich}, Chen and {Ho}, Shirley and {Krause}, Elisabeth and {Padmanabhan}, Nikhil and {Spergel}, David and {Teplitz}, Harry I.},
        title = "{The High Latitude Spectroscopic Survey on the Nancy Grace Roman Space Telescope}",
      journal = {\apj},
         year = 2022,
        month = mar,
       volume = {928},
       number = {1},
          eid = {1},
        pages = {1},
          doi = {10.3847/1538-4357/ac4973},
archivePrefix = {arXiv},
       eprint = {2110.01829},
 primaryClass = {astro-ph.CO},
       adsurl = {https://ui.adsabs.harvard.edu/abs/2022ApJ...928....1W}
}

@ARTICLE{TheLynxTeam2018,
       author = {{The Lynx Team}},
        title = "{The Lynx Mission Concept Study Interim Report}",
      journal = {arXiv e-prints},
         year = 2018,
        month = sep,
          eid = {arXiv:1809.09642},
        pages = {arXiv:1809.09642},
          doi = {10.48550/arXiv.1809.09642},
archivePrefix = {arXiv},
       eprint = {1809.09642},
 primaryClass = {astro-ph.IM},
       adsurl = {https://ui.adsabs.harvard.edu/abs/2018arXiv180909642T}
}

@INPROCEEDINGS{Mushotzky2018,
       author = {{Mushotzky}, R.},
        title = "{AXIS: a probe class next generation high angular resolution x-ray imaging satellite}",
    booktitle = {Space Telescopes and Instrumentation 2018: Ultraviolet to Gamma Ray},
         year = 2018,
       editor = {{den Herder}, Jan-Willem A. and {Nikzad}, Shouleh and {Nakazawa}, Kazuhiro},
       series = {Society of Photo-Optical Instrumentation Engineers (SPIE) Conference Series},
       volume = {10699},
        month = jul,
          eid = {1069929},
        pages = {1069929},
          doi = {10.1117/12.2310003},
archivePrefix = {arXiv},
       eprint = {1807.02122},
 primaryClass = {astro-ph.HE},
       adsurl = {https://ui.adsabs.harvard.edu/abs/2018SPIE10699E..29M}
}

@ARTICLE{Marsan2017,
       author = {{Marsan}, Z. Cemile and {Marchesini}, Danilo and {Brammer}, Gabriel B. and {Geier}, Stefan and {Kado-Fong}, Erin and {Labb{\'e}}, Ivo and {Muzzin}, Adam and {Stefanon}, Mauro},
        title = "{A Spectroscopic Follow-up Program of Very Massive Galaxies at 3 < z < 4: Confirmation of Spectroscopic Redshifts, and a High Fraction of Powerful AGNs}",
      journal = {\apj},
         year = 2017,
        month = jun,
       volume = {842},
       number = {1},
          eid = {21},
        pages = {21},
          doi = {10.3847/1538-4357/aa7206},
archivePrefix = {arXiv},
       eprint = {1606.05350},
 primaryClass = {astro-ph.GA},
       adsurl = {https://ui.adsabs.harvard.edu/abs/2017ApJ...842...21M}
}

@ARTICLE{Torrey2018,
       author = {{Torrey}, Paul and {Vogelsberger}, Mark and {Hernquist}, Lars and {McKinnon}, Ryan and {Marinacci}, Federico and {Simcoe}, Robert A. and {Springel}, Volker and {Pillepich}, Annalisa and {Naiman}, Jill and {Pakmor}, R{\"u}diger and {Weinberger}, Rainer and {Nelson}, Dylan and {Genel}, Shy},
        title = "{Similar star formation rate and metallicity variability time-scales drive the fundamental metallicity relation}",
      journal = {\mnras},
         year = 2018,
        month = jun,
       volume = {477},
       number = {1},
        pages = {L16-L20},
          doi = {10.1093/mnrasl/sly031},
archivePrefix = {arXiv},
       eprint = {1711.11039},
 primaryClass = {astro-ph.GA},
       adsurl = {https://ui.adsabs.harvard.edu/abs/2018MNRAS.477L..16T}
}

@ARTICLE{Agazie2023_nanograv_pta,
       author = {{Agazie}, Gabriella and {Anumarlapudi}, Akash and {Archibald}, Anne M. and {Arzoumanian}, Zaven and {Baker}, Paul T. and {B{\'e}csy}, Bence and {Blecha}, Laura and {Brazier}, Adam and {Brook}, Paul R. and {Burke-Spolaor}, Sarah and {Burnette}, Rand and {Case}, Robin and {Charisi}, Maria and {Chatterjee}, Shami and {Chatziioannou}, Katerina and {Cheeseboro}, Belinda D. and {Chen}, Siyuan and {Cohen}, Tyler and {Cordes}, James M. and {Cornish}, Neil J. and {Crawford}, Fronefield and {Cromartie}, H. Thankful and {Crowter}, Kathryn and {Cutler}, Curt J. and {Decesar}, Megan E. and {Degan}, Dallas and {Demorest}, Paul B. and {Deng}, Heling and {Dolch}, Timothy and {Drachler}, Brendan and {Ellis}, Justin A. and {Ferrara}, Elizabeth C. and {Fiore}, William and {Fonseca}, Emmanuel and {Freedman}, Gabriel E. and {Garver-Daniels}, Nate and {Gentile}, Peter A. and {Gersbach}, Kyle A. and {Glaser}, Joseph and {Good}, Deborah C. and {G{\"u}ltekin}, Kayhan and {Hazboun}, Jeffrey S. and {Hourihane}, Sophie and {Islo}, Kristina and {Jennings}, Ross J. and {Johnson}, Aaron D. and {Jones}, Megan L. and {Kaiser}, Andrew R. and {Kaplan}, David L. and {Kelley}, Luke Zoltan and {Kerr}, Matthew and {Key}, Joey S. and {Klein}, Tonia C. and {Laal}, Nima and {Lam}, Michael T. and {Lamb}, William G. and {Lazio}, T. Joseph W. and {Lewandowska}, Natalia and {Littenberg}, Tyson B. and {Liu}, Tingting and {Lommen}, Andrea and {Lorimer}, Duncan R. and {Luo}, Jing and {Lynch}, Ryan S. and {Ma}, Chung-Pei and {Madison}, Dustin R. and {Mattson}, Margaret A. and {McEwen}, Alexander and {McKee}, James W. and {McLaughlin}, Maura A. and {McMann}, Natasha and {Meyers}, Bradley W. and {Meyers}, Patrick M. and {Mingarelli}, Chiara M.~F. and {Mitridate}, Andrea and {Natarajan}, Priyamvada and {Ng}, Cherry and {Nice}, David J. and {Ocker}, Stella Koch and {Olum}, Ken D. and {Pennucci}, Timothy T. and {Perera}, Benetge B.~P. and {Petrov}, Polina and {Pol}, Nihan S. and {Radovan}, Henri A. and {Ransom}, Scott M. and {Ray}, Paul S. and {Romano}, Joseph D. and {Sardesai}, Shashwat C. and {Schmiedekamp}, Ann and {Schmiedekamp}, Carl and {Schmitz}, Kai and {Schult}, Levi and {Shapiro-Albert}, Brent J. and {Siemens}, Xavier and {Simon}, Joseph and {Siwek}, Magdalena S. and {Stairs}, Ingrid H. and {Stinebring}, Daniel R. and {Stovall}, Kevin and {Sun}, Jerry P. and {Susobhanan}, Abhimanyu and {Swiggum}, Joseph K. and {Taylor}, Jacob and {Taylor}, Stephen R. and {Turner}, Jacob E. and {Unal}, Caner and {Vallisneri}, Michele and {van Haasteren}, Rutger and {Vigeland}, Sarah J. and {Wahl}, Haley M. and {Wang}, Qiaohong and {Witt}, Caitlin A. and {Young}, Olivia and {Nanograv Collaboration}},
        title = "{The NANOGrav 15 yr Data Set: Evidence for a Gravitational-wave Background}",
      journal = {\apjl},
         year = 2023,
        month = jul,
       volume = {951},
       number = {1},
          eid = {L8},
        pages = {L8},
          doi = {10.3847/2041-8213/acdac6},
archivePrefix = {arXiv},
       eprint = {2306.16213},
 primaryClass = {astro-ph.HE},
       adsurl = {https://ui.adsabs.harvard.edu/abs/2023ApJ...951L...8A}
}

@ARTICLE{Yang2023,
       author = {{Yang}, G. and {Caputi}, K.~I. and {Papovich}, C. and {Arrabal Haro}, P. and {Bagley}, M.~B. and {Behroozi}, P. and {Bell}, E.~F. and {Bisigello}, L. and {Buat}, V. and {Burgarella}, D. and {Cheng}, Y. and {Cleri}, N.~J. and {Dav{\'e}}, R. and {Dickinson}, M. and {Elbaz}, D. and {Ferguson}, H.~C. and {Finkelstein}, S.~L. and {Grogin}, N.~A. and {Hathi}, N.~P. and {Hirschmann}, M. and {Holwerda}, B.~W. and {Huertas-Company}, M. and {Hutchison}, T.~A. and {Iani}, E. and {Kartaltepe}, J.~S. and {Kirkpatrick}, A. and {Kocevski}, D.~D. and {Koekemoer}, A.~M. and {Kokorev}, V. and {Larson}, R.~L. and {Lucas}, R.~A. and {P{\'e}rez-Gonz{\'a}lez}, P.~G. and {Rinaldi}, P. and {Shen}, L. and {Trump}, J.~R. and {de la Vega}, A. and {Yung}, L.~Y.~A. and {Zavala}, J.~A.},
        title = "{CEERS Key Paper. VI. JWST/MIRI Uncovers a Large Population of Obscured AGN at High Redshifts}",
      journal = {\apjl},
         year = 2023,
        month = jun,
       volume = {950},
       number = {1},
          eid = {L5},
        pages = {L5},
          doi = {10.3847/2041-8213/acd639},
archivePrefix = {arXiv},
       eprint = {2303.11736},
 primaryClass = {astro-ph.GA},
       adsurl = {https://ui.adsabs.harvard.edu/abs/2023ApJ...950L...5Y}
}

@ARTICLE{Vito2018,
       author = {{Vito}, F. and {Brandt}, W.~N. and {Yang}, G. and {Gilli}, R. and {Luo}, B. and {Vignali}, C. and {Xue}, Y.~Q. and {Comastri}, A. and {Koekemoer}, A.~M. and {Lehmer}, B.~D. and {Liu}, T. and {Paolillo}, M. and {Ranalli}, P. and {Schneider}, D.~P. and {Shemmer}, O. and {Volonteri}, M. and {Wang}, J.},
        title = "{High-redshift AGN in the Chandra Deep Fields: the obscured fraction and space density of the sub-L$_{*}$ population}",
      journal = {\mnras},
         year = 2018,
        month = jan,
       volume = {473},
       number = {2},
        pages = {2378-2406},
          doi = {10.1093/mnras/stx2486},
archivePrefix = {arXiv},
       eprint = {1709.07892},
 primaryClass = {astro-ph.GA},
       adsurl = {https://ui.adsabs.harvard.edu/abs/2018MNRAS.473.2378V}
}

@ARTICLE{Muzzin2013,
       author = {{Muzzin}, Adam and {Marchesini}, Danilo and {Stefanon}, Mauro and {Franx}, Marijn and {McCracken}, Henry J. and {Milvang-Jensen}, Bo and {Dunlop}, James S. and {Fynbo}, J.~P.~U. and {Brammer}, Gabriel and {Labb{\'e}}, Ivo and {van Dokkum}, Pieter G.},
        title = "{The Evolution of the Stellar Mass Functions of Star-forming and Quiescent Galaxies to z = 4 from the COSMOS/UltraVISTA Survey}",
      journal = {\apj},
         year = 2013,
        month = nov,
       volume = {777},
       number = {1},
          eid = {18},
        pages = {18},
          doi = {10.1088/0004-637X/777/1/18},
archivePrefix = {arXiv},
       eprint = {1303.4409},
 primaryClass = {astro-ph.CO},
       adsurl = {https://ui.adsabs.harvard.edu/abs/2013ApJ...777...18M}
}

@ARTICLE{Weaver2023,
       author = {{Weaver}, J.~R. and {Davidzon}, I. and {Toft}, S. and {Ilbert}, O. and {McCracken}, H.~J. and {Gould}, K.~M.~L. and {Jespersen}, C.~K. and {Steinhardt}, C. and {Lagos}, C.~D.~P. and {Capak}, P.~L. and {Casey}, C.~M. and {Chartab}, N. and {Faisst}, A.~L. and {Hayward}, C.~C. and {Kartaltepe}, J.~S. and {Kauffmann}, O.~B. and {Koekemoer}, A.~M. and {Kokorev}, V. and {Laigle}, C. and {Liu}, D. and {Long}, A. and {Magdis}, G.~E. and {McPartland}, C.~J.~R. and {Milvang-Jensen}, B. and {Mobasher}, B. and {Moneti}, A. and {Peng}, Y. and {Sanders}, D.~B. and {Shuntov}, M. and {Sneppen}, A. and {Valentino}, F. and {Zalesky}, L. and {Zamorani}, G.},
        title = "{COSMOS2020: The galaxy stellar mass function. The assembly and star formation cessation of galaxies at 0.2< z {\ensuremath{\leq}} 7.5}",
      journal = {\aap},
         year = 2023,
        month = sep,
       volume = {677},
          eid = {A184},
        pages = {A184},
          doi = {10.1051/0004-6361/202245581},
archivePrefix = {arXiv},
       eprint = {2212.02512},
 primaryClass = {astro-ph.GA},
       adsurl = {https://ui.adsabs.harvard.edu/abs/2023A&A...677A.184W}
}

@ARTICLE{Woo2008,
       author = {{Woo}, Joanna and {Courteau}, St{\'e}phane and {Dekel}, Avishai},
        title = "{Scaling relations and the fundamental line of the local group dwarf galaxies}",
      journal = {\mnras},
         year = 2008,
        month = nov,
       volume = {390},
       number = {4},
        pages = {1453-1469},
          doi = {10.1111/j.1365-2966.2008.13770.x},
archivePrefix = {arXiv},
       eprint = {0807.1331},
 primaryClass = {astro-ph},
       adsurl = {https://ui.adsabs.harvard.edu/abs/2008MNRAS.390.1453W}
}

@ARTICLE{Pakmor2023,
       author = {{Pakmor}, R{\"u}diger and {Springel}, Volker and {Coles}, Jonathan P. and {Guillet}, Thomas and {Pfrommer}, Christoph and {Bose}, Sownak and {Barrera}, Monica and {Delgado}, Ana Maria and {Ferlito}, Fulvio and {Frenk}, Carlos and {Hadzhiyska}, Boryana and {Hern{\'a}ndez-Aguayo}, C{\'e}sar and {Hernquist}, Lars and {Kannan}, Rahul and {White}, Simon D.~M.},
        title = "{The MillenniumTNG Project: the hydrodynamical full physics simulation and a first look at its galaxy clusters}",
      journal = {\mnras},
         year = 2023,
        month = sep,
       volume = {524},
       number = {2},
        pages = {2539-2555},
          doi = {10.1093/mnras/stac3620},
archivePrefix = {arXiv},
       eprint = {2210.10060},
 primaryClass = {astro-ph.CO},
       adsurl = {https://ui.adsabs.harvard.edu/abs/2023MNRAS.524.2539P}
}

@ARTICLE{Nelson2018,
       author = {{Nelson}, Dylan and {Pillepich}, Annalisa and {Springel}, Volker and {Weinberger}, Rainer and {Hernquist}, Lars and {Pakmor}, R{\"u}diger and {Genel}, Shy and {Torrey}, Paul and {Vogelsberger}, Mark and {Kauffmann}, Guinevere and {Marinacci}, Federico and {Naiman}, Jill},
        title = "{First results from the IllustrisTNG simulations: the galaxy colour bimodality}",
      journal = {\mnras},
         year = 2018,
        month = mar,
       volume = {475},
       number = {1},
        pages = {624-647},
          doi = {10.1093/mnras/stx3040},
archivePrefix = {arXiv},
       eprint = {1707.03395},
 primaryClass = {astro-ph.GA},
       adsurl = {https://ui.adsabs.harvard.edu/abs/2018MNRAS.475..624N}
}

@ARTICLE{Chabrier2003,
       author = {{Chabrier}, Gilles},
        title = "{Galactic Stellar and Substellar Initial Mass Function}",
      journal = {\pasp},
         year = 2003,
        month = jul,
       volume = {115},
       number = {809},
        pages = {763-795},
          doi = {10.1086/376392},
archivePrefix = {arXiv},
       eprint = {astro-ph/0304382},
 primaryClass = {astro-ph},
       adsurl = {https://ui.adsabs.harvard.edu/abs/2003PASP..115..763C}
}

@ARTICLE{Chen2023_dual,
       author = {{Chen}, Nianyi and {Di Matteo}, Tiziana and {Ni}, Yueying and {Tremmel}, Michael and {DeGraf}, Colin and {Shen}, Yue and {Holgado}, A. Miguel and {Bird}, Simeon and {Croft}, Rupert and {Feng}, Yu},
        title = "{Properties and evolution of dual and offset AGN in the ASTRID simulation at z   2}",
      journal = {\mnras},
         year = 2023,
        month = jun,
       volume = {522},
       number = {2},
        pages = {1895-1913},
          doi = {10.1093/mnras/stad834},
archivePrefix = {arXiv},
       eprint = {2208.04970},
 primaryClass = {astro-ph.GA},
       adsurl = {https://ui.adsabs.harvard.edu/abs/2023MNRAS.522.1895C}
}

@ARTICLE{Sato-Polito2025,
       author = {{Sato-Polito}, Gabriela and {Zaldarriaga}, Matias},
        title = "{Distribution of the gravitational-wave background from supermassive black holes}",
      journal = {\prd},
         year = 2025,
        month = jan,
       volume = {111},
       number = {2},
          eid = {023043},
        pages = {023043},
          doi = {10.1103/PhysRevD.111.023043},
archivePrefix = {arXiv},
       eprint = {2406.17010},
 primaryClass = {astro-ph.CO},
       adsurl = {https://ui.adsabs.harvard.edu/abs/2025PhRvD.111b3043S}
}

@ARTICLE{Zhou2025_PTA_CW,
       author = {{Zhou}, Yihao and {Di Matteo}, Tiziana and {Chen}, Nianyi and {Kelley}, Luke Zoltan and {Blecha}, Laura and {Ni}, Yueying and {Bird}, Simeon and {Yang}, Yanhui and {Croft}, Rupert},
        title = "{Central Cluster Galaxies: A Hotspot for Detectable Gravitational Waves from Black Hole Mergers}",
      journal = {arXiv e-prints},
         year = 2025,
        month = feb,
          eid = {arXiv:2502.01845},
        pages = {arXiv:2502.01845},
          doi = {10.48550/arXiv.2502.01845},
archivePrefix = {arXiv},
       eprint = {2502.01845},
 primaryClass = {astro-ph.GA},
       adsurl = {https://ui.adsabs.harvard.edu/abs/2025arXiv250201845Z}
}

@ARTICLE{Dutton2011,
       author = {{Dutton}, Aaron A. and {van den Bosch}, Frank C. and {Faber}, Sandra M. and {Simard}, Luc and {Kassin}, Susan A. and {Koo}, David C. and {Bundy}, Kevin and {Huang}, Jiasheng and {Weiner}, Benjamin J. and {Cooper}, Michael C. and {Newman}, Jeffrey A. and {Mozena}, Mark and {Koekemoer}, Anton M.},
        title = "{On the evolution of the velocity-mass-size relations of disc-dominated galaxies over the past 10 billion years}",
      journal = {\mnras},
         year = 2011,
        month = jan,
       volume = {410},
       number = {3},
        pages = {1660-1676},
          doi = {10.1111/j.1365-2966.2010.17555.x},
archivePrefix = {arXiv},
       eprint = {1006.3558},
 primaryClass = {astro-ph.GA},
       adsurl = {https://ui.adsabs.harvard.edu/abs/2011MNRAS.410.1660D}
}

@ARTICLE{Kirby2013,
       author = {{Kirby}, Evan N. and {Cohen}, Judith G. and {Guhathakurta}, Puragra and {Cheng}, Lucy and {Bullock}, James S. and {Gallazzi}, Anna},
        title = "{The Universal Stellar Mass-Stellar Metallicity Relation for Dwarf Galaxies}",
      journal = {\apj},
         year = 2013,
        month = dec,
       volume = {779},
       number = {2},
          eid = {102},
        pages = {102},
          doi = {10.1088/0004-637X/779/2/102},
archivePrefix = {arXiv},
       eprint = {1310.0814},
 primaryClass = {astro-ph.GA},
       adsurl = {https://ui.adsabs.harvard.edu/abs/2013ApJ...779..102K}
}

@ARTICLE{Bauer2013,
       author = {{Bauer}, Amanda E. and {Hopkins}, Andrew M. and {Gunawardhana}, Madusha and {Taylor}, Edward N. and {Baldry}, Ivan and {Bamford}, Steven P. and {Bland-Hawthorn}, Joss and {Brough}, Sarah and {Brown}, Michael J.~I. and {Cluver}, Michelle E. and {Colless}, Matthew and {Conselice}, Christopher J. and {Croom}, Scott and {Driver}, Simon and {Foster}, Caroline and {Jones}, D. Heath and {Lara-Lopez}, Maritza A. and {Liske}, Jochen and {L{\'o}pez-S{\'a}nchez}, {\'A}ngel R. and {Loveday}, Jon and {Norberg}, Peder and {Owers}, Matt S. and {Pimbblet}, Kevin and {Robotham}, Aaron and {Sansom}, Anne E. and {Sharp}, Rob},
        title = "{Galaxy And Mass Assembly (GAMA): linking star formation histories and stellar mass growth}",
      journal = {\mnras},
         year = 2013,
        month = sep,
       volume = {434},
       number = {1},
        pages = {209-221},
          doi = {10.1093/mnras/stt1011},
archivePrefix = {arXiv},
       eprint = {1306.2424},
 primaryClass = {astro-ph.CO},
       adsurl = {https://ui.adsabs.harvard.edu/abs/2013MNRAS.434..209B}
}

@ARTICLE{vanderWel2014,
       author = {{van der Wel}, A. and {Franx}, M. and {van Dokkum}, P.~G. and {Skelton}, R.~E. and {Momcheva}, I.~G. and {Whitaker}, K.~E. and {Brammer}, G.~B. and {Bell}, E.~F. and {Rix}, H. -W. and {Wuyts}, S. and {Ferguson}, H.~C. and {Holden}, B.~P. and {Barro}, G. and {Koekemoer}, A.~M. and {Chang}, Yu-Yen and {McGrath}, E.~J. and {H{\"a}ussler}, B. and {Dekel}, A. and {Behroozi}, P. and {Fumagalli}, M. and {Leja}, J. and {Lundgren}, B.~F. and {Maseda}, M.~V. and {Nelson}, E.~J. and {Wake}, D.~A. and {Patel}, S.~G. and {Labb{\'e}}, I. and {Faber}, S.~M. and {Grogin}, N.~A. and {Kocevski}, D.~D.},
        title = "{3D-HST+CANDELS: The Evolution of the Galaxy Size-Mass Distribution since z = 3}",
      journal = {\apj},
         year = 2014,
        month = jun,
       volume = {788},
       number = {1},
          eid = {28},
        pages = {28},
          doi = {10.1088/0004-637X/788/1/28},
archivePrefix = {arXiv},
       eprint = {1404.2844},
 primaryClass = {astro-ph.GA},
       adsurl = {https://ui.adsabs.harvard.edu/abs/2014ApJ...788...28V}
}

@ARTICLE{Lange2015,
       author = {{Lange}, Rebecca and {Driver}, Simon P. and {Robotham}, Aaron S.~G. and {Kelvin}, Lee S. and {Graham}, Alister W. and {Alpaslan}, Mehmet and {Andrews}, Stephen K. and {Baldry}, Ivan K. and {Bamford}, Steven and {Bland-Hawthorn}, Joss and {Brough}, Sarah and {Cluver}, Michelle E. and {Conselice}, Christopher J. and {Davies}, Luke J.~M. and {Haeussler}, Boris and {Konstantopoulos}, Iraklis S. and {Loveday}, Jon and {Moffett}, Amanda J. and {Norberg}, Peder and {Phillipps}, Steven and {Taylor}, Edward N. and {L{\'o}pez-S{\'a}nchez}, {\'A}ngel R. and {Wilkins}, Stephen M.},
        title = "{Galaxy And Mass Assembly (GAMA): mass-size relations of z < 0.1 galaxies subdivided by S{\'e}rsic index, colour and morphology}",
      journal = {\mnras},
         year = 2015,
        month = mar,
       volume = {447},
       number = {3},
        pages = {2603-2630},
          doi = {10.1093/mnras/stu2467},
archivePrefix = {arXiv},
       eprint = {1411.6355},
 primaryClass = {astro-ph.GA},
       adsurl = {https://ui.adsabs.harvard.edu/abs/2015MNRAS.447.2603L}
}

@ARTICLE{Shen2003,
       author = {{Shen}, Shiyin and {Mo}, H.~J. and {White}, Simon D.~M. and {Blanton}, Michael R. and {Kauffmann}, Guinevere and {Voges}, Wolfgang and {Brinkmann}, J. and {Csabai}, Istvan},
        title = "{The size distribution of galaxies in the Sloan Digital Sky Survey}",
      journal = {\mnras},
         year = 2003,
        month = aug,
       volume = {343},
       number = {3},
        pages = {978-994},
          doi = {10.1046/j.1365-8711.2003.06740.x},
archivePrefix = {arXiv},
       eprint = {astro-ph/0301527},
 primaryClass = {astro-ph},
       adsurl = {https://ui.adsabs.harvard.edu/abs/2003MNRAS.343..978S}
}

@ARTICLE{Moustakas2013,
       author = {{Moustakas}, John and {Coil}, Alison L. and {Aird}, James and {Blanton}, Michael R. and {Cool}, Richard J. and {Eisenstein}, Daniel J. and {Mendez}, Alexander J. and {Wong}, Kenneth C. and {Zhu}, Guangtun and {Arnouts}, St{\'e}phane},
        title = "{PRIMUS: Constraints on Star Formation Quenching and Galaxy Merging, and the Evolution of the Stellar Mass Function from z = 0-1}",
      journal = {\apj},
         year = 2013,
        month = apr,
       volume = {767},
       number = {1},
          eid = {50},
        pages = {50},
          doi = {10.1088/0004-637X/767/1/50},
archivePrefix = {arXiv},
       eprint = {1301.1688},
 primaryClass = {astro-ph.CO},
       adsurl = {https://ui.adsabs.harvard.edu/abs/2013ApJ...767...50M}
}

@ARTICLE{Bouwens2015,
       author = {{Bouwens}, R.~J. and {Illingworth}, G.~D. and {Oesch}, P.~A. and {Trenti}, M. and {Labb{\'e}}, I. and {Bradley}, L. and {Carollo}, M. and {van Dokkum}, P.~G. and {Gonzalez}, V. and {Holwerda}, B. and {Franx}, M. and {Spitler}, L. and {Smit}, R. and {Magee}, D.},
        title = "{UV Luminosity Functions at Redshifts z {\ensuremath{\sim}} 4 to z {\ensuremath{\sim}} 10: 10,000 Galaxies from HST Legacy Fields}",
      journal = {\apj},
         year = 2015,
        month = apr,
       volume = {803},
       number = {1},
          eid = {34},
        pages = {34},
          doi = {10.1088/0004-637X/803/1/34},
archivePrefix = {arXiv},
       eprint = {1403.4295},
 primaryClass = {astro-ph.CO},
       adsurl = {https://ui.adsabs.harvard.edu/abs/2015ApJ...803...34B}
}

@ARTICLE{Madau2014,
       author = {{Madau}, Piero and {Dickinson}, Mark},
        title = "{Cosmic Star-Formation History}",
      journal = {\araa},
         year = 2014,
        month = aug,
       volume = {52},
        pages = {415-486},
          doi = {10.1146/annurev-astro-081811-125615},
archivePrefix = {arXiv},
       eprint = {1403.0007},
 primaryClass = {astro-ph.CO},
       adsurl = {https://ui.adsabs.harvard.edu/abs/2014ARA&A..52..415M}
}

@ARTICLE{Sijacki2015,
       author = {{Sijacki}, Debora and {Vogelsberger}, Mark and {Genel}, Shy and {Springel}, Volker and {Torrey}, Paul and {Snyder}, Gregory F. and {Nelson}, Dylan and {Hernquist}, Lars},
        title = "{The Illustris simulation: the evolving population of black holes across cosmic time}",
      journal = {\mnras},
         year = 2015,
        month = sep,
       volume = {452},
       number = {1},
        pages = {575-596},
          doi = {10.1093/mnras/stv1340},
archivePrefix = {arXiv},
       eprint = {1408.6842},
 primaryClass = {astro-ph.GA},
       adsurl = {https://ui.adsabs.harvard.edu/abs/2015MNRAS.452..575S}
}

@ARTICLE{Grazian2015,
       author = {{Grazian}, A. and {Fontana}, A. and {Santini}, P. and {Dunlop}, J.~S. and {Ferguson}, H.~C. and {Castellano}, M. and {Amorin}, R. and {Ashby}, M.~L.~N. and {Barro}, G. and {Behroozi}, P. and {Boutsia}, K. and {Caputi}, K.~I. and {Chary}, R.~R. and {Dekel}, A. and {Dickinson}, M.~E. and {Faber}, S.~M. and {Fazio}, G.~G. and {Finkelstein}, S.~L. and {Galametz}, A. and {Giallongo}, E. and {Giavalisco}, M. and {Grogin}, N.~A. and {Guo}, Y. and {Kocevski}, D. and {Koekemoer}, A.~M. and {Koo}, D.~C. and {Lee}, K. -S. and {Lu}, Y. and {Merlin}, E. and {Mobasher}, B. and {Nonino}, M. and {Papovich}, C. and {Paris}, D. and {Pentericci}, L. and {Reddy}, N. and {Renzini}, A. and {Salmon}, B. and {Salvato}, M. and {Sommariva}, V. and {Song}, M. and {Vanzella}, E.},
        title = "{The galaxy stellar mass function at 3.5 {\ensuremath{\leq}}z {\ensuremath{\leq}} 7.5 in the CANDELS/UDS, GOODS-South, and HUDF fields}",
      journal = {\aap},
         year = 2015,
        month = mar,
       volume = {575},
          eid = {A96},
        pages = {A96},
          doi = {10.1051/0004-6361/201424750},
archivePrefix = {arXiv},
       eprint = {1412.0532},
 primaryClass = {astro-ph.GA},
       adsurl = {https://ui.adsabs.harvard.edu/abs/2015A&A...575A..96G}
}

@ARTICLE{Akino2022,
       author = {{Akino}, Daichi and {Eckert}, Dominique and {Okabe}, Nobuhiro and {Sereno}, Mauro and {Umetsu}, Keiichi and {Oguri}, Masamune and {Gastaldello}, Fabio and {Chiu}, I. -Non and {Ettori}, Stefano and {Evrard}, August E. and {Farahi}, Arya and {Maughan}, Ben and {Pierre}, Marguerite and {Ricci}, Marina and {Valtchanov}, Ivan and {McCarthy}, Ian and {McGee}, Sean and {Miyazaki}, Satoshi and {Nishizawa}, Atsushi J. and {Tanaka}, Masayuki},
        title = "{HSC-XXL: Baryon budget of the 136 XXL groups and clusters}",
      journal = {\pasj},
         year = 2022,
        month = feb,
       volume = {74},
       number = {1},
        pages = {175-208},
          doi = {10.1093/pasj/psab115},
archivePrefix = {arXiv},
       eprint = {2111.10080},
 primaryClass = {astro-ph.CO},
       adsurl = {https://ui.adsabs.harvard.edu/abs/2022PASJ...74..175A}
}

@ARTICLE{Aird2015,
       author = {{Aird}, J. and {Coil}, A.~L. and {Georgakakis}, A. and {Nandra}, K. and {Barro}, G. and {P{\'e}rez-Gonz{\'a}lez}, P.~G.},
        title = "{The evolution of the X-ray luminosity functions of unabsorbed and absorbed AGNs out to z{\ensuremath{\sim}} 5}",
      journal = {\mnras},
         year = 2015,
        month = aug,
       volume = {451},
       number = {2},
        pages = {1892-1927},
          doi = {10.1093/mnras/stv1062},
archivePrefix = {arXiv},
       eprint = {1503.01120},
 primaryClass = {astro-ph.HE},
       adsurl = {https://ui.adsabs.harvard.edu/abs/2015MNRAS.451.1892A}
}

@ARTICLE{Volonteri2016,
       author = {{Volonteri}, Marta and {Reines}, Amy E.},
        title = "{Inferences on the Relations Between Central Black Hole Mass and Total Galaxy Stellar Mass in the High-redshift Universe}",
      journal = {\apjl},
         year = 2016,
        month = mar,
       volume = {820},
       number = {1},
          eid = {L6},
        pages = {L6},
          doi = {10.3847/2041-8205/820/1/L6},
archivePrefix = {arXiv},
       eprint = {1602.05711},
 primaryClass = {astro-ph.GA},
       adsurl = {https://ui.adsabs.harvard.edu/abs/2016ApJ...820L...6V}
}

@ARTICLE{EPTACollaboration2023_EPTA_GWB,
       author = {{EPTA Collaboration} and {InPTA Collaboration} and {Antoniadis}, J. and {Arumugam}, P. and {Arumugam}, S. and {Babak}, S. and {Bagchi}, M. and {Bak Nielsen}, A. -S. and {Bassa}, C.~G. and {Bathula}, A. and {Berthereau}, A. and {Bonetti}, M. and {Bortolas}, E. and {Brook}, P.~R. and {Burgay}, M. and {Caballero}, R.~N. and {Chalumeau}, A. and {Champion}, D.~J. and {Chanlaridis}, S. and {Chen}, S. and {Cognard}, I. and {Dandapat}, S. and {Deb}, D. and {Desai}, S. and {Desvignes}, G. and {Dhanda-Batra}, N. and {Dwivedi}, C. and {Falxa}, M. and {Ferdman}, R.~D. and {Franchini}, A. and {Gair}, J.~R. and {Goncharov}, B. and {Gopakumar}, A. and {Graikou}, E. and {Grie{\ss}meier}, J. -M. and {Guillemot}, L. and {Guo}, Y.~J. and {Gupta}, Y. and {Hisano}, S. and {Hu}, H. and {Iraci}, F. and {Izquierdo-Villalba}, D. and {Jang}, J. and {Jawor}, J. and {Janssen}, G.~H. and {Jessner}, A. and {Joshi}, B.~C. and {Kareem}, F. and {Karuppusamy}, R. and {Keane}, E.~F. and {Keith}, M.~J. and {Kharbanda}, D. and {Kikunaga}, T. and {Kolhe}, N. and {Kramer}, M. and {Krishnakumar}, M.~A. and {Lackeos}, K. and {Lee}, K.~J. and {Liu}, K. and {Liu}, Y. and {Lyne}, A.~G. and {McKee}, J.~W. and {Maan}, Y. and {Main}, R.~A. and {Mickaliger}, M.~B. and {Ni{\c{t}}u}, I.~C. and {Nobleson}, K. and {Paladi}, A.~K. and {Parthasarathy}, A. and {Perera}, B.~B.~P. and {Perrodin}, D. and {Petiteau}, A. and {Porayko}, N.~K. and {Possenti}, A. and {Prabu}, T. and {Quelquejay Leclere}, H. and {Rana}, P. and {Samajdar}, A. and {Sanidas}, S.~A. and {Sesana}, A. and {Shaifullah}, G. and {Singha}, J. and {Speri}, L. and {Spiewak}, R. and {Srivastava}, A. and {Stappers}, B.~W. and {Surnis}, M. and {Susarla}, S.~C. and {Susobhanan}, A. and {Takahashi}, K. and {Tarafdar}, P. and {Theureau}, G. and {Tiburzi}, C. and {van der Wateren}, E. and {Vecchio}, A. and {Venkatraman Krishnan}, V. and {Verbiest}, J.~P.~W. and {Wang}, J. and {Wang}, L. and {Wu}, Z.},
        title = "{The second data release from the European Pulsar Timing Array. III. Search for gravitational wave signals}",
      journal = {\aap},
         year = 2023,
        month = oct,
       volume = {678},
          eid = {A50},
        pages = {A50},
          doi = {10.1051/0004-6361/202346844},
archivePrefix = {arXiv},
       eprint = {2306.16214},
 primaryClass = {astro-ph.HE},
       adsurl = {https://ui.adsabs.harvard.edu/abs/2023A&A...678A..50E}
}

@ARTICLE{Xu2023_CPTA,
       author = {{Xu}, Heng and {Chen}, Siyuan and {Guo}, Yanjun and {Jiang}, Jinchen and {Wang}, Bojun and {Xu}, Jiangwei and {Xue}, Zihan and {Nicolas Caballero}, R. and {Yuan}, Jianping and {Xu}, Yonghua and {Wang}, Jingbo and {Hao}, Longfei and {Luo}, Jingtao and {Lee}, Kejia and {Han}, Jinlin and {Jiang}, Peng and {Shen}, Zhiqiang and {Wang}, Min and {Wang}, Na and {Xu}, Renxin and {Wu}, Xiangping and {Manchester}, Richard and {Qian}, Lei and {Guan}, Xin and {Huang}, Menglin and {Sun}, Chun and {Zhu}, Yan},
        title = "{Searching for the Nano-Hertz Stochastic Gravitational Wave Background with the Chinese Pulsar Timing Array Data Release I}",
      journal = {Research in Astronomy and Astrophysics},
         year = 2023,
        month = jul,
       volume = {23},
       number = {7},
          eid = {075024},
        pages = {075024},
          doi = {10.1088/1674-4527/acdfa5},
archivePrefix = {arXiv},
       eprint = {2306.16216},
 primaryClass = {astro-ph.HE},
       adsurl = {https://ui.adsabs.harvard.edu/abs/2023RAA....23g5024X}
}

@ARTICLE{Burke-Spolaor2019,
       author = {{Burke-Spolaor}, Sarah and {Taylor}, Stephen R. and {Charisi}, Maria and {Dolch}, Timothy and {Hazboun}, Jeffrey S. and {Holgado}, A. Miguel and {Kelley}, Luke Zoltan and {Lazio}, T. Joseph W. and {Madison}, Dustin R. and {McMann}, Natasha and {Mingarelli}, Chiara M.~F. and {Rasskazov}, Alexander and {Siemens}, Xavier and {Simon}, Joseph J. and {Smith}, Tristan L.},
        title = "{The astrophysics of nanohertz gravitational waves}",
      journal = {\aapr},
         year = 2019,
        month = jun,
       volume = {27},
       number = {1},
          eid = {5},
        pages = {5},
          doi = {10.1007/s00159-019-0115-7},
archivePrefix = {arXiv},
       eprint = {1811.08826},
 primaryClass = {astro-ph.HE},
       adsurl = {https://ui.adsabs.harvard.edu/abs/2019A&ARv..27....5B}
}

@ARTICLE{Rosado2015,
       author = {{Rosado}, Pablo A. and {Sesana}, Alberto and {Gair}, Jonathan},
        title = "{Expected properties of the first gravitational wave signal detected with pulsar timing arrays}",
      journal = {\mnras},
         year = 2015,
        month = aug,
       volume = {451},
       number = {3},
        pages = {2417-2433},
          doi = {10.1093/mnras/stv1098},
archivePrefix = {arXiv},
       eprint = {1503.04803},
 primaryClass = {astro-ph.HE},
       adsurl = {https://ui.adsabs.harvard.edu/abs/2015MNRAS.451.2417R}
}

@ARTICLE{Kelley2018_SS,
       author = {{Kelley}, Luke Zoltan and {Blecha}, Laura and {Hernquist}, Lars and {Sesana}, Alberto and {Taylor}, Stephen R.},
        title = "{Single sources in the low-frequency gravitational wave sky: properties and time to detection by pulsar timing arrays}",
      journal = {\mnras},
         year = 2018,
        month = jun,
       volume = {477},
       number = {1},
        pages = {964-976},
          doi = {10.1093/mnras/sty689},
archivePrefix = {arXiv},
       eprint = {1711.00075},
 primaryClass = {astro-ph.HE},
       adsurl = {https://ui.adsabs.harvard.edu/abs/2018MNRAS.477..964K}
}

@ARTICLE{Ueda2014,
       author = {{Ueda}, Yoshihiro and {Akiyama}, Masayuki and {Hasinger}, G{\"u}nther and {Miyaji}, Takamitsu and {Watson}, Michael G.},
        title = "{Toward the Standard Population Synthesis Model of the X-Ray Background: Evolution of X-Ray Luminosity and Absorption Functions of Active Galactic Nuclei Including Compton-thick Populations}",
      journal = {\apj},
         year = 2014,
        month = may,
       volume = {786},
       number = {2},
          eid = {104},
        pages = {104},
          doi = {10.1088/0004-637X/786/2/104},
archivePrefix = {arXiv},
       eprint = {1402.1836},
 primaryClass = {astro-ph.CO},
       adsurl = {https://ui.adsabs.harvard.edu/abs/2014ApJ...786..104U}
}

@ARTICLE{Habouzit2022,
       author = {{Habouzit}, M{\'e}lanie and {Somerville}, Rachel S. and {Li}, Yuan and {Genel}, Shy and {Aird}, James and {Angl{\'e}s-Alc{\'a}zar}, Daniel and {Dav{\'e}}, Romeel and {Georgiev}, Iskren Y. and {McAlpine}, Stuart and {Rosas-Guevara}, Yetli and {Dubois}, Yohan and {Nelson}, Dylan and {Banados}, Eduardo and {Hernquist}, Lars and {Peirani}, S{\'e}bastien and {Vogelsberger}, Mark},
        title = "{Supermassive black holes in cosmological simulations - II: the AGN population and predictions for upcoming X-ray missions}",
      journal = {\mnras},
         year = 2022,
        month = jan,
       volume = {509},
       number = {2},
        pages = {3015-3042},
          doi = {10.1093/mnras/stab3147},
archivePrefix = {arXiv},
       eprint = {2111.01802},
 primaryClass = {astro-ph.GA},
       adsurl = {https://ui.adsabs.harvard.edu/abs/2022MNRAS.509.3015H}
}

@ARTICLE{Barchiesi2025,
       author = {{Barchiesi}, Luigi and {Carrera}, F.~J. and {Vignali}, C. and {Pozzi}, F. and {Marchetti}, L. and {Gruppioni}, C. and {Delvecchio}, I. and {Bisigello}, L. and {Calura}, F. and {Aird}, J. and {Vaccari}, M.},
        title = "{A New Hope for Obscured AGN: The PRIMA-NewAthena Alliance}",
      journal = {arXiv e-prints},
         year = 2025,
        month = mar,
          eid = {arXiv:2503.19915},
        pages = {arXiv:2503.19915},
          doi = {10.48550/arXiv.2503.19915},
archivePrefix = {arXiv},
       eprint = {2503.19915},
 primaryClass = {astro-ph.GA},
       adsurl = {https://ui.adsabs.harvard.edu/abs/2025arXiv250319915B}
}

@ARTICLE{Hopkins2004,
       author = {{Hopkins}, A.~M.},
        title = "{On the Evolution of Star-forming Galaxies}",
      journal = {\apj},
         year = 2004,
        month = nov,
       volume = {615},
       number = {1},
        pages = {209-221},
          doi = {10.1086/424032},
archivePrefix = {arXiv},
       eprint = {astro-ph/0407170},
 primaryClass = {astro-ph},
       adsurl = {https://ui.adsabs.harvard.edu/abs/2004ApJ...615..209H}
}

@ARTICLE{Hopkins2007,
       author = {{Hopkins}, Philip F. and {Richards}, Gordon T. and {Hernquist}, Lars},
        title = "{An Observational Determination of the Bolometric Quasar Luminosity Function}",
      journal = {\apj},
         year = 2007,
        month = jan,
       volume = {654},
       number = {2},
        pages = {731-753},
          doi = {10.1086/509629},
archivePrefix = {arXiv},
       eprint = {astro-ph/0605678},
 primaryClass = {astro-ph},
       adsurl = {https://ui.adsabs.harvard.edu/abs/2007ApJ...654..731H}
}

@ARTICLE{Churazov2005,
       author = {{Churazov}, E. and {Sazonov}, S. and {Sunyaev}, R. and {Forman}, W. and {Jones}, C. and {B{\"o}hringer}, H.},
        title = "{Supermassive black holes in elliptical galaxies: switching from very bright to very dim}",
      journal = {\mnras},
         year = 2005,
        month = oct,
       volume = {363},
       number = {1},
        pages = {L91-L95},
          doi = {10.1111/j.1745-3933.2005.00093.x},
archivePrefix = {arXiv},
       eprint = {astro-ph/0507073},
 primaryClass = {astro-ph},
       adsurl = {https://ui.adsabs.harvard.edu/abs/2005MNRAS.363L..91C}
}

@ARTICLE{Xiao2011,
       author = {{Xiao}, Ting and {Barth}, Aaron J. and {Greene}, Jenny E. and {Ho}, Luis C. and {Bentz}, Misty C. and {Ludwig}, Randi R. and {Jiang}, Yanfei},
        title = "{Exploring the Low-mass End of the M $_{BH}$-{\ensuremath{\sigma}}$_{*}$ Relation with Active Galaxies}",
      journal = {\apj},
         year = 2011,
        month = sep,
       volume = {739},
       number = {1},
          eid = {28},
        pages = {28},
          doi = {10.1088/0004-637X/739/1/28},
archivePrefix = {arXiv},
       eprint = {1106.6232},
 primaryClass = {astro-ph.CO},
       adsurl = {https://ui.adsabs.harvard.edu/abs/2011ApJ...739...28X}
}

@ARTICLE{Greene2006,
       author = {{Greene}, Jenny E. and {Ho}, Luis C.},
        title = "{The M$_{BH}$-{\ensuremath{\sigma}}$_{*}$ Relation in Local Active Galaxies}",
      journal = {\apjl},
         year = 2006,
        month = apr,
       volume = {641},
       number = {1},
        pages = {L21-L24},
          doi = {10.1086/500507},
archivePrefix = {arXiv},
       eprint = {astro-ph/0512461},
 primaryClass = {astro-ph},
       adsurl = {https://ui.adsabs.harvard.edu/abs/2006ApJ...641L..21G}
}

@ARTICLE{Bentz2018,
       author = {{Bentz}, Misty C. and {Manne-Nicholas}, Emily},
        title = "{Black Hole-Galaxy Scaling Relationships for Active Galactic Nuclei with Reverberation Masses}",
      journal = {\apj},
         year = 2018,
        month = sep,
       volume = {864},
       number = {2},
          eid = {146},
        pages = {146},
          doi = {10.3847/1538-4357/aad808},
archivePrefix = {arXiv},
       eprint = {1808.01329},
 primaryClass = {astro-ph.GA},
       adsurl = {https://ui.adsabs.harvard.edu/abs/2018ApJ...864..146B}
}

@ARTICLE{Tremmel2017,
       author = {{Tremmel}, M. and {Karcher}, M. and {Governato}, F. and {Volonteri}, M. and {Quinn}, T.~R. and {Pontzen}, A. and {Anderson}, L. and {Bellovary}, J.},
        title = "{The Romulus cosmological simulations: a physical approach to the formation, dynamics and accretion models of SMBHs}",
      journal = {\mnras},
         year = 2017,
        month = sep,
       volume = {470},
       number = {1},
        pages = {1121-1139},
          doi = {10.1093/mnras/stx1160},
archivePrefix = {arXiv},
       eprint = {1607.02151},
 primaryClass = {astro-ph.GA},
       adsurl = {https://ui.adsabs.harvard.edu/abs/2017MNRAS.470.1121T}
}

@ARTICLE{Bellovary2011,
       author = {{Bellovary}, Jillian and {Volonteri}, Marta and {Governato}, Fabio and {Shen}, Sijing and {Quinn}, Thomas and {Wadsley}, James},
        title = "{The First Massive Black Hole Seeds and Their Hosts}",
      journal = {\apj},
         year = 2011,
        month = nov,
       volume = {742},
       number = {1},
          eid = {13},
        pages = {13},
          doi = {10.1088/0004-637X/742/1/13},
archivePrefix = {arXiv},
       eprint = {1104.3858},
 primaryClass = {astro-ph.CO},
       adsurl = {https://ui.adsabs.harvard.edu/abs/2011ApJ...742...13B}
}

@ARTICLE{Weinberger2017_tng,
       author = {{Weinberger}, Rainer and {Springel}, Volker and {Hernquist}, Lars and {Pillepich}, Annalisa and {Marinacci}, Federico and {Pakmor}, R{\"u}diger and {Nelson}, Dylan and {Genel}, Shy and {Vogelsberger}, Mark and {Naiman}, Jill and {Torrey}, Paul},
        title = "{Simulating galaxy formation with black hole driven thermal and kinetic feedback}",
      journal = {\mnras},
         year = 2017,
        month = mar,
       volume = {465},
       number = {3},
        pages = {3291-3308},
          doi = {10.1093/mnras/stw2944},
archivePrefix = {arXiv},
       eprint = {1607.03486},
 primaryClass = {astro-ph.GA},
       adsurl = {https://ui.adsabs.harvard.edu/abs/2017MNRAS.465.3291W}
}

@ARTICLE{BondiHoyle1944,
   author = {{Bondi}, H. and {Hoyle}, F.},
    title = "{On the mechanism of accretion by stars}",
  journal = {\mnras},
     year = 1944,
   volume = 104,
    pages = {273-+},
   adsurl = {http://adsabs.harvard.edu/abs/1944MNRAS.104..273B}
}

@ARTICLE{Springel2001,
       author = {{Springel}, Volker and {White}, Simon D.~M. and {Tormen}, Giuseppe and {Kauffmann}, Guinevere},
        title = "{Populating a cluster of galaxies - I. Results at z=0}",
      journal = {\mnras},
         year = 2001,
        month = dec,
       volume = {328},
       number = {3},
        pages = {726-750},
          doi = {10.1046/j.1365-8711.2001.04912.x},
archivePrefix = {arXiv},
       eprint = {astro-ph/0012055},
 primaryClass = {astro-ph},
       adsurl = {https://ui.adsabs.harvard.edu/abs/2001MNRAS.328..726S}
}

@ARTICLE{Davis1985,
       author = {{Davis}, M. and {Efstathiou}, G. and {Frenk}, C.~S. and {White}, S.~D.~M.},
        title = "{The evolution of large-scale structure in a universe dominated by cold dark matter}",
      journal = {\apj},
         year = 1985,
        month = may,
       volume = {292},
        pages = {371-394},
          doi = {10.1086/163168},
       adsurl = {https://ui.adsabs.harvard.edu/abs/1985ApJ...292..371D}
}

@ARTICLE{Weinmann2011,
       author = {{Weinmann}, Simone M. and {Lisker}, Thorsten and {Guo}, Qi and {Meyer}, Hagen T. and {Janz}, Joachim},
        title = "{Dwarf galaxy populations in present-day galaxy clusters - I. Abundances and red fractions}",
      journal = {\mnras},
         year = 2011,
        month = sep,
       volume = {416},
       number = {2},
        pages = {1197-1214},
          doi = {10.1111/j.1365-2966.2011.19118.x},
archivePrefix = {arXiv},
       eprint = {1105.0674},
 primaryClass = {astro-ph.CO},
       adsurl = {https://ui.adsabs.harvard.edu/abs/2011MNRAS.416.1197W}
}

@ARTICLE{Chen2025,
       author = {{Chen}, Nianyi and {Di Matteo}, Tiziana and {Zhou}, Yihao and {Kelley}, Luke Zoltan and {Blecha}, Laura and {Ni}, Yueying and {Bird}, Simeon and {Yang}, Yanhui and {Croft}, Rupert},
        title = "{The Gravitational Wave Background from Massive Black Holes in the ASTRID Simulation}",
      journal = {arXiv e-prints},
         year = 2025,
        month = feb,
          eid = {arXiv:2502.01024},
        pages = {arXiv:2502.01024},
          doi = {10.48550/arXiv.2502.01024},
archivePrefix = {arXiv},
       eprint = {2502.01024},
 primaryClass = {astro-ph.GA},
       adsurl = {https://ui.adsabs.harvard.edu/abs/2025arXiv250201024C}
}

@ARTICLE{Wang2025,
       author = {{Wang}, Bonny Y. and {Zhou}, Yihao and {Chen}, William and {Chen}, Nianyi and {Di Matteo}, Tiziana and {Croft}, Rupert and {Bird}, Simeon and {Ni}, Yueying},
        title = "{Gravitational Waves from Massive Black Hole Mergers in ASTRID: Predictions for LISA}",
      journal = {arXiv e-prints},
         year = 2025,
        month = mar,
          eid = {arXiv:2503.24304},
        pages = {arXiv:2503.24304},
          doi = {10.48550/arXiv.2503.24304},
archivePrefix = {arXiv},
       eprint = {2503.24304},
 primaryClass = {astro-ph.GA},
       adsurl = {https://ui.adsabs.harvard.edu/abs/2025arXiv250324304W}
}

@ARTICLE{Naiman2018,
       author = {{Naiman}, Jill P. and {Pillepich}, Annalisa and {Springel}, Volker and {Ramirez-Ruiz}, Enrico and {Torrey}, Paul and {Vogelsberger}, Mark and {Pakmor}, R{\"u}diger and {Nelson}, Dylan and {Marinacci}, Federico and {Hernquist}, Lars and {Weinberger}, Rainer and {Genel}, Shy},
        title = "{First results from the IllustrisTNG simulations: a tale of two elements - chemical evolution of magnesium and europium}",
      journal = {\mnras},
         year = 2018,
        month = jun,
       volume = {477},
       number = {1},
        pages = {1206-1224},
          doi = {10.1093/mnras/sty618},
archivePrefix = {arXiv},
       eprint = {1707.03401},
 primaryClass = {astro-ph.GA},
       adsurl = {https://ui.adsabs.harvard.edu/abs/2018MNRAS.477.1206N}
}

@ARTICLE{Bigwood2025_xfable,
       author = {{Bigwood}, Leah and {Bourne}, Martin A. and {Ir{\v{s}}i{\v{c}}}, Vid and {Amon}, Alexandra and {Sijacki}, Debora},
        title = "{The case for large-scale AGN feedback in galaxy formation simulations: insights from XFABLE}",
      journal = {\mnras},
         year = 2025,
        month = sep,
          doi = {10.1093/mnras/staf1435},
archivePrefix = {arXiv},
       eprint = {2501.16983},
 primaryClass = {astro-ph.CO},
       adsurl = {https://ui.adsabs.harvard.edu/abs/2025MNRAS.tmp.1393B}
}

@ARTICLE{Bressan2012_parsec,
       author = {{Bressan}, Alessandro and {Marigo}, Paola and {Girardi}, L{\'e}o. and {Salasnich}, Bernardo and {Dal Cero}, Claudia and {Rubele}, Stefano and {Nanni}, Ambra},
        title = "{PARSEC: stellar tracks and isochrones with the PAdova and TRieste Stellar Evolution Code}",
      journal = {\mnras},
         year = 2012,
        month = nov,
       volume = {427},
       number = {1},
        pages = {127-145},
          doi = {10.1111/j.1365-2966.2012.21948.x},
archivePrefix = {arXiv},
       eprint = {1208.4498},
 primaryClass = {astro-ph.SR},
       adsurl = {https://ui.adsabs.harvard.edu/abs/2012MNRAS.427..127B}
}

@ARTICLE{Sanchez-Blazquez2006_miles,
       author = {{S{\'a}nchez-Bl{\'a}zquez}, P. and {Peletier}, R.~F. and {Jim{\'e}nez-Vicente}, J. and {Cardiel}, N. and {Cenarro}, A.~J. and {Falc{\'o}n-Barroso}, J. and {Gorgas}, J. and {Selam}, S. and {Vazdekis}, A.},
        title = "{Medium-resolution Isaac Newton Telescope library of empirical spectra}",
      journal = {\mnras},
         year = 2006,
        month = sep,
       volume = {371},
       number = {2},
        pages = {703-718},
          doi = {10.1111/j.1365-2966.2006.10699.x},
archivePrefix = {arXiv},
       eprint = {astro-ph/0607009},
 primaryClass = {astro-ph},
       adsurl = {https://ui.adsabs.harvard.edu/abs/2006MNRAS.371..703S}
}

@ARTICLE{Kauffmann2003,
       author = {{Kauffmann}, Guinevere and {Heckman}, Timothy M. and {White}, Simon D.~M. and {Charlot}, St{\'e}phane and {Tremonti}, Christy and {Brinchmann}, Jarle and {Bruzual}, Gustavo and {Peng}, Eric W. and {Seibert}, Mark and {Bernardi}, Mariangela and {Blanton}, Michael and {Brinkmann}, Jon and {Castander}, Francisco and {Cs{\'a}bai}, Istvan and {Fukugita}, Masataka and {Ivezic}, Zeljko and {Munn}, Jeffrey A. and {Nichol}, Robert C. and {Padmanabhan}, Nikhil and {Thakar}, Aniruddha R. and {Weinberg}, David H. and {York}, Donald},
        title = "{Stellar masses and star formation histories for {}10$^{5}$ galaxies from the Sloan Digital Sky Survey}",
      journal = {\mnras},
         year = 2003,
        month = may,
       volume = {341},
       number = {1},
        pages = {33-53},
          doi = {10.1046/j.1365-8711.2003.06291.x},
archivePrefix = {arXiv},
       eprint = {astro-ph/0204055},
 primaryClass = {astro-ph},
       adsurl = {https://ui.adsabs.harvard.edu/abs/2003MNRAS.341...33K}
}

@ARTICLE{Baldry2006,
       author = {{Baldry}, I.~K. and {Balogh}, M.~L. and {Bower}, R.~G. and {Glazebrook}, K. and {Nichol}, R.~C. and {Bamford}, S.~P. and {Budavari}, T.},
        title = "{Galaxy bimodality versus stellar mass and environment}",
      journal = {\mnras},
         year = 2006,
        month = dec,
       volume = {373},
       number = {2},
        pages = {469-483},
          doi = {10.1111/j.1365-2966.2006.11081.x},
archivePrefix = {arXiv},
       eprint = {astro-ph/0607648},
 primaryClass = {astro-ph},
       adsurl = {https://ui.adsabs.harvard.edu/abs/2006MNRAS.373..469B}
}

@ARTICLE{Henden2018_fable,
       author = {{Henden}, Nicholas A. and {Puchwein}, Ewald and {Shen}, Sijing and {Sijacki}, Debora},
        title = "{The FABLE simulations: a feedback model for galaxies, groups, and clusters}",
      journal = {\mnras},
         year = 2018,
        month = oct,
       volume = {479},
       number = {4},
        pages = {5385-5412},
          doi = {10.1093/mnras/sty1780},
archivePrefix = {arXiv},
       eprint = {1804.05064},
 primaryClass = {astro-ph.GA},
       adsurl = {https://ui.adsabs.harvard.edu/abs/2018MNRAS.479.5385H}
}

@ARTICLE{Marinacci2018,
       author = {{Marinacci}, Federico and {Vogelsberger}, Mark and {Pakmor}, R{\"u}diger and {Torrey}, Paul and {Springel}, Volker and {Hernquist}, Lars and {Nelson}, Dylan and {Weinberger}, Rainer and {Pillepich}, Annalisa and {Naiman}, Jill and {Genel}, Shy},
        title = "{First results from the IllustrisTNG simulations: radio haloes and magnetic fields}",
      journal = {\mnras},
         year = 2018,
        month = nov,
       volume = {480},
       number = {4},
        pages = {5113-5139},
          doi = {10.1093/mnras/sty2206},
archivePrefix = {arXiv},
       eprint = {1707.03396},
 primaryClass = {astro-ph.CO},
       adsurl = {https://ui.adsabs.harvard.edu/abs/2018MNRAS.480.5113M}
}

@ARTICLE{Dolag2017,
       author = {{Dolag}, Klaus and {Mevius}, Emilio and {Remus}, Rhea-Silvia},
        title = "{Distribution and Evolution of Metals in the Magneticum Simulations}",
      journal = {Galaxies},
         year = 2017,
        month = aug,
       volume = {5},
       number = {3},
          eid = {35},
        pages = {35},
          doi = {10.3390/galaxies5030035},
archivePrefix = {arXiv},
       eprint = {1708.00027},
 primaryClass = {astro-ph.GA},
       adsurl = {https://ui.adsabs.harvard.edu/abs/2017Galax...5...35D}
}

@ARTICLE{Schaye2023,
       author = {{Schaye}, Joop and {Kugel}, Roi and {Schaller}, Matthieu and {Helly}, John C. and {Braspenning}, Joey and {Elbers}, Willem and {McCarthy}, Ian G. and {van Daalen}, Marcel P. and {Vandenbroucke}, Bert and {Frenk}, Carlos S. and {Kwan}, Juliana and {Salcido}, Jaime and {Bah{\'e}}, Yannick M. and {Borrow}, Josh and {Chaikin}, Evgenii and {Hahn}, Oliver and {Hu{\v{s}}ko}, Filip and {Jenkins}, Adrian and {Lacey}, Cedric G. and {Nobels}, Folkert S.~J.},
        title = "{The FLAMINGO project: cosmological hydrodynamical simulations for large-scale structure and galaxy cluster surveys}",
      journal = {\mnras},
         year = 2023,
        month = dec,
       volume = {526},
       number = {4},
        pages = {4978-5020},
          doi = {10.1093/mnras/stad2419},
archivePrefix = {arXiv},
       eprint = {2306.04024},
 primaryClass = {astro-ph.CO},
       adsurl = {https://ui.adsabs.harvard.edu/abs/2023MNRAS.526.4978S}
}

@ARTICLE{Li2009,
       author = {{Li}, Cheng and {White}, Simon D.~M.},
        title = "{The distribution of stellar mass in the low-redshift Universe}",
      journal = {\mnras},
         year = 2009,
        month = oct,
       volume = {398},
       number = {4},
        pages = {2177-2187},
          doi = {10.1111/j.1365-2966.2009.15268.x},
archivePrefix = {arXiv},
       eprint = {0901.0706},
 primaryClass = {astro-ph.CO},
       adsurl = {https://ui.adsabs.harvard.edu/abs/2009MNRAS.398.2177L}
}

@ARTICLE{Bernardi2013,
       author = {{Bernardi}, M. and {Meert}, A. and {Sheth}, R.~K. and {Vikram}, V. and {Huertas-Company}, M. and {Mei}, S. and {Shankar}, F.},
        title = "{The massive end of the luminosity and stellar mass functions: dependence on the fit to the light profile}",
      journal = {\mnras},
         year = 2013,
        month = nov,
       volume = {436},
       number = {1},
        pages = {697-704},
          doi = {10.1093/mnras/stt1607},
archivePrefix = {arXiv},
       eprint = {1304.7778},
 primaryClass = {astro-ph.CO},
       adsurl = {https://ui.adsabs.harvard.edu/abs/2013MNRAS.436..697B}
}

@ARTICLE{Rodriguez-Puebla2017,
       author = {{Rodr{\'\i}guez-Puebla}, Aldo and {Primack}, Joel R. and {Avila-Reese}, Vladimir and {Faber}, S.~M.},
        title = "{Constraining the galaxy-halo connection over the last 13.3 Gyr: star formation histories, galaxy mergers and structural properties}",
      journal = {\mnras},
         year = 2017,
        month = sep,
       volume = {470},
       number = {1},
        pages = {651-687},
          doi = {10.1093/mnras/stx1172},
archivePrefix = {arXiv},
       eprint = {1703.04542},
 primaryClass = {astro-ph.GA},
       adsurl = {https://ui.adsabs.harvard.edu/abs/2017MNRAS.470..651R}
}

@ARTICLE{Shankar2009,
       author = {{Shankar}, Francesco and {Weinberg}, David H. and {Miralda-Escud{\'e}}, Jordi},
        title = "{Self-Consistent Models of the AGN and Black Hole Populations: Duty Cycles, Accretion Rates, and the Mean Radiative Efficiency}",
      journal = {\apj},
         year = 2009,
        month = jan,
       volume = {690},
       number = {1},
        pages = {20-41},
          doi = {10.1088/0004-637X/690/1/20},
archivePrefix = {arXiv},
       eprint = {0710.4488},
 primaryClass = {astro-ph},
       adsurl = {https://ui.adsabs.harvard.edu/abs/2009ApJ...690...20S}
}

@ARTICLE{Mingarelli2017,
       author = {{Mingarelli}, Chiara M.~F. and {Lazio}, T. Joseph W. and {Sesana}, Alberto and {Greene}, Jenny E. and {Ellis}, Justin A. and {Ma}, Chung-Pei and {Croft}, Steve and {Burke-Spolaor}, Sarah and {Taylor}, Stephen R.},
        title = "{The local nanohertz gravitational-wave landscape from supermassive black hole binaries}",
      journal = {Nature Astronomy},
         year = 2017,
        month = nov,
       volume = {1},
        pages = {886-892},
          doi = {10.1038/s41550-017-0299-6},
archivePrefix = {arXiv},
       eprint = {1708.03491},
 primaryClass = {astro-ph.GA},
       adsurl = {https://ui.adsabs.harvard.edu/abs/2017NatAs...1..886M}
}

@ARTICLE{Katz2020,
       author = {{Katz}, Michael L. and {Kelley}, Luke Zoltan and {Dosopoulou}, Fani and {Berry}, Samantha and {Blecha}, Laura and {Larson}, Shane L.},
        title = "{Probing massive black hole binary populations with LISA}",
      journal = {\mnras},
         year = 2020,
        month = jan,
       volume = {491},
       number = {2},
        pages = {2301-2317},
          doi = {10.1093/mnras/stz3102},
archivePrefix = {arXiv},
       eprint = {1908.05779},
 primaryClass = {astro-ph.HE},
       adsurl = {https://ui.adsabs.harvard.edu/abs/2020MNRAS.491.2301K}
}

@ARTICLE{Vogelsberger2020,
       author = {{Vogelsberger}, Mark and {Marinacci}, Federico and {Torrey}, Paul and {Puchwein}, Ewald},
        title = "{Cosmological simulations of galaxy formation}",
      journal = {Nature Reviews Physics},
         year = 2020,
        month = jan,
       volume = {2},
       number = {1},
        pages = {42-66},
          doi = {10.1038/s42254-019-0127-2},
archivePrefix = {arXiv},
       eprint = {1909.07976},
 primaryClass = {astro-ph.GA},
       adsurl = {https://ui.adsabs.harvard.edu/abs/2020NatRP...2...42V}
}

@ARTICLE{Dubois2021,
       author = {{Dubois}, Yohan and {Beckmann}, Ricarda and {Bournaud}, Fr{\'e}d{\'e}ric and {Choi}, Hoseung and {Devriendt}, Julien and {Jackson}, Ryan and {Kaviraj}, Sugata and {Kimm}, Taysun and {Kraljic}, Katarina and {Laigle}, Clotilde and {Martin}, Garreth and {Park}, Min-Jung and {Peirani}, S{\'e}bastien and {Pichon}, Christophe and {Volonteri}, Marta and {Yi}, Sukyoung K.},
        title = "{Introducing the NEWHORIZON simulation: Galaxy properties with resolved internal dynamics across cosmic time}",
      journal = {\aap},
         year = 2021,
        month = jul,
       volume = {651},
          eid = {A109},
        pages = {A109},
          doi = {10.1051/0004-6361/202039429},
archivePrefix = {arXiv},
       eprint = {2009.10578},
 primaryClass = {astro-ph.GA},
       adsurl = {https://ui.adsabs.harvard.edu/abs/2021A&A...651A.109D}
}

@ARTICLE{Dubois2016,
       author = {{Dubois}, Yohan and {Peirani}, S{\'e}bastien and {Pichon}, Christophe and {Devriendt}, Julien and {Gavazzi}, Rapha{\"e}l and {Welker}, Charlotte and {Volonteri}, Marta},
        title = "{The HORIZON-AGN simulation: morphological diversity of galaxies promoted by AGN feedback}",
      journal = {\mnras},
         year = 2016,
        month = dec,
       volume = {463},
       number = {4},
        pages = {3948-3964},
          doi = {10.1093/mnras/stw2265},
archivePrefix = {arXiv},
       eprint = {1606.03086},
 primaryClass = {astro-ph.GA},
       adsurl = {https://ui.adsabs.harvard.edu/abs/2016MNRAS.463.3948D}
}

@ARTICLE{Springel2018,
       author = {{Springel}, Volker and {Pakmor}, R{\"u}diger and {Pillepich}, Annalisa and {Weinberger}, Rainer and {Nelson}, Dylan and {Hernquist}, Lars and {Vogelsberger}, Mark and {Genel}, Shy and {Torrey}, Paul and {Marinacci}, Federico and {Naiman}, Jill},
        title = "{First results from the IllustrisTNG simulations: matter and galaxy clustering}",
      journal = {\mnras},
         year = 2018,
        month = mar,
       volume = {475},
       number = {1},
        pages = {676-698},
          doi = {10.1093/mnras/stx3304},
archivePrefix = {arXiv},
       eprint = {1707.03397},
 primaryClass = {astro-ph.GA},
       adsurl = {https://ui.adsabs.harvard.edu/abs/2018MNRAS.475..676S}
}

@ARTICLE{Genel2018,
       author = {{Genel}, Shy and {Nelson}, Dylan and {Pillepich}, Annalisa and {Springel}, Volker and {Pakmor}, R{\"u}diger and {Weinberger}, Rainer and {Hernquist}, Lars and {Naiman}, Jill and {Vogelsberger}, Mark and {Marinacci}, Federico and {Torrey}, Paul},
        title = "{The size evolution of star-forming and quenched galaxies in the IllustrisTNG simulation}",
      journal = {\mnras},
         year = 2018,
        month = mar,
       volume = {474},
       number = {3},
        pages = {3976-3996},
          doi = {10.1093/mnras/stx3078},
archivePrefix = {arXiv},
       eprint = {1707.05327},
 primaryClass = {astro-ph.GA},
       adsurl = {https://ui.adsabs.harvard.edu/abs/2018MNRAS.474.3976G}
}

@ARTICLE{Shen2020,
       author = {{Shen}, Xuejian and {Hopkins}, Philip F. and {Faucher-Gigu{\`e}re}, Claude-Andr{\'e} and {Alexander}, D.~M. and {Richards}, Gordon T. and {Ross}, Nicholas P. and {Hickox}, R.~C.},
        title = "{The bolometric quasar luminosity function at z = 0-7}",
      journal = {\mnras},
         year = 2020,
        month = jan,
       volume = {495},
       number = {3},
        pages = {3252-3275},
          doi = {10.1093/mnras/staa1381},
archivePrefix = {arXiv},
       eprint = {2001.02696},
 primaryClass = {astro-ph.GA},
       adsurl = {https://ui.adsabs.harvard.edu/abs/2020MNRAS.495.3252S}
}

@ARTICLE{Driver2022,
       author = {{Driver}, Simon P. and {Bellstedt}, Sabine and {Robotham}, Aaron S.~G. and {Baldry}, Ivan K. and {Davies}, Luke J. and {Liske}, Jochen and {Obreschkow}, Danail and {Taylor}, Edward N. and {Wright}, Angus H. and {Alpaslan}, Mehmet and {Bamford}, Steven P. and {Bauer}, Amanda E. and {Bland-Hawthorn}, Joss and {Bilicki}, Maciej and {Bravo}, Mat{\'\i}as and {Brough}, Sarah and {Casura}, Sarah and {Cluver}, Michelle E. and {Colless}, Matthew and {Conselice}, Christopher J. and {Croom}, Scott M. and {de Jong}, Jelte and {D'Eugenio}, Franceso and {De Propris}, Roberto and {Dogruel}, Burak and {Drinkwater}, Michael J. and {Dvornik}, Andrej and {Farrow}, Daniel J. and {Frenk}, Carlos S. and {Giblin}, Benjamin and {Graham}, Alister W. and {Grootes}, Meiert W. and {Gunawardhana}, Madusha L.~P. and {Hashemizadeh}, Abdolhosein and {H{\"a}u{\ss}ler}, Boris and {Heymans}, Catherine and {Hildebrandt}, Hendrik and {Holwerda}, Benne W. and {Hopkins}, Andrew M. and {Jarrett}, Tom H. and {Heath Jones}, D. and {Kelvin}, Lee S. and {Koushan}, Soheil and {Kuijken}, Konrad and {Lara-L{\'o}pez}, Maritza A. and {Lange}, Rebecca and {L{\'o}pez-S{\'a}nchez}, {\'A}ngel R. and {Loveday}, Jon and {Mahajan}, Smriti and {Meyer}, Martin and {Moffett}, Amanda J. and {Napolitano}, Nicola R. and {Norberg}, Peder and {Owers}, Matt S. and {Radovich}, Mario and {Raouf}, Mojtaba and {Peacock}, John A. and {Phillipps}, Steven and {Pimbblet}, Kevin A. and {Popescu}, Cristina and {Said}, Khaled and {Sansom}, Anne E. and {Seibert}, Mark and {Sutherland}, Will J. and {Thorne}, Jessica E. and {Tuffs}, Richard J. and {Turner}, Ryan and {van der Wel}, Arjen and {van Kampen}, Eelco and {Wilkins}, Steve M.},
        title = "{Galaxy And Mass Assembly (GAMA): Data Release 4 and the z < 0.1 total and z < 0.08 morphological galaxy stellar mass functions}",
      journal = {\mnras},
         year = 2022,
        month = jun,
       volume = {513},
       number = {1},
        pages = {439-467},
          doi = {10.1093/mnras/stac472},
archivePrefix = {arXiv},
       eprint = {2203.08539},
 primaryClass = {astro-ph.GA},
       adsurl = {https://ui.adsabs.harvard.edu/abs/2022MNRAS.513..439D}
}

@ARTICLE{Buchner2015,
       author = {{Buchner}, Johannes and {Georgakakis}, Antonis and {Nandra}, Kirpal and {Brightman}, Murray and {Menzel}, Marie-Luise and {Liu}, Zhu and {Hsu}, Li-Ting and {Salvato}, Mara and {Rangel}, Cyprian and {Aird}, James and {Merloni}, Andrea and {Ross}, Nicholas},
        title = "{Obscuration-dependent Evolution of Active Galactic Nuclei}",
      journal = {\apj},
         year = 2015,
        month = apr,
       volume = {802},
       number = {2},
          eid = {89},
        pages = {89},
          doi = {10.1088/0004-637X/802/2/89},
archivePrefix = {arXiv},
       eprint = {1501.02805},
 primaryClass = {astro-ph.HE},
       adsurl = {https://ui.adsabs.harvard.edu/abs/2015ApJ...802...89B}
}

@ARTICLE{Ragusa2023,
       author = {{Ragusa}, R. and {Iodice}, E. and {Spavone}, M. and {Montes}, M. and {Forbes}, D.~A. and {Brough}, S. and {Mirabile}, M. and {Cantiello}, M. and {Paolillo}, M. and {Schipani}, P.},
        title = "{Does the virial mass drive the intra-cluster light?. Relationship between the ICL and M$_{vir}$ from VEGAS}",
      journal = {\aap},
         year = 2023,
        month = feb,
       volume = {670},
          eid = {L20},
        pages = {L20},
          doi = {10.1051/0004-6361/202245530},
archivePrefix = {arXiv},
       eprint = {2212.06164},
 primaryClass = {astro-ph.GA},
       adsurl = {https://ui.adsabs.harvard.edu/abs/2023A&A...670L..20R}
}

@ARTICLE{Ragusa2021,
       author = {{Ragusa}, Rossella and {Spavone}, Marilena and {Iodice}, Enrichetta and {Brough}, Sarah and {Raj}, Maria Angela and {Paolillo}, Maurizio and {Cantiello}, Michele and {Forbes}, Duncan A. and {La Marca}, Antonio and {D'Ago}, Giuseppe and {Rampazzo}, Roberto and {Schipani}, Pietro},
        title = "{VEGAS: A VST Early-type GAlaxy Survey. VI. Diffuse light in HCG 86 as seen from the ultra-deep VEGAS images}",
      journal = {\aap},
         year = 2021,
        month = jul,
       volume = {651},
          eid = {A39},
        pages = {A39},
          doi = {10.1051/0004-6361/202039921},
archivePrefix = {arXiv},
       eprint = {2105.06970},
 primaryClass = {astro-ph.GA},
       adsurl = {https://ui.adsabs.harvard.edu/abs/2021A&A...651A..39R}
}

@ARTICLE{Cooray2002,
       author = {{Cooray}, Asantha and {Sheth}, Ravi},
        title = "{Halo models of large scale structure}",
      journal = {\physrep},
         year = 2002,
        month = dec,
       volume = {372},
       number = {1},
        pages = {1-129},
          doi = {10.1016/S0370-1573(02)00276-4},
archivePrefix = {arXiv},
       eprint = {astro-ph/0206508},
 primaryClass = {astro-ph},
       adsurl = {https://ui.adsabs.harvard.edu/abs/2002PhR...372....1C}
}

@ARTICLE{Kluge2021,
       author = {{Kluge}, M. and {Bender}, R. and {Riffeser}, A. and {Goessl}, C. and {Hopp}, U. and {Schmidt}, M. and {Ries}, C.},
        title = "{Photometric Dissection of Intracluster Light and Its Correlations with Host Cluster Properties}",
      journal = {\apjs},
         year = 2021,
        month = feb,
       volume = {252},
       number = {2},
          eid = {27},
        pages = {27},
          doi = {10.3847/1538-4365/abcda6},
archivePrefix = {arXiv},
       eprint = {2011.12992},
 primaryClass = {astro-ph.GA},
       adsurl = {https://ui.adsabs.harvard.edu/abs/2021ApJS..252...27K}
}

@ARTICLE{Montes2021,
       author = {{Montes}, Mireia and {Brough}, Sarah and {Owers}, Matt S. and {Santucci}, Giulia},
        title = "{The Buildup of the Intracluster Light of A85 as Seen by Subaru's Hyper Suprime-Cam}",
      journal = {\apj},
         year = 2021,
        month = mar,
       volume = {910},
       number = {1},
          eid = {45},
        pages = {45},
          doi = {10.3847/1538-4357/abddb6},
archivePrefix = {arXiv},
       eprint = {2101.08290},
 primaryClass = {astro-ph.GA},
       adsurl = {https://ui.adsabs.harvard.edu/abs/2021ApJ...910...45M}
}
\bibliographystyle{aasjournal}


\end{CJK*}
\end{document}